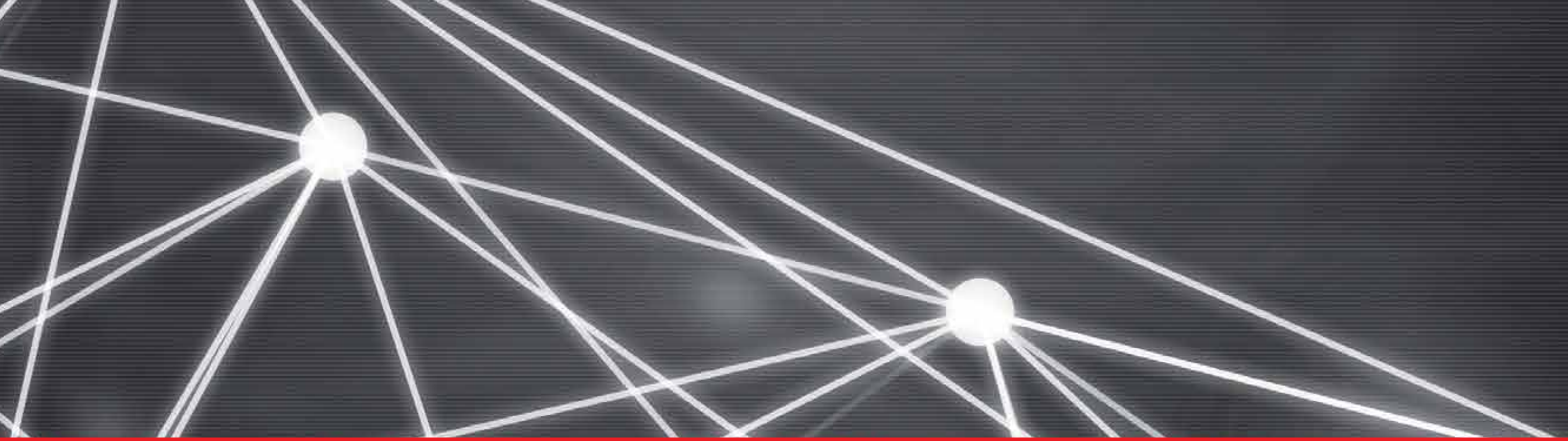

IntechOpen

# Information Security and Privacy in the Digital World

## Some Selected Topics

*Edited by Jaydip Sen and Joceli Mayer*

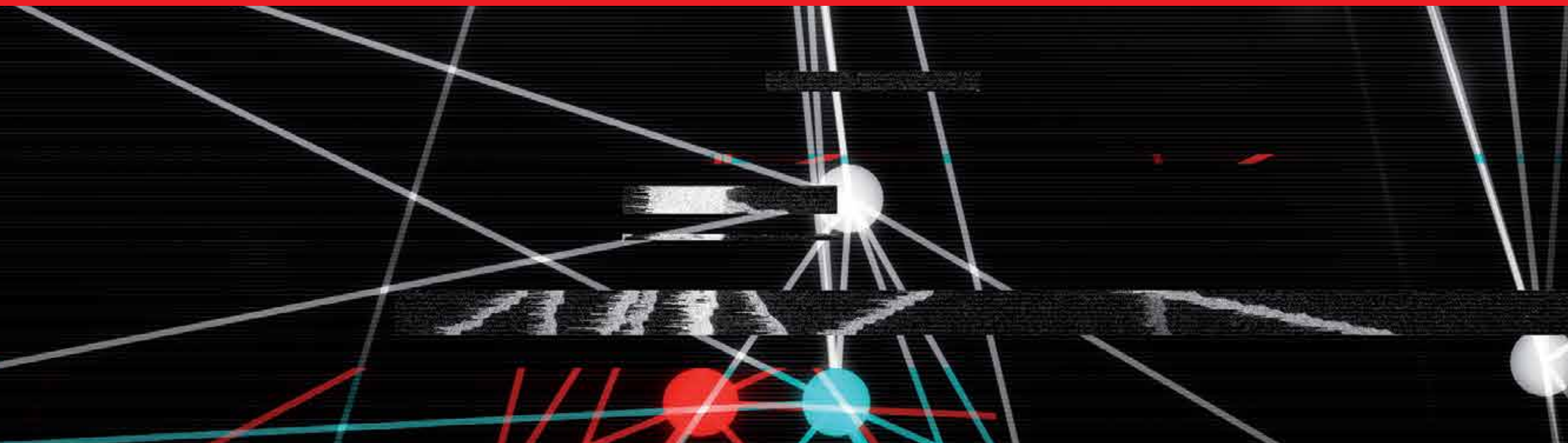

# Information Security and Privacy in the Digital World - Some Selected Topics

*Edited by Jaydip Sen and Joceli Mayer*

Contributors
Abderrahim Bajit, Abdelhadi EI Allali, Ilham Morino, Salma AIT Oussous, Siham Beloualid, Ahmed Tamtaoui, Joceli Mayer, Preethi Nanjundan, Jossy P. George, Pinaki Mitra, Subrata Nandi, Srinivasan Krishnaswamy, Naga Prasanthi Kundeti, Chandrasekhara Rao MVP, Sudha Sree Chekuri, SeshuBabu Pallapothu, Mahendra Pratap Singh, Jaydip Sen, Subhasis Dasgupta

# Meet the editors

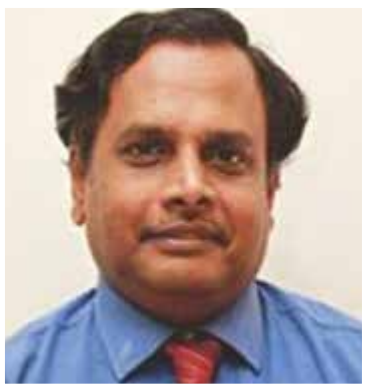

Jaydip Sen is a professor in the Department of Data Science, Praxis Business School, Kolkata, India. His research areas include security and privacy issues in computing and communication, intrusion detection systems, machine learning, deep learning, and artificial intelligence in the financial domain. He has published more than 200 papers in reputed indexed journals, refereed international conference proceedings, and 18 book chapters. He has authored three books and edited ten volumes. He is the editor of *Knowledge Decision Support Systems in Finance* and served on the technical program committees of several high-ranked international conferences of the Institute of Electrical and Electronics Engineers (IEEE) and the Association for Computing Machinery (ACM). He has contributed to several standardization efforts of the IEEE, USA, including the 802.16m standards and the 3GPP LTE standards. Prof. Sen has been listed among the top 2% of scientists in the world by Stanford University for the last four consecutive years (2019–2022).

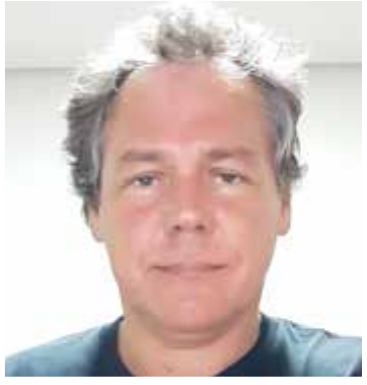

Dr. Joceli Mayer graduated in electrical engineering from the Universidade Federal de Santa Catarina (UFSC), Brazil, in 1998. He received a master's degree in Electrical Engineering from UFSC in 1991, and a master's degree in Computer Engineering and a DPhil from the University of California at Santa Cruz (UCSC), USA, in 1998 and 1999, respectively. He received the Best Student Paper Award from the IEEE International Conference on Image Processing and IBM in 2006 and became an IEEE senior member in 2012. Currently, Dr. Mayer is a Full Professor of Electrical Engineering at UFSC. He has published more than 100 articles in conferences and periodicals, authored two books and two chapters, and advised undergraduate and graduate students on research projects. He has developed and supervised projects on super-resolution, speech compression, VOIP systems, image processing, digital watermarking, hardcopy document authentication, and assistive technology applications for hearing, speech, and mobility-disabled people with the Internet of Things and speech recognition technologies. His research has been supported by industry and government agencies including FINEP, CNPq, Hewlett Packard, and Intelbras, among others.

# Contents

# Preface

With the exponential growth of wireless communications, the Internet of Things (IoT), cloud computing, and the increasingly important roles played by electronic commerce in business and industry, information security and privacy in communication, computing, and storage are increasingly becoming the most critical challenges in technology innovation. Information security is the key prerequisite for the sustained development and successful exploitation of information technology since only a robust and secure information system can ensure access control, the confidentiality of data, user authentication, integrity, non-repudiation, and privacy. However, with the mind-boggling evolution and transformation of the communication and computing world, there has been a paradigm change in the field of information security as well. The context and purview of security have moved from a narrow and bounded environment encompassing a known and disciplined user community to one of worldwide scope with a body of users that is largely unknown and not necessarily trusted. Importantly, security control now must deal with circumstances over which there is largely no control. In this regard, information security exhibits a similarity with liability assurance. They both operate in a threat environment generally known at the top level, including attacks over a broad spectrum of sources. However, the real challenge they face is that the exact details and time of an attack are unknown until an incident occurs. It is not difficult to understand why the field of security has become increasingly complex with the evolution of new communication and computing technologies.

On the other hand, data privacy protection has the objective of safeguarding sensitive personal information from any unauthorized access or accidental disclosure. While access control mechanisms may serve as privacy preservation methods, the issue of data privacy is far more complex, involving ethics and trust. In the era of ubiquitous and pervasive computing and communication, and with the advent of artificial intelligence, applications have the increasing capability of tracking, analyzing, predicting, and even manipulating the behavior of humans. This has created an unprecedented challenge to the technology and systems for privacy protection. For protecting the privacy of sensitive user data, Privacy by Design is a key principle that organizations have started implementing so that the technical and organizational measures and controls are taken care of right at the design phase of any information processing system.

The chapters of this volume highlight different schemes and methods of cryptography, privacy preservation of data, watermarking methods to defend against various attacks on digital artworks, and role-based access controls. The subject matter discussed in these chapters illustrates the complexities involved in the algorithms, protocols, and architectures of different security and privacy protection systems and their various applications in the real world.

In Chapter 1, "Introductory Chapter: Data Privacy Preservation on the Internet of Things", Jaydip Sen and Subhasis Dasgupta present a short survey of the existing machine learning-based approaches and mechanisms for the privacy preservation of data in the IoT. The authors discuss various systems built on centralized architectures and distributed encryption mechanisms for protecting the privacy of sensitive user data in the IoT environment. The chapter also discusses some emerging trends and future directions of research in the field of data privacy.

In Chapter 2, "Adversarial Attacks on Image Classification Models: FGSM and Patch Attacks and Their Impact", Jaydip Sen and Subhasis Dasgupta discuss the concept of adversarial attacks on convolutional neural network-based image classification models. The authors focus on two specific attacks, the fast gradient sign method (FGSM) and the adversarial patch attack, and illustrate how adversely these attacks can affect the accuracy of three powerful pre-trained image classification models.

In Chapter 3, "Recent Results on Some Word Oriented Stream Ciphers: SNOW 1.0, SNOW 2.0 and SNOW 3G", Nandi et al. propose three word-oriented stream ciphers: SNOW 1.0, SNOW 2.0, and SNOW 3G. The authors discuss each protocol's working principle, implementation details, and security analysis. The security analysis made by the authors also includes all possible security vulnerabilities to which these three protocols are exposed. Finally, the authors argue that since SNOW 3G is the protocol for ensuring confidentiality and integrity protection to the users and messages in the telecom networks including 3G, 4G, and 5G systems, a thorough understanding of its security vulnerabilities and crypt-analysis is vital.

In Chapter 4, "Role of Access Control in Information Security: A Security Analysis Approach", Mahendra Pratap Singh presents a propositional logic-based framework for analyzing a role-based access control (RBAC) system that uses machine learning algorithms. The approach proposed by the author establishes relationships between RBAC policies and security policies by mapping them through propositional rules. The chapter demonstrates the effectiveness of the proposition on various datasets with defined RBAC policies.

In Chapter 5, "Enhanced Hybrid Privacy Preserving Data Mining Technique", Kundeti et al. propose a privacy-preserving data mining scheme called enhanced hybrid privacy preserving data mining (EHPPDM). The proposed approach integrates various approaches to privacy protection such as randomization, perturbation, anonymization, and so on. The experimental results show that the proposed hybrid approach is more effective in protecting the privacy of the users compared to the individual approaches.

In Chapter 6, "Review on Watermarking Techniques Aiming Authentication of Digital Image Artistic Works Minted as NFTs into Blockchains", Joceli Mayer presents an overview of various watermarking techniques for defending against attacks on digital artworks such as images, videos, and animations. Digital piracy attacks infringe on the authenticity of such artworks and can have a very detrimental effect. Because of this, the author analyzes the watermarking mechanisms from the point of view of their transparency, robustness, and payload. In particular, the author discusses the fragile watermarking techniques, spread spectrum, and least significant bits technique. Moreover, the author discusses a secure certification protocol for watermarking that is used with NFT minting for ensuring more robustness and security.

In Chapter 7, "Perspective Chapter: Text Watermark Analysis – Concept, Technique, and Applications", Preethi Nanjundan and Jossy P. George discuss various theories and methods involved in text watermarking and their applications. Some of the research challenges in this field including protection of information integrity, information accessibility, preservation of originality of the content, information security, and preservation of the privacy of data are also highlighted.

Finally, in Chapter 8, "Application of Computational Intelligence in Visual Quality Optimization Watermarking and Coding Tools to Improve the Medical IoT Platforms Using ECC Cybersecurity Based CoAP Protocol", Allali et al. investigate the possibility and feasibility of a soft computing-based image watermarking method for an IoT framework in the medical and healthcare domain. The proposed scheme can accommodate multiple concurrent users in securely accessing their privacy-sensitive data in public places like stadiums and marketplaces. The scheme, built of digital watermarking and elliptic curve cryptography (ECC), ensures high data communication and storage security while optimizing execution time, bandwidth needed in communication, energy, and memory requirement.

I am hopeful that researchers, engineers, doctoral students, and faculty members of graduate schools and universities working in cryptography, data privacy protection, authentication, and data integrity will find this volume useful. However, this is not a text for beginners in the fields of cryptography, security, and privacy. The chapters in the volume deal with advanced topics and the readers are expected to have the requisite background knowledge of the topics covered in the book.

I sincerely thank all authors for their valuable contributions. It is their cooperation and scholastic contributions that have made the publication of this book possible. I express my thanks to Publishing Process Manager Mr. Dominik Samardzija at IntechOpen for his support, patience, and cooperation during the long period of the publication of the volume. My sincere thanks also go to Commissioning Editor Ms. Mirena Calmic for having faith in me and delegating to me the critical responsibility of editorship of yet another academic volume. I would also like to acknowledge the support and cooperation I received from my faculty colleagues and graduate students of the School of Data Science, Praxis Business School, Kolkata, India. My family members have always been the sources of my inspiration and motivation. I dedicate this volume to my beloved sister, Ms. Nabanita Sen, who unfortunately left us on 27 September 2021 due to the deadly disease of cancer. My sister had always been my pillar of strength. Finally, I gratefully acknowledge the support and motivation I received from my wife Ms. Nalanda Sen, my daughter Ms. Ritabrata Sen, and my mother Ms. Krishna Sen. Without their support, motivation, and inspiration the publication of this volume would not have been possible.

**Jaydip Sen**
Professor,
Department of Data Science,
Praxis Business School,
Kolkata, India

**Joceli Mayer**
Professor,
Department of Electrical Engineering,
Federal University of Santa Catarina,
Florianópolis, Brazil

Section 1

# Attacks and Defense Mechanisms

**Chapter 1**

# Introductory Chapter: Data Privacy Preservation on the Internet of Things

*Jaydip Sen and Subhasis Dasgupta*

## 1. Introduction

Recent developments in hardware and information technology have enabled the emergence of billions of connected, intelligent devices around the world exchanging information with minimal human involvement. This paradigm, known as the Internet of Things (IoT), is progressing quickly, with an estimated 27 billion devices by 2025 (almost four devices per person) [1, 2]. These smart devices help improve our quality of life, with wearables to monitor health, vehicles that interact with traffic centers and other vehicles to ensure safety, and various home appliances offering comfort. This increase in the number of IoT devices and successful IoT services has generated tremendous data. The International Data Corporation report estimates that by 2025 this data will grow from 4 to 140 zettabytes [3].

However, this humongous volume of data poses growing concerns for user privacy. Gartner predicts approximately 15 billion connected devices will be linked to computing networks by 2022 [4]. These gadgets could be vulnerable, and the massive amounts of unsecured online data create a liability. In addition, users having difficulty controlling the data from their devices has highlighted privacy as a major issue. To guarantee high levels of user data protection, IoT systems must adhere to regulations such as the European Union's *general data protection regulation* (GDRP) of 2018 [5]. GDPR is a law enacted in the European Union that specifies rules for how organizations and companies must use personal data without violating their integrity. These regulation policies focus on giving users control over what is collected, when, and for what purpose. By 2023, the regulators will demand organizations protect consumer privacy rights for more than 5 billion citizens and comply with more than 70% of the GDPR requirements [5].

Traditional privacy protection schemes are insufficient for IoT applications which necessitate new techniques such as distributed cybersecurity controls, models, and decisions that take into account vulnerabilities in system development platforms as well as malicious users and attack surfaces. Machine learning techniques can provide improved detection of novel cyberattacks when dealing with large volumes of data in IoT systems. Furthermore, they can enhance how sensitive data are shared between components to keep them secure. Machine learning-based schemes thus improve the operations related to privacy protection and more effectively comply with the regulations. This chapter presents a survey on the currently existing machine learning-based approaches for the privacy preservation of data in the IoT.

The rest of the chapter is organized as follows. Section 2 identifies some of the existing surveys on the privacy preservation of data in the IoT. Section 3 discusses current privacy schemes for IoT based on a centralized architecture. Section 4 highlights the existing schemes working on the principles of distributed learning. In Section 5, some well-known privacy schemes on distributed encryption mechanisms are discussed. The concept of differential privacy and some schemes working on this principle are presented in Section 6. Finally, the chapter is concluded in Section 7, highlighting some emerging trends in the field of privacy in the IoT.

## 2. Some existing survey works on privacy issues in the IoT

In the literature, several studies have reviewed privacy issues in IoT environments, focusing mostly on threats and attacks on such systems. A comprehensive survey is carried out on various threat models and the classification of various attack types in the context of IoT [6]. The study found that the training dataset used for building the machine learning model for designing the privacy protection system is the most vulnerable to attack. Other sensitive assets are the model, the parameters and hyper-parameters involved, and the model architecture. On the other hand, the sensitive actors are the owners of data, the owners of the model, and the users of the model. Another important observation of this study is that among the machine learning models, the ordinary least square regression model, decision tree, and support vector machine model are the most vulnerable ones. Another recently published paper presented a comprehensive survey on various machine learning and deep learning-based approaches used for protecting user data privacy in the IoT [7].

Many surveys focus on reviewing the mechanisms and models for preserving data privacy. Various issues include differential privacy, homomorphic encryption, and learning architectures and models. In one study, the threats and vulnerabilities of privacy protection systems in IoT are classified into four groups (i) attacks on authentication, (ii) attacks on the components of edge computing, (iii) attacks on the anatomization and perturbation schemes, and (iv) attacks on data summarization [8]. In another survey work, the existing privacy protection systems with centralized architectures and machine learning approaches are analyzed by categorizing the data generated at different layers [9]. Kounoudes and Kapitsaki [10] analyzed several privacy-preservation solutions to determine basic characteristics. The authors proposed a mix of machine learning techniques for providing user protection, along with the policy languages to set user privacy preferences and negotiation techniques that improve services while preserving user rights. Zhu et al.'s survey work included several approaches, including differential privacy, secure multi-party computing, and homomorphic encryption for training models [11]. The authors classified the models based on collaborative or aggregated scenarios to protect user identity or information. Ouadrhiri et al. analyzed the current methods within federated learning environments classifying them into three distinct groups: (i) $k$-anonymity, (ii) $l$-diversity, and (iii) $t$-closeness to protect datasets [12]. The authors also observe that differential privacy-based technologies are mostly used for training the privacy models. This approach, however, suffers from a high computational complexity for the encryption and decryption operations.

## 3. Centralized architecture-based encryption schemes

The data privacy mechanisms and systems under this category use encryption techniques such as homomorphic encryption, attribute access control, multi-party computation, and lightweight cryptography. These approaches are usually resource hungry and involve high computational resources and large memory spaces. On the other hand, homomorphic encryption systems provide a very high level of privacy even when deployed in third-party computations. Researchers designed several variants of homomorphic encryption systems such as partially homomorphic and somewhat homomorphic encryption [13–15]. While somewhat homomorphic encryption systems minimize communication overhead by using a smaller key size, partially homomorphic encryption systems are suitable for lightweight protocols of IoT since they yield shorter ciphertexts.

In building privacy models, the modelers encounter a difficult challenge. While data owners do not want what to expose their sensitive information to untrusted and potentially malicious models, the model owners prefer not to share information about their models as they are valuable assets. As such, classification protocols utilize machine learning classifiers over encrypted data to protect privacy on both sides. Bost et al., De Cock et al., Rahulamathavan et al., Wang et al., Zhu et al., and Jiang et al. all proposed several protocols for privacy-preserving classification using different datasets and models including hyperplane decision, naive Bayes, decision trees, support vector machines, multilayer extreme learning machine among others. These models have yielded an accuracy of results varying between 86% and 98% [16–21]. These efforts also reduce training and execution times compared to traditional deep learning models like convolutional neural networks.

## 4. Distributed learning-based solutions

Of late, privacy protection of data using distributed machine learning [22, 23] has gained considerable popularity in the context of the IoT. Distributed machine learning allows the learning models to be generated at each participant device, while the central server acting as the coordinator, creates a global model and distributes the knowledge to the participating nodes. Shokri and Shmatikov proposed a collaborative computing system that works on deep learning to protect a user's sensitive data while utilizing the information content of nonsensitive data of other users in the system [22]. The deep learning algorithms use the stochastic gradient descent algorithm because of the parallelization and asynchronous execution capability of the latter. The privacy model yields a very high accuracy on the test dataset. A distributed learning-based mechanism for data privacy preservation on IoT devices is proposed by Servia-Rodriguez et al. [23]. The scheme does not involve any data communication to the cloud environment. The system works in two phases. In the first step, the model is trained on data voluntarily shared by some users and possibly not containing any privacy-sensitive information. Once the model is trained, no further user data are shared. The model, tested on a public dataset, yields high accuracy. This scheme assumes user data privacy preservation since the original data never leaves their device. However, this is incorrect as the distributed machine learning models are vulnerable to privacy inference attacks that attempt to access privacy-sensitive data or model inversion attacks recovering original data [24, 25]. This enforces protection techniques such as encryption or differential privacy into distributed learning systems.

## 5. Distributed learning and encryption

Encryption techniques are integrated into distributed machine learning to boost data privacy in IoT applications. The most commonly used encryption method used is homomorphic encryption, in which the user data is encrypted before being sent to the computing nodes. A privacy protection system has been proposed based on the joint operation of a multilayer perceptron and a convolutional neural network model [26]. The model has been tested on the *modified national institute of standards and technology* (MNIST) and *street view house number* (SVHN) datasets [27]. A secure information system for healthcare applications in the IoT environment has been proposed [28]. The proposed model uses Pallier additive homomorphic encryption [29]. Another privacy system based on the Pallier system has been presented that works on blockchain technology [30]. The authors tested the system on two datasets of the University of California, Irvine (UCI) data repository [31, 32]. Homomorphic encryption systems offer increased privacy compared to differential privacy-based ones. However, fully homomorphic encryption can be costly in terms of computation overload, while partial homomorphic encryption can only be used for carrying out single operations. Moreover, partial homomorphic encryption methods require trusted third parties in place, or they work on simpler models approximating complex equations using single mathematical operations. A mechanism is proposed for protecting the privacy of data for the Industrial Internet of Things (IIoT) built on the principles of distributed learning [33]. The scheme works on a variational autoencoder model trained using homomorphic encryption. The accuracy of the model is found to be high, while its execution time is low. A hybrid framework for privacy protection is proposed by Osia et al. [34]. The scheme utilizes Siamese architecture and can perform efficient privacy-preserving analytics splitting a neural network IoT devices and cloud [35]. The feature extraction is done at the device, while the classification is carried out in the cloud. The scheme uses a convolutional neural network model evaluated with gender classification datasets *Internet movie database* (IMDB-Wiki) [36] and *labeled faces in the wild* (LFW) [37], achieving an accuracy of 94% and 93%, respectively. A data privacy-preserving scheme known named CP-ABPRE is presented by Zhou et al. that works on a policy-based encryption approach [38]. The scheme is found to be robust against privacy attacks and has a low computational overhead required for its encryption and decryption processes.

## 6. Distributed learning and differential privacy

In the differential privacy approach, the privacy of data is protected through the addition of some random perturbations into the original data. In other words, a perturbation in the data is done with a predetermined measure of the error caused by modifications to the data [39]. Several well-known techniques of perturbation include swapping, randomized response, micro-aggregation, additive perturbation, and condensation. However, perturbations reduce the quality of the data for analysis as the original data are modified. Privacy models work on a trade-off between the utility of data and its associated privacy level. In the privacy-utility trade-off, several algorithms and approaches exist in the literature. In the context of differential privacy, Abadi et al. presented a scheme involving training a neural network with differential privacy to prevent the disclosure of sensitive information [40]. The scheme is proved to be highly effective in preserving the privacy of sensitive data, as observed from its performance on the test dataset. Another scheme for privacy preservation of sensitive data is proposed in which a subset of parameters is shared and obfuscated using differential privacy as the training of the deep learning structures is carried out locally [41]. While the differential privacy-based schemes do not need high computational resources, they may be inaccurate since perturbations can reduce training quality. Moreover, these schemes cannot fully protect data privacy (i.e., there is always a trade-off between the model's accuracy-privacy). Wang et al. [42] enhanced the performance of the distributed machine learning system with differential privacy in an IoT environment via their Arden framework [42]. The scheme proposed by the authors involves protecting sensitive information using nullification or noise addition [27]. The model is tested on the MNIST/SVHN datasets and has yielded high accuracy while considerably reducing resource consumption [27]. The scheme proposed by Zhang et al. focused on distributed sensing systems where an obfuscate function was used to preserve training data privacy when shared with third parties [43].

Lyu et al. proposed a privacy mechanism using the random projection method to perturb the original data and embedding fog computing into deep learning [44]. This scheme is able to reduce communication overhead and computation load. The novel method of privacy protection, known as the fog-embedded privacy-preserving deep learning framework, can preserve the privacy of data using a robust defense method. First, a random perturbation is used to preserve the original data's statistical characteristics. Then, differentially private stochastic gradient descent is used to train the fog-level models with a multilayer perceptron model. The multilayer perceptron model consists of two hidden layers equipped with the *rectified linear unit* (ReLU) activation function. The accuracy yielded by the scheme on the test data is quite acceptable, although it is slightly lower compared to models with centralized architecture. However, the communication and computation overheads are significantly reduced.

Some privacy-preservation schemes utilize Gaussian projections to implement collaborative learning environments [45] efficiently. In these schemes, the resource-constrained IoT devices participate collaboratively and randomly apply multiplicative Gaussian projections on the training data records. This process obfuscates the privacy-sensitive input data. The coordinator node applies a deep learning-based model to learn from the complex patterns of the obfuscated data supplied by the Gaussian random projections. The performance results of the scheme demonstrated its efficiency and effectiveness in data privacy protection.

Among other approaches, obfuscation-based methods are also used in distributed machine learning to control the computation overhead involved in the encryption procedures in massively large-sized data. A scheme proposed by Alguliyev et al. protects big data in the context of IoT [46]. The mechanism involves the transformation of sensitive data into data that can be publicly shared. The proposed method works in two phases. In the first phase, data is transformed through a denoising type autoencoder. The parameter for designating the sparsity parameter of the autoencoder is specified for minimizing the loss in the autoencoder objective function during the data compression process. In the second phase, the transformed data from the output of the denoising autoencoder is classified using a convolutional neural network model. The proposed scheme was tested on several disease datasets and was found to be highly accurate in its prediction. Du et al. proposed a novel privacy-preserving scheme for big data in IoT deployed in edge computing applications [47]. The mechanism is based on a differential privacy approach built on machine learning models, which can improve query accuracy while minimizing the exposure of sensitive data to the public. The working mechanism involves two steps. In the first step, a Laplacian noise is added to the output data to carry out perturbation, while in the second step, random noise is added to the objective function that reduces the disturbance to the objective values. The data perturbation is carried out before transferring the data to the edge nodes. The model is tested on four diverse datasets and is found to be highly accurate in its performance. The machine learning models used in the scheme are stochastic gradient descent and generative adversarial networks.

Speech recognition systems, commonly found in IoT services, are susceptible to breaching user privacy as voice information is generally transmitted as plaintext and sometimes used for authentication purposes. To address this issue, Rouhani et al. proposed a scheme called *deepsecure* [48]. The working principle of the scheme is based on the garbled circuit protocol of Yao [49], and it executes much faster than the homomorphic encryption-based schemes. However, the proposition suffers from issues related to reusability and difficulty in implementation [50]. Differential privacy has been utilized in work by adding perturbations to user data [40]. However, the proposed scheme has a lower level of accuracy. Ma et al. [51] have thereby improved upon this by proposing a secret-sharing-based method that improves accuracy and reduces the computation and communication overhead for both linear and nonlinear operations using a long-and-short-term memory network model with interactive protocols for each gate. The proposed scheme was tested on a private dataset yielding a very high accuracy. Although privacy-preservation approaches based on obfuscation methods, in most cases, overcome the shortcomings of distributed machine learning and encryption-based distributed machine learning methods, these schemes are found to be vulnerable to some attacks [52–54].

## 7. Conclusion

This introductory chapter has presented a brief survey of some of the existing data privacy-preservation schemes proposed by researchers in the field of the Internet of Things. However, the design of privacy protection schemes in resource-constrained devices is still in its early stages. Reducing the latency and throughput of neural network training on encrypted data for privacy protection is a big challenge. Most of the existing schemes deploy their deep learning tasks to some external resources with adequate computing resources and storage spaces while keeping user data protected, making the schemes computationally efficient. New approaches should explore alternatives, such as quantum computing techniques, for designing more efficient and precise systems. In terms of future possibilities, parallel learning and cost optimization are being pursued, like network pruning and how different malicious activities interact. The relevant standard bodies should also make effective standardization efforts for all privacy protection schemes [55]. Finally, evaluating and assessing privacy solutions in real-world scenarios is tough, especially when considering the balance between IoT quality-of-service and privacy protection.

## Author details

Jaydip Sen* and Subhasis Dasgupta
Department of Data Science, Praxis Business School, Kolkata, India

*Address all correspondence to: jaydip.sen@acm.org

**Chapter 2**

# Adversarial Attacks on Image Classification Models: FGSM and Patch Attacks and Their Impact

*Jaydip Sen and Subhasis Dasgupta*

## Abstract

This chapter introduces the concept of adversarial attacks on image classification models built on convolutional neural networks (CNN). CNNs are very popular deep-learning models which are used in image classification tasks. However, very powerful and pre-trained CNN models working very accurately on image datasets for image classification tasks may perform disastrously when the networks are under adversarial attacks. In this work, two very well-known adversarial attacks are discussed and their impact on the performance of image classifiers is analyzed. These two adversarial attacks are the fast gradient sign method (FGSM) and adversarial patch attack. These attacks are launched on three powerful pre-trained image classifier architectures, ResNet-34, GoogleNet, and DenseNet-161. The classification accuracy of the models in the absence and presence of the two attacks are computed on images from the publicly accessible ImageNet dataset. The results are analyzed to evaluate the impact of the attacks on the image classification task.

## 1. Introduction

Szegedy et al. observed that a number of machine-learning models, even cutting-edge neural networks, are susceptible to adversarial samples [1]. In other words, these machine learning models categorize incorrectly cases that differ by a marginal amount from examples that are correctly classified and taken from the distribution of data. The same adversarial example is most often classified incorrectly by a wide range of models of varied architecture which are built on different sub-samples of the training data. This shows that fundamental flaws in our training algorithms are exposed by adversarial samples. It was unclear what caused these adversarial cases. However, speculative explanations have indicated that it may be related to the extreme non-linearity property of deep neural networks in combination with inadequate model averaging and insufficient regularization of the supervised learning problem that the models attempt to handle.

However, Goodfellow et al. disprove the need for these speculative hypotheses [2]. The authors argued that only linear behavior in high-dimensional domains is needed to produce adversarial cases. With the help of this viewpoint, it is possible to quickly create adversarial examples, which makes adversarial training feasible. The authors have also demonstrated that in addition to the regularization benefits offered by the techniques such as dropout, adversarial training can also regularize deep learning models [3]. Changing to nonlinear model families like RBF networks can significantly reduce a model's vulnerability to adversarial examples compared to generic regularization procedures like dropout, pretraining, and model averaging.

One may consider deep learning, which is frequently employed in autonomous (driverless) automobiles, to see why such misclassification is risky [4]. To recognize signs or other cars on the road, systems based on DNNs are utilized [5]. The automobile might not stop and end up in a collision, which might have disastrous repercussions, if tampering with the input of such systems, for as by significantly changing the body of the car, stops DNNs from correctly recognizing it as a moving vehicle. When an enemy may gain by avoiding detection or having their information misclassified, there is a significant threat. These kinds of attacks are frequent in non-DL classification systems nowadays [6–10].

Goodfellow et al. argue that there is a fundamental incompatibility between building simple-to-train linear models and building models that use nonlinear effects to withstand hostile disruption [2]. By creating more effective optimization methods that can successfully train more nonlinear models in the long run, this trade-off may be avoided.

While the bulk of adversarial attacks has concentrated on slightly altering each pixel of an image, there are examples of attacks that are not limited to barely discernible alterations in the image. An approach that is based on creating an image-independent patch and positioning it to cover a tiny area of the image was demonstrated by Brown et al. [11]. The classifier will reliably predict a particular class for the image in the presence of this patch based on the attacker's preference. This assault is significantly more dangerous than pixel-based attacks like FGSM because it can potentially cause even more damage and because the attacker does not need to know what image they are attacking when they are building the attack. An adversarial patch might then be produced and disseminated for use by more attackers. The conventional defense strategies, which concentrate on protecting against minor perturbations, may not be robust to larger disturbances like these since the attack involves a massive perturbation.

This chapter discusses various adversarial attacks on image classification models and focuses particularly on two specific attacks, the *fast gradient sign method* (FGSM), and the *adversarial patch attack*. The impact of these two attacks on image classification accuracy is analyzed and extensive results are presented. The rest of the chapter is organized as follows. Section 2 presents a few related works. Some theoretical background information on adversarial attacks and pre-trained image classification models is discussed in Section 3. Section 4 presents detailed results and their analysis. Finally, the chapter is concluded in Section 5 highlighting some future works.

## 2. Related work

Deep learning systems are generally prone to adversarial instances. These instances are deliberately selected inputs that influence the network to alter its output without being obvious to a human [5, 12]. Several optimization techniques, including L-BFGS [1], Fast Gradient Sign Method (FGSM) [2], DeepFool [13], and Projected Gradient Descent (PGD) [14] can be used to find these adversarial examples, which typically change each pixel by only a small amount. Other attack strategies aim to change only a small portion of the image's pixels (Jacobian-based saliency map [15]), or a small patch at a predetermined location [16].

A wide range of fascinating traits of neural networks and related models were demonstrated by Szegedy et al. [1]. The following are some of the important observations of the study: (1) Box-constrained L-BFGS can consistently discover adversarial cases. (2) The adversarial instances in ImageNet [17] data are so similar to the original examples that it is impossible for a human to distinguish between the two. (3) The same adversarial example is commonly classified incorrectly by a large number of classification models, each of which is trained using a different sample of the training data. (4) Adversarial events sometimes make shallow Softmax regression models less robust. (5) Training on adversarial examples can lead to a better regularization of the classification models.

By printing out a huge poster that resembles a stop sign or by applying various stickers to a stop sign, Eykholt et al. [12] demonstrated numerous techniques for creating stop signs that are misclassified by models.

These results suggest that classifiers developed using modern machine learning methods do not actually learn the underlying principles that determine the appropriate output label, even if they perform exceptionally well on the test data. The classification algorithms working for these models perform flawlessly with naturally occurring data. However, their classification accuracy drastically reduces for points that have a low probability in the underlying data distribution. This poses a big challenge to image classification since convolutional neural networks used in the classification work on the computation of perceptual feature similarity based on Euclidean distance. However, the resemblance found in this approach is false if images with unrealistically small perceptual distances actually belong to different classes as per the representation of the neural network.

The problem discussed above is particularly relevant to deep neural networks although linear classifiers are not immune to this problem. No model has yet been able to resist adversarial perturbation while preserving state-of-the-art accuracy on clean inputs. However, several approaches to defending against small perturbations-based adversarial attacks and some novel training approaches have been proposed by researchers [14, 15, 18–26]. Some of these works proposing methods to defend against adversarial attacks are briefly presented in the following.

Madry et al. designed and trained deep neural networks on the MNIST and CIFAR10 image set that are robust to a wide range of adversarial attacks [14]. The authors formulated an approach to identify a saddle point for optimizing the error function and used a projected gradient descent (PGD) as the adversary. The proposed approach was found to yield a classification accuracy of 89% against the strongest adversary in the test data.

Papernot et al. proposed a novel method to create adversarial samples based on a thorough comprehension of the mapping between inputs and outputs of deep neural networks [15]. In a computer vision application, the authors demonstrated that, while only changing an average of 4.02% of the input characteristics of each sample, their proposed method can consistently create samples that were correctly classified by humans but incorrectly classified in certain targets by a deep neural network with a 97% adversarial success rate. Then, by designing a hardness metric, the authors assessed the susceptibility of various sample classes to adversarial perturbations and outlined a defense mechanism against adversarial samples.

Tramer et al. observed that adversarial attacks are more impactful in a black-box setup, in which perturbations are computed and transferred on undefended models [18]. Adversarial attacks are also very effective when they are launched in a single step that escapes the non-smooth neighborhood of the input data through a short random step. The authors proposed an ensemble adversarial training, a method that adds perturbations obtained from other models to training data. The proposed approach is found to be resistant to black-box adversarial attacks on the ImageNet dataset.

For assessing adversarial resilience on image classification tasks, Dong et al. developed a reliable benchmark [21]. The authors made some important useful observations including the following. First, adversarial training is one of the most effective defense strategies because it can generalize across different threat models. Second. model robustness ness curves are useful in the evaluation of the adversarial robustness of models. Finally, the randomization-based defenses are more resistant to query-based black-box attacks.

Chen et al. examined and evaluated the features and effectiveness of several defense strategies against adversarial attacks [22]. The authors considered the evaluation from four different perspectives: (i) gradient masking, (ii) adversarial training, (iii) adversarial examples detection, and (iv) input modifications. The authors presented several benefits and drawbacks of various defense mechanisms against adversarial attacks and explored the future trends in designing robust methods to defend against such attacks on image classification models.

## 3. Background concepts

In this section, for the benefit of the readers, some background theories are discussed. The concepts of adversarial attack, fast gradient sign method (FGSM) attack, and three pre-trained convolutional neural network (CNN)-based deep neural network models, ResNet-34, GoogleNet, and DenseNet-161, are briefly introduced in this section.

### 3.1 Adversarial attacks

Many different adversarial attack plans have been put out, all of which aim to significantly affect the model's prediction by slightly changing the data or picture input. How can we modify the image of a goldfish so that a classification model that could correctly classify the image before would no longer recognize it? On the other hand, a human would still categorize the image as a goldfish without any doubt, hence the label of the image should not change at the same time. The generator network's goal under the framework for generative adversarial networks is the same as this one: try to trick another network (a discriminator) by altering its input.

### 3.2 Fast gradient sign method

The Fast Gradient Sign Method (FGSM), created by Ian Goodfellow et al., is one of the initial attack tactics suggested [2]. The FGSM uses a neural network's gradients to produce an adversarial image. Essentially, the adversarial image is produced by FGSM by computing the gradients of a loss function (such as mean-square error or category cross-entropy) with respect to the input image and using the sign of the gradients to produce a new image (i.e., the adversarial image) that maximizes the loss. The end result is an output image that, to human sight, appears just like the original, but it causes the neural network to anticipate something different than it should have. The FGSM is represented in (1).

$$adv_x = x + \varepsilon * sign(\nabla_x J(\theta, x, y) \quad (1)$$

The symbols used in (1) have the following significance:

$adv_x$ the adversarial image as the output

$x$ the original image as the input

$y$ the actual class (i.e., the ground-truth label) of the input image

ε the noise intensity expressed as a small fractional value by which the signed gradients are multiplied to create perturbations. The perturbations should be small enough so that the human eye cannot distinguish the adversarial image from the original image.

θ the neural network model used for image classification

$J$ the loss function

The FGSM attack on an image involves the following three steps.

1. The value of the loss function is computed after the forward propagation in the network.

2. The gradients are computed with respect to pixels in the original (i.e., input) image.

3. The pixels of the input image are perturbed slightly in the direction of the computed gradients so that the value of the loss function is maximized.

Most often, in machine learning, determining the loss after forward propagation is frequently the initial step. To determine how closely the model's prediction matches the actual class, a negative likelihood loss function is used. Gradients are used to choose the direction in which to move the weights in order to lower the value of the loss function when training neural networks. However, calculating gradients in relation to an image's pixels is not a usual task. In FGSM, the pixels in the input image are moved in the direction of the gradient to maximize the value of the loss function.

### 3.3 ResNet-34 architecture

He et al. presented a cutting-edge image classification neural network model containing 34 layers [27]. This deep convolutional neural network is known as the ResNet-34 model. The ImageNet dataset, which includes more than 100,000 images in 200 different classes, served as the pre-training data for ResNet-34. Similar to residual neural networks used for text prediction, ResNet architecture differs from typical neural networks in that it uses the residuals from each layer in the connected layers that follow.

### 3.4 GoogleNet architecture

Szegedy et al. introduced GoogleNet (also known as Inception V1) in their paper titled "Going Deeper with Convolutions" [28]. In the 2014 ILSVRC image classification competition, this architecture was the winner. This architecture employs methods like global average pooling and 1–1 convolution in the middle of the architecture. A network may experience the issue of overfitting if it is constructed with very deep layers. To address this issue, the GoogleNet architecture was developed with the idea of having filters of various sizes that could function at the same level [28]. The network actually gets bigger with this concept rather than deeper. The architecture has a total of 22 layers, including 27 pooling layers. There are nine linearly stacked inception components that are connected to the global average pooling layer. The readers may refer to the work of Szegedy et al. for more details [28].

### 3.5 DenseNet-161 architecture

A class of CNN called DenseNets uses dense connections between network layers for matching convolution operation feature-map sizes [29]. These dense connections are called dense blocks. Each layer receives extra inputs from all earlier layers and transmits its own feature maps to all later layers in order to maintain the feed-forward character of the system. Huang et al. demonstrated that a variant of DenseNet architecture called DensseNet-161with $k$ = 48 features per layer and having 29 million parameters can achieve a classification accuracy of 77.8% (i.e., top-1 classification accuracy) on the ImageNet ILSVRC classification dataset. As its name implies, the DenseNet-161 architecture contains 161 layers of nodes. More details on DenseNet-161 architecture may be found in [29, 30].

## 4. Image classification results and analysis

Experiments are conducted to analyze the effect of two types of adversarial attacks on three well-known pre-trained CNN architectures. Two adversarial attacks considered in the study are the *FGSM attack* and the *adversarial patch attack* on a set of images. Three pre-trained architectures on which the attacks are simulated are ResNet-34, GoogleNet, and DenseNet-161. The images are chosen from the ImageNet dataset [17]. The pre-trained CNN models of ResNet-34, GoogleNet, and DenseNet-161 integrated into PyTorch's *torchvision* package, are used in the experiments.

### 4.1 Classification results in the absence of an attack

Before we study the impact of adversarial attacks on the image classification models, we analyze the classification accuracy of the models in the absence of any attack. Since the ImageNet dataset includes 1000 classes, it is not prudent to evaluate a model's performance just on the basis of its classification accuracy alone. Consider a model that consistently predicts the true label of an input image as the second-highest class using the *Softmax* activation function. Despite the fact that we would say it recognizes the object in the image, its accuracy is zero. There is not always one distinct label we can assign an image to in ImageNet's 1000 classes. This is why "Top-5 accuracy" is a popular alternative metric for picture classification over a large number of classes. It shows how often the real label has been within the model's top 5 most likely predictions. Since the three pre-trained architectures perform very well on the images in the ImageNet dataset, instead of accuracy, the error, i.e., (1- accuracy) values are presented in the results.

**Table 1** presents the performance results of three classification models on the whole ImageNet dataset containing 1000 classes of images. It is evident that all three models are highly accurate as depicted by their Top-% error percentage values. The DeepNet-161 model has yielded the highest level of accuracy and the least error among the three architectures. The Top-5 and Top-1 error rates for this model are found to be 2.30% and 15.10%, respectively.

| Metric | ResNet34 model | GoogleNet model | DenseNet161 model |
|---|---|---|---|
| Top-1 error (%) | 19.10 | 25.26 | 15.10 |
| Top-5 error (%) | 4.30 | 7.74 | 2.30 |

**Table 1.**
*Classification accuracy of ResNet-34, GoogleNet, and Densenet-161 CNN models on the ImageNet data.*

After evaluating the overall performance of the three models, we investigate some specific images in the dataset. For this purpose, the images with the indices 0, 6, 13, and 18 are randomly chosen and how the model has classified these images are checked. The images corresponding to the four indices chosen belong to the classes "tench", "goldfish", "great white shark" and "tiger shark", respectively.

**Table 2** presents the performance of the ResNet-34 model on the classification task for the four images. It is evident that the model has been very accurate in classification as the confidence associated with the true class of each of the four images is more than 90%. It may be noted that confidence here means the probability value that the model associates with the corresponding class. For example, the ResNet-34 model has yielded a confidence value of 0.9817 for the image whose true class is "tench" with the predicted class "tench", implying that the model has associated a probability of 0.9817 with its classification of the image to the class "tench".

**Figure 1** depicts the classification results of the ResNet-34 model on the four images. In **Figure 1**, the input image is shown on the left and the confidence values of the model for the top five classes for the image are shown on the right. The confidence values are shown in the form of horizontal bars.

| Image index | Image true class | Top-5 predicted classes and their confidence | |
|---|---|---|---|
| | | Class | Confidence |
| 0 | tench | **tench** | **0.9817** |
| | | barracouta | 0.0095 |
| | | coho | 0.0085 |
| | | gar | 0.0002 |
| | | sturgeon | 0.0001 |
| 6 | goldfish | **goldfish** | **0.9982** |
| | | tench | 0.0005 |
| | | barracouta | 0.0005 |
| | | tailed frog | 0.0003 |
| | | puffer | 0.0002 |
| 13 | great white shark | **great white shark** | **0.9855** |
| | | tiger shark | 0.0109 |
| | | submarine | 0.0007 |
| | | sturgeon | 0.0006 |
| | | hammerhead | 0.0005 |

| Image index | Image true class | Top-5 predicted classes and their confidence | |
|---|---|---|---|
| | | Class | Confidence |
| 18 | tiger shark | **tiger shark** | **0.9118** |
| | | sturgeon | 0.0251 |
| | | great white shark | 0.0202 |
| | | puffer | 0.0192 |
| | | electric ray | 0.0038 |

**Table 2.**
*The classification results of the ResNet-34 model for the chosen images.*

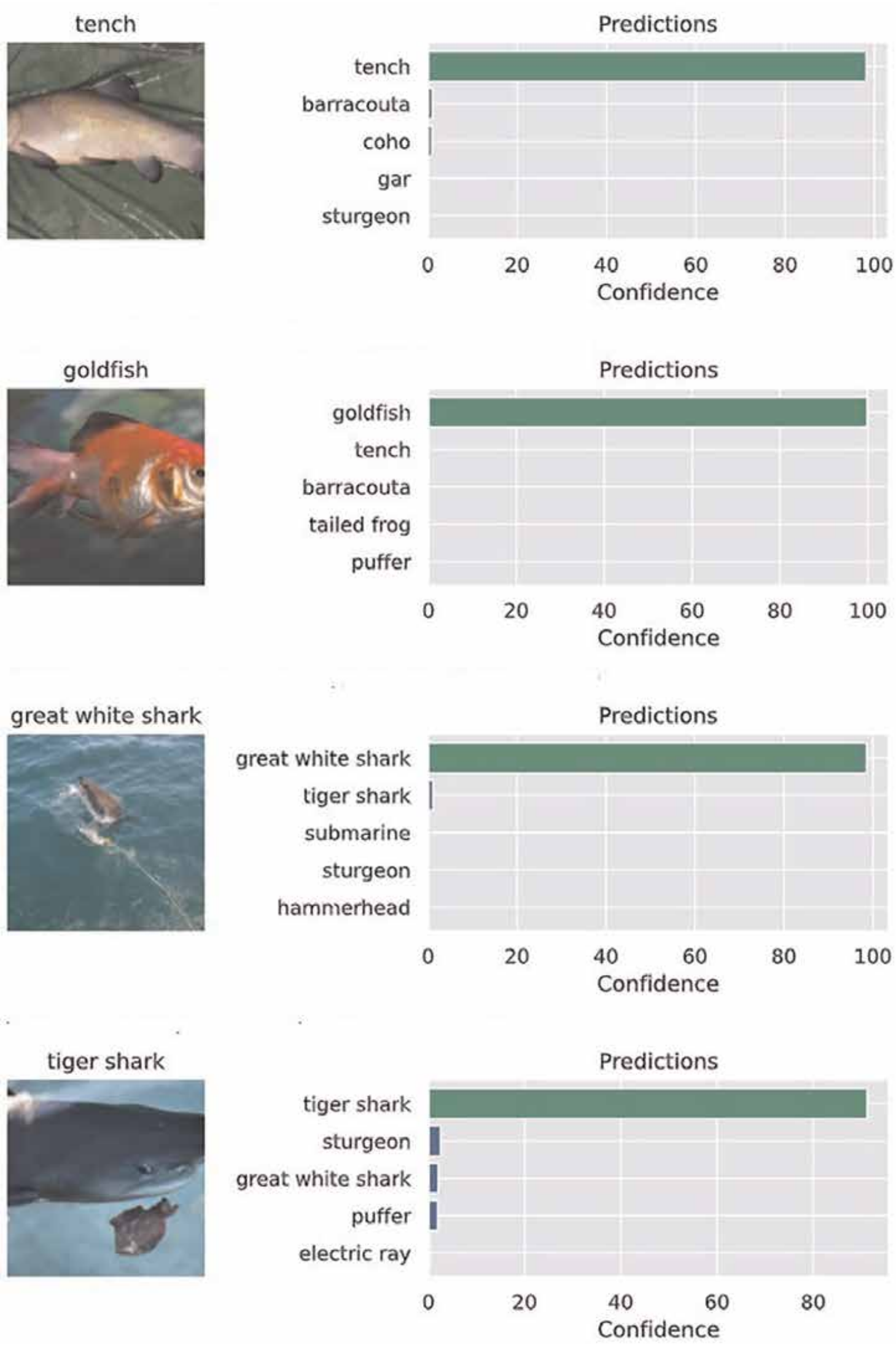

**Figure 1.**
*The classification results of the ResNet34 model for the chosen images.*

| Image index | Image true class | Top-5 predicted classes and their confidence | |
|---|---|---|---|
| | | Class | Confidence |
| 0 | tench | **tench** | **0.9826** |
| | | coho | 0.0075 |
| | | barracouta | 0.0034 |
| | | goldfish | 0.0008 |
| | | gar | 0.0005 |
| 6 | goldfish | **goldfish** | **0.9617** |
| | | tench | 0.0129 |
| | | loggerhead | 0.0018 |
| | | barracouta | 0.0014 |
| | | coho | 0.0010 |
| 13 | great white shark | **great white shark** | **0.8188** |
| | | sea lion | 0.0532 |
| | | gray whale | 0.0376 |
| | | tiger shark | 0.0144 |
| | | loggerhead | 0.0082 |
| 18 | tiger shark | **tiger shark** | **0.3484** |
| | | platypus | 0.1978 |
| | | hammerhead | 0.0277 |
| | | sturgeon | 0.0245 |
| | | great white shark | 0.0166 |

**Table 3.**
*The classification results of the GoogleNet model for the chosen images.*

**Table 3** presents the performance of the GoogleNet model on the classification task for the four images. It is observed that the model has been very accurate in the classification task for the “tench” and “goldfish” images. While its accuracy for the image “great white shark” class is high, the model has performed poorly for the image “tiger shark”. However, for the “tiger shark” image the model has still associated the highest confidence value for the correct class, although the confidence is quite low, i.e., 0.3484.

**Figure 2** depicts the classification results of the GoogleNet model on the four images. In **Figure 2**, the input image is shown on the left and the confidence values of the model for the top five classes for the image are shown on the right. The confidence values are shown in the form of horizontal bars.

**Table 4** presents the performance of the DenseNet-161 model on the classification task for the four images. It is observed that the performance of the model on the classification task has been excellent. For all four images, the confidence values computed by the model for the true class have been higher than 94. The results also show that among the three architectures, DenseNet-161 has been the most accurate model for the classification of the four images chosen for analysis.

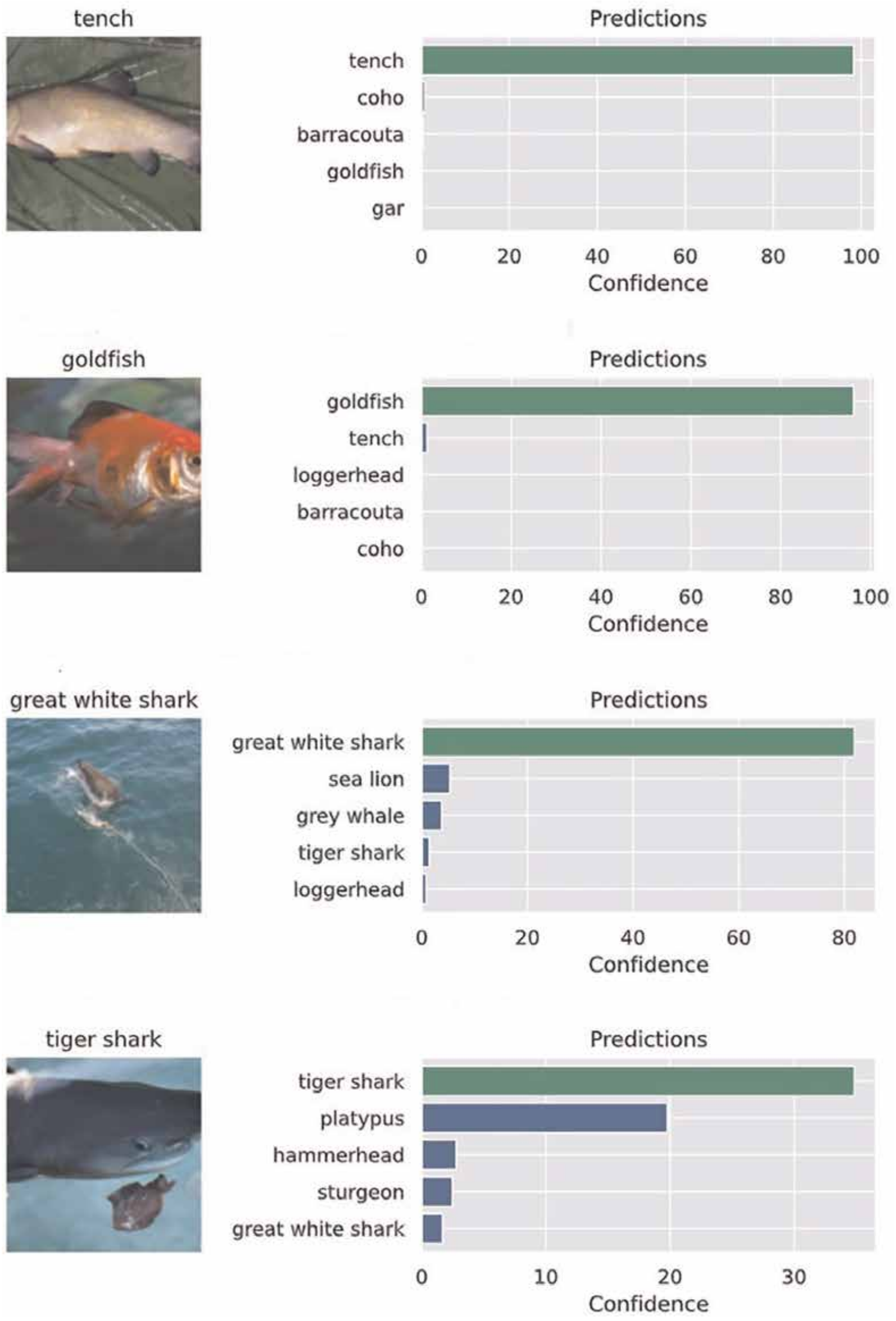

**Figure 2.**
*The classification results of the GoogleNet model for the chosen images.*

| Image index | Image true class | Top-5 predicted classes and their confidence | |
|---|---|---|---|
| | | Class | Confidence |
| 0 | tench | **tench** | **0.9993** |
| | | barracouta | 0.0003 |
| | | coho | 0.0002 |
| | | gar | 0.0001 |
| | | platypus | 0.0001 |

| Image index | Image true class | Top-5 predicted classes and their confidence | |
|---|---|---|---|
| | | **Class** | **Confidence** |
| 6 | goldfish | **goldfish** | **0.9999** |
| | | barracouta | 0.0001 |
| | | tench | 0.0001 |
| | | coho | 0.0001 |
| | | gar | 0.0001 |
| 13 | great white shark | **great white shark** | **0.9490** |
| | | tiger shark | 0.0177 |
| | | dugong | 0.0127 |
| | | sea lion | 0.0113 |
| | | gray whale | 0.0074 |
| 18 | tiger shark | **tiger shark** | **0.9932** |
| | | great white shark | 0.0047 |
| | | gar | 0.0008 |
| | | sturgeon | 0.0002 |
| | | hammerhead | 0.0001 |

**Table 4.**
*The classification results of the DenseNet-161 model for the chosen images.*

**Figure 3** depicts the classification results of the DenseNet-161 model on the four images. In **Figure 3**, the input image is shown on the left and the confidence values of the model for the top five classes for the image are shown on the right. The confidence values are shown in the form of horizontal bars.

### 4.2 Classification results in the presence of the FGSM attack

After observing the performance of the three CNN architectures for the image classification tasks on the images in the ImageNet dataset, the impact of the adversarial attacks on the classifier models is studied. We start with the FGSM attack with a value of 0.02 for epsilon ($\varepsilon$). The value of $\varepsilon = 0.02$ indicates that the values of pixels are changed by an amount of 1 (approximately) in the range of 0 to 255 – the range over which a pixel value can change. This change is so small that it will be impossible to distinguish the adversarial image from the original one. The performance results of the three models in the presence of FGSM attack with $\varepsilon = 0.02$ have been presented in **Tables 5–7**. The results are pictorially depicted in **Figures 4–6**.

It is evident that all three models are adversely affected by the FGSM attack even with a value of $\varepsilon$ as low as 0.02. While the adversarial images are impossible to distinguish from the original ones, none of the models could correctly classify any of the four images as the highest confidence values were assigned to incorrect classes (**Tables 8–10**).

The value of the parameter $\varepsilon$ is increased from 0.01 to 0.10 by a step of 0.01. It is observed that except for a few cases, the classification error increased consistently with ε till ε reaches a value in the range of 0.08–0.09. The impact of the FGSM attack is so severe that the classification error for the ResNet-34 model in the presence of this attack reaches as high values as 97.00% (Top-1 error) and 77.68% (Top-5 error). The corresponding values for the GoogleNet model are 95.46% (Top-1 error) and 79.24% (Top-5 error), and for the DenseNet-161 are 94.42% (Top-1 error) and 66.94% (Top-5 error). Among the three models, DenseNet-161 looked to be the most robust against the FGSM attack.

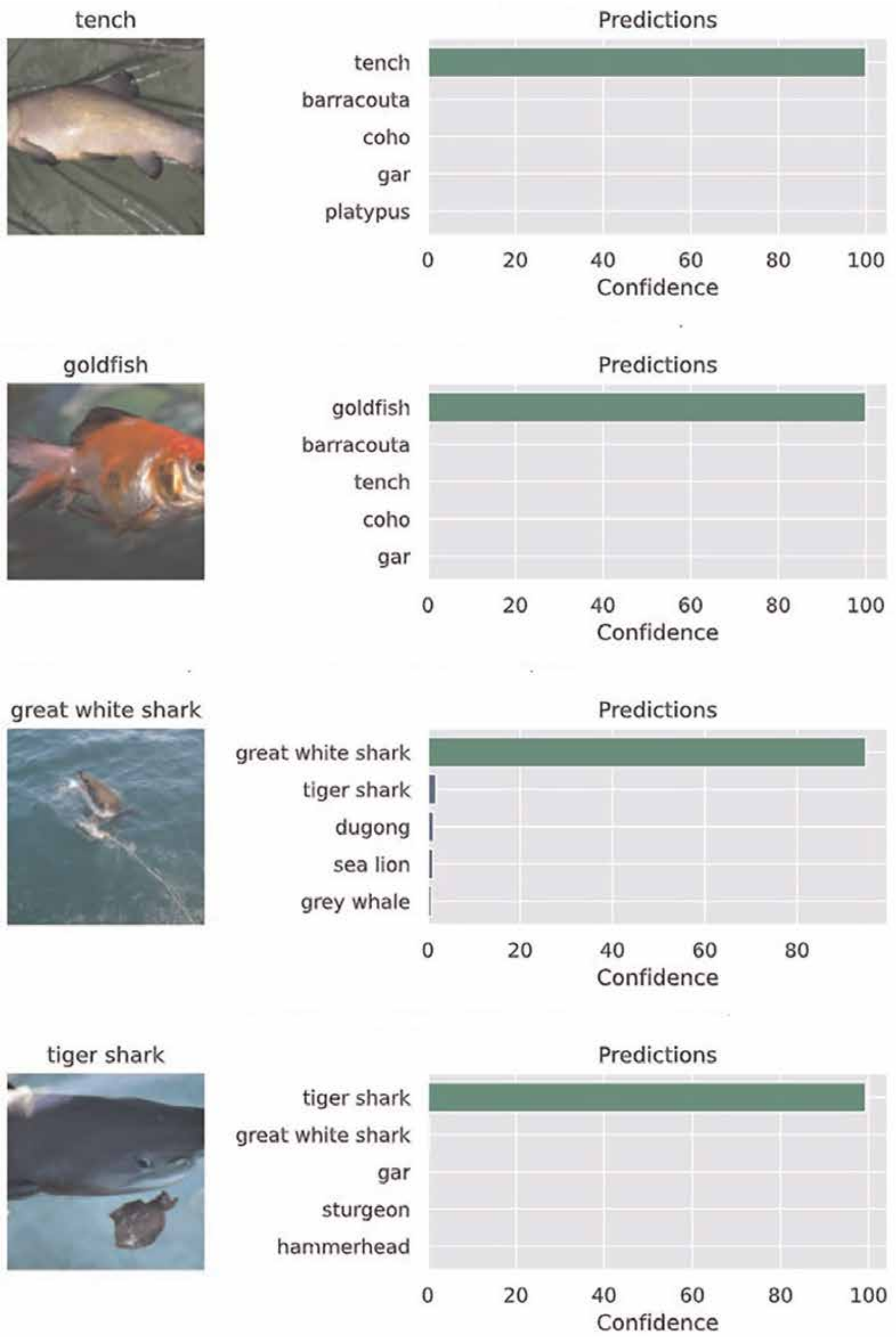

**Figure 3.**
*The classification results of the DenseNet161 model for the chosen images.*

| **Image true class** | **Top-5 predicted classes and their confidence** | |
|---|---|---|
| | **Class** | **Confidence** |
| tench | coho | 0.6684 |
| | barracouta | 0.2799 |
| | gar | 0.0321 |
| | **tench** | **0.0128** |
| | sturgeon | 0.0066 |

| Image true class | Top-5 predicted classes and their confidence | |
|---|---|---|
| | Class | Confidence |
| goldfish | barracouta | 0.5885 |
| | tench | 0.1392 |
| | gar | 0.0945 |
| | tailed frog | 0.0462 |
| | coho | 0.0434 |
| great white shark | dugong | 0.2661 |
| | tiger shark | 0.2575 |
| | gray whale | 0.0578 |
| | **great white shark** | **0.0537** |
| | submarine | 0.0303 |
| tiger shark | otter | 0.2462 |
| | puffer | 0.1666 |
| | beaver | 0.1543 |
| | platypus | 0.1083 |
| | sea lion | 0.0565 |

**Table 5.**
*The performance of ResNet-34 model under FGSM attack with ε = 0.02.*

| Image true class | Top-5 predicted classes and their confidence | |
|---|---|---|
| | Class | Confidence |
| tench | coho | 0.2652 |
| | **tench** | **0.2116** |
| | barracouta | 0.1275 |
| | gar | 0.0219 |
| | sturgeon | 0.0153 |
| goldfish | **goldfish** | **0.0553** |
| | tench | 0.0305 |
| | barracouta | 0.0218 |
| | gar | 0.0127 |
| | great white shark | 0.0115 |
| great white shark | weasel | 0.1880 |
| | sea lion | 0.1489 |
| | otter | 0.1402 |
| | platypus | 0.0605 |
| | tailed frog | 0.0400 |

| Image true class | Top-5 predicted classes and their confidence | |
|---|---|---|
| | Class | Confidence |
| tiger shark | platypus | 0.5450 |
| | beaver | 0.0336 |
| | American coot | 0.0265 |
| | terrapin | 0.0142 |
| | otter | 0.0127 |

**Table 6.**
*The performance of GoogleNet model under FGSM attack with ε = 0.02.*

| Image true class | Top-5 predicted classes and their confidence | |
|---|---|---|
| | Class | Confidence |
| tench | coho | 0.6793 |
| | **tench** | **0.1645** |
| | gar | 0.0466 |
| | barracouta | 0.0373 |
| | sturgeon | 0.0273 |
| goldfish | barracouta | 0.7139 |
| | tench | 0.1357 |
| | coho | 0.0665 |
| | gar | 0.0645 |
| | **goldfish** | **0.0137** |
| great white shark | sea lion | 0.4830 |
| | dugong | 0.4388 |
| | gray whale | 0.0255 |
| | tiger shark | 0.0099 |
| | snorkel | 0.0023 |
| tiger shark | great white shark | 0.9025 |
| | gar | 0.0606 |
| | barracouta | 0.0059 |
| | **tiger shark** | **0.0058** |
| | coho | 0.0058 |

**Table 7.**
*The performance of DenseNet-161 model under FGSM attack with ε 0.02.*

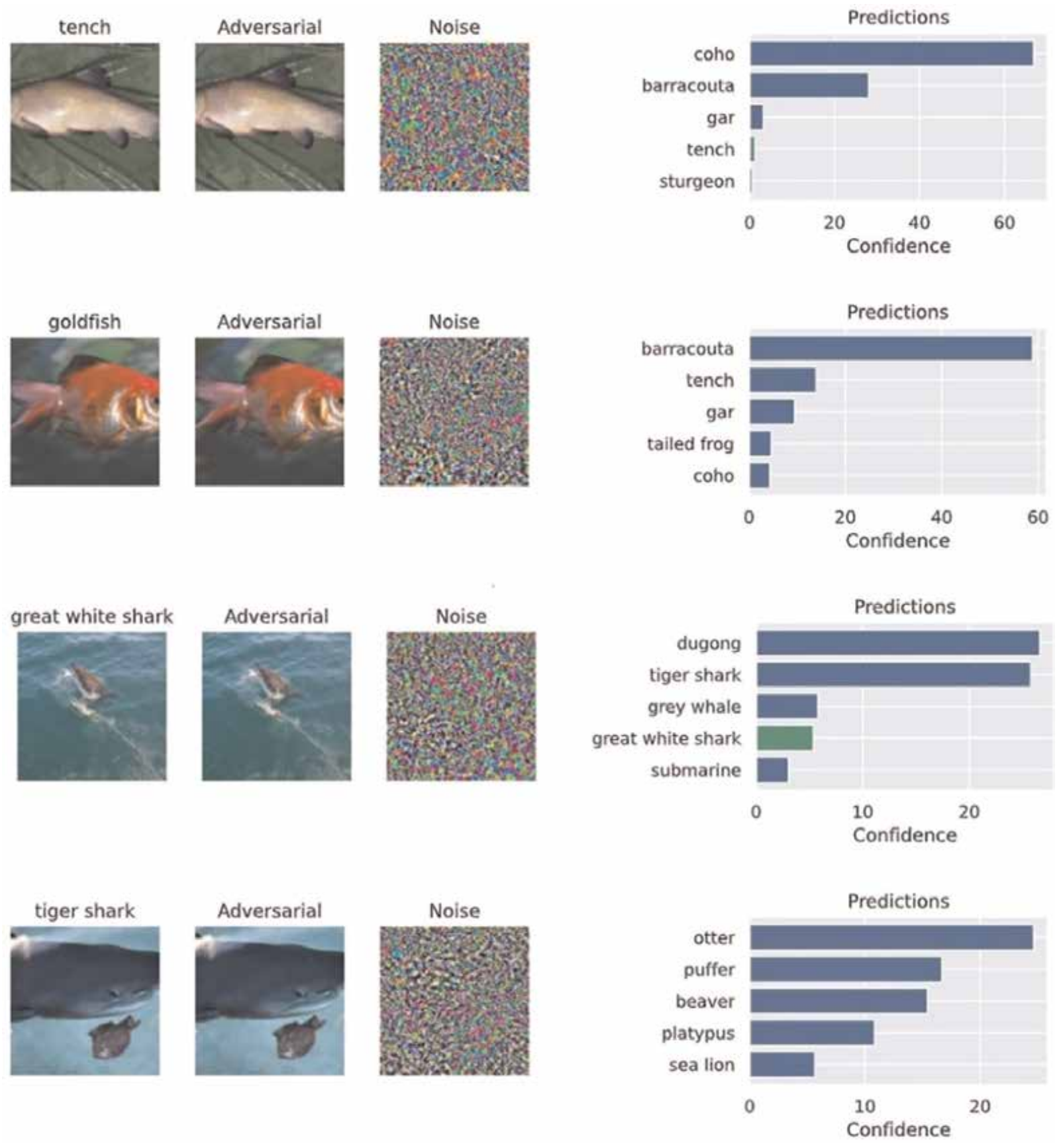

**Figure 4.**
*The performance of the ResNet-34 model under FGSM attack with ε = 0.02.*

### 4.3 Classification results in the presence of the adversarial patch attack

As mentioned in Section 1, an attack can also be launched on image classification models by introducing adversarial patches [11]. In this attack, the strategy is to transform a small portion of the image into a desired form and shape instead of the FGSM's approach of slightly altering some pixels. This will be able to deceive the classification model and force it to predict a certain pre-determined class. In practical applications, this type of attack poses a greater hazard than FGSM. Consider an autonomous vehicle network that receives a real-time image from a camera. To trick this vehicle into thinking that an automobile is actually a pedestrian, another driver may print out a certain design and stick it on the rear part of the vehicle.

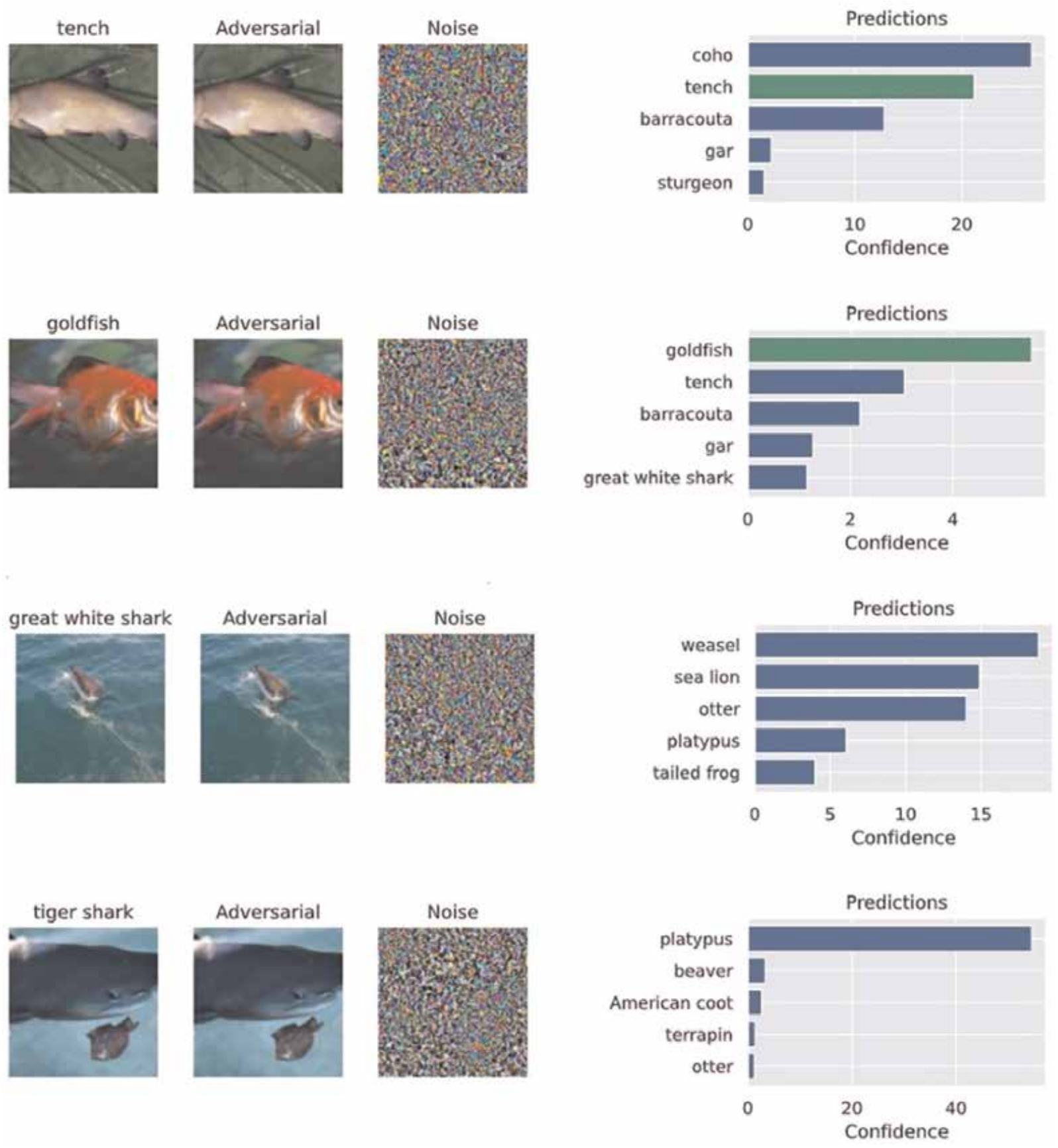

**Figure 5.**
*The performance of the GoogleNet model under FGSM attack with ε = 0.02.*

For simulating the adversarial patch attack on the same four images on which the FGSM attack was launched, at first, five images are chosen randomly which will be used as the patches. As shown in **Figure 7**, the five patch images are (i) cock, (ii) balloon, (iii) computer keyboard, (iv) electric guitar, and (v) radio. For the purpose of studying the effect of the sizes of the patch images on the accuracies of the classification models, three different sizes are considered for each patch image. The three sizes are (i) 32*32, (ii) 48*48, and (iii) 64*64. The sizes are expressed in terms of the number of pixels along the x and y dimensions. **Tables 11–16** present the accuracies (Top 1% and Top 5%) of the models for different sizes of different patch images. Here, accuracy refers to the percentage of cases in which the images have been classified as the target class (i.e., patch class) with the highest confidence. **Figures 8–10** depict the performance of the classification models in the presence of a “balloon” patch of size 64*64. The pictures for other patch images and other sizes are not presented for the sake of brevity.

The following observations are made on the results of the adversarial patch attack.

1. For the same patch image, all three models, ResNet-34, GoogleNet, and DenseNet-161, exhibited higher accuracy in deceiving the models into the wrong classification for a bigger patch size. In other words, for all three models, the attack effectiveness is the highest for the patch size 64*64, for a given patch image.

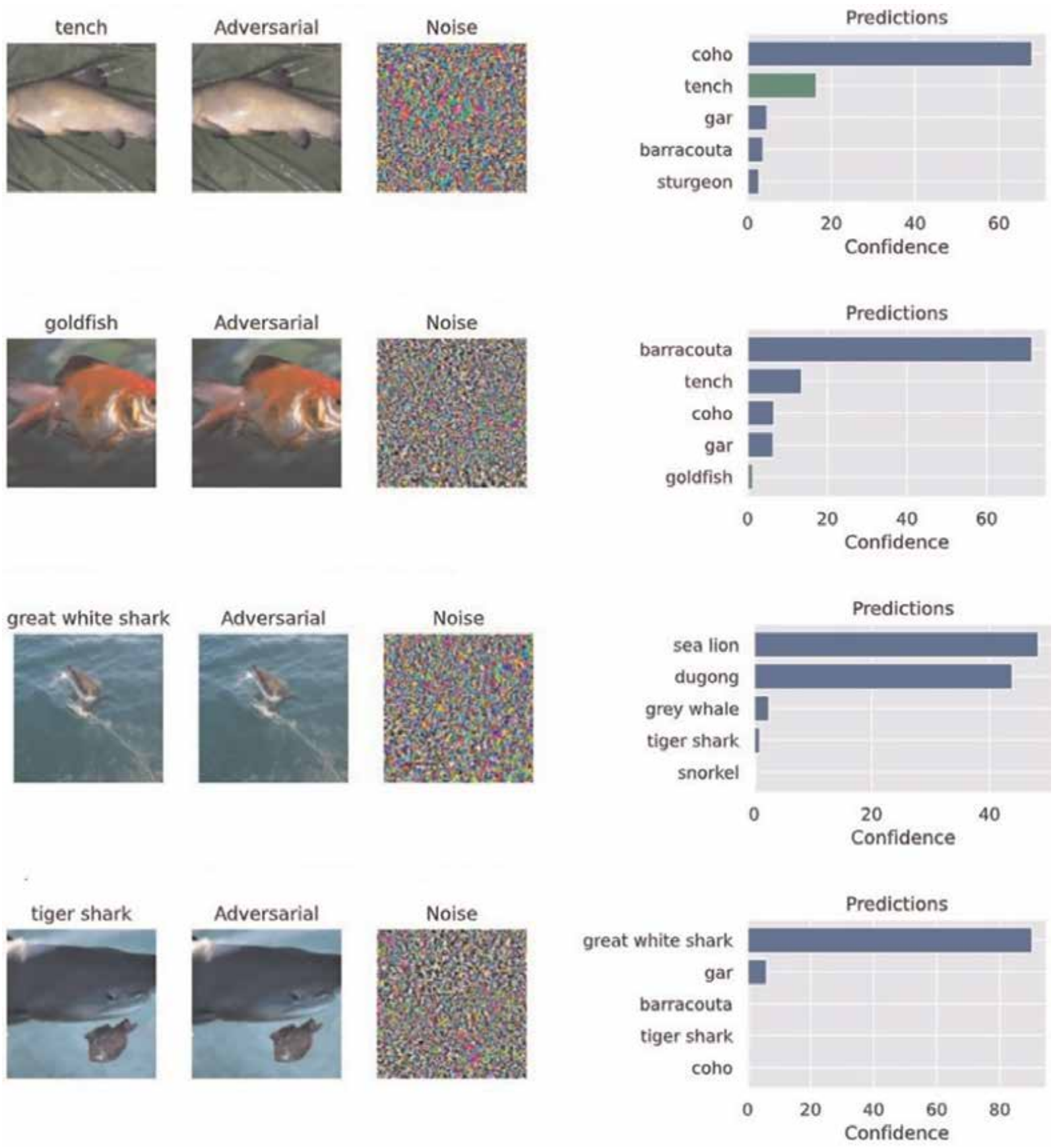

**Figure 6.**
*The performance of DenseNet-161 model under FGSM attack with ε = 0.02.*

2. For most of the patch images and patch sizes, the effectiveness of the attack on three models is found to be the most for the patch image of “balloon”. However, for the ResNet-34 model, with a patch image of “cock” the attack yielded the maximum effectiveness for the patch size of 64*64. For the GoogleNet model, along with the “balloon” patch image, the “electric guitar” patch of size 64*64 also produced the maximum Top-1 accuracy. For the DenseNet-161 model, the attack exhibited the highest effectiveness for the patch image of “computer keyboard” of size 64*64 for both Top-1 and Top-5 cases.

3. For obvious reasons, the attack effectiveness (i.e., the accuracy of the attack) is found to be always higher for the Top-5 case than its corresponding Top-1 counterpart.

| Noise level (ε) | Classification error in percent | |
|---|---|---|
| | **Top-1 error** | **Top-5 error** |
| 0.01 | 83.44 | 43.76 |
| 0.02 | 93.56 | 60.54 |
| 0.03 | 95.66 | 68.60 |
| 0.04 | 96.24 | 72.42 |
| 0.05 | 96.76 | 74.78 |
| **0.06** | **97.00** | 76.18 |
| 0.07 | 96.98 | 76.92 |
| 0.08 | 97.00 | 77.54 |
| **0.09** | 96.94 | **77.68** |
| 0.10 | 96.92 | 77.56 |

**Table 8.**
*Performance of ResNet34 under FGSM attack for different values of ε.*

| Noise level (ε) | Classification error in percent | |
|---|---|---|
| | **Top-1 error** | **Top-5 error** |
| 0.01 | 82.76 | 49.52 |
| 0.02 | 91.10 | 65.86 |
| 0.03 | 93.72 | 72.68 |
| 0.04 | 94.66 | 75.86 |
| 0.05 | 95.14 | 77.62 |
| 0.06 | 95.26 | 78.40 |
| 0.07 | 95.36 | 78.96 |
| 0.08 | 95.40 | 79.04 |
| **0.09** | **95.46** | **79.24** |
| 0.10 | 95.40 | 79.20 |

**Table 9.**
*Performance of GoogleNet under FGSM attack for different values of ε.*

| Noise level (ε) | Classification error in percent | |
|---|---|---|
| | **Top-1 error** | **Top-5 error** |
| 0.01 | 79.08 | 33.10 |
| 0.02 | 90.08 | 50.64 |
| 0.03 | 92.98 | 58.64 |
| 0.04 | 94.04 | 62.88 |

| Noise level (ε) | Classification error in percent | |
|---|---|---|
| | **Top-1 error** | **Top-5 error** |
| 0.05 | 94.38 | 65.12 |
| 0.06 | 94.38 | 66.38 |
| 0.07 | 94.34 | 66.76 |
| **0.08** | **94.42** | **66.94** |
| 0.09 | 94.18 | 66.68 |
| 0.10 | 94.10 | 66.70 |

**Table 10.**
*Performance of DenseNet161 under FGSM attack for different values of ε.*

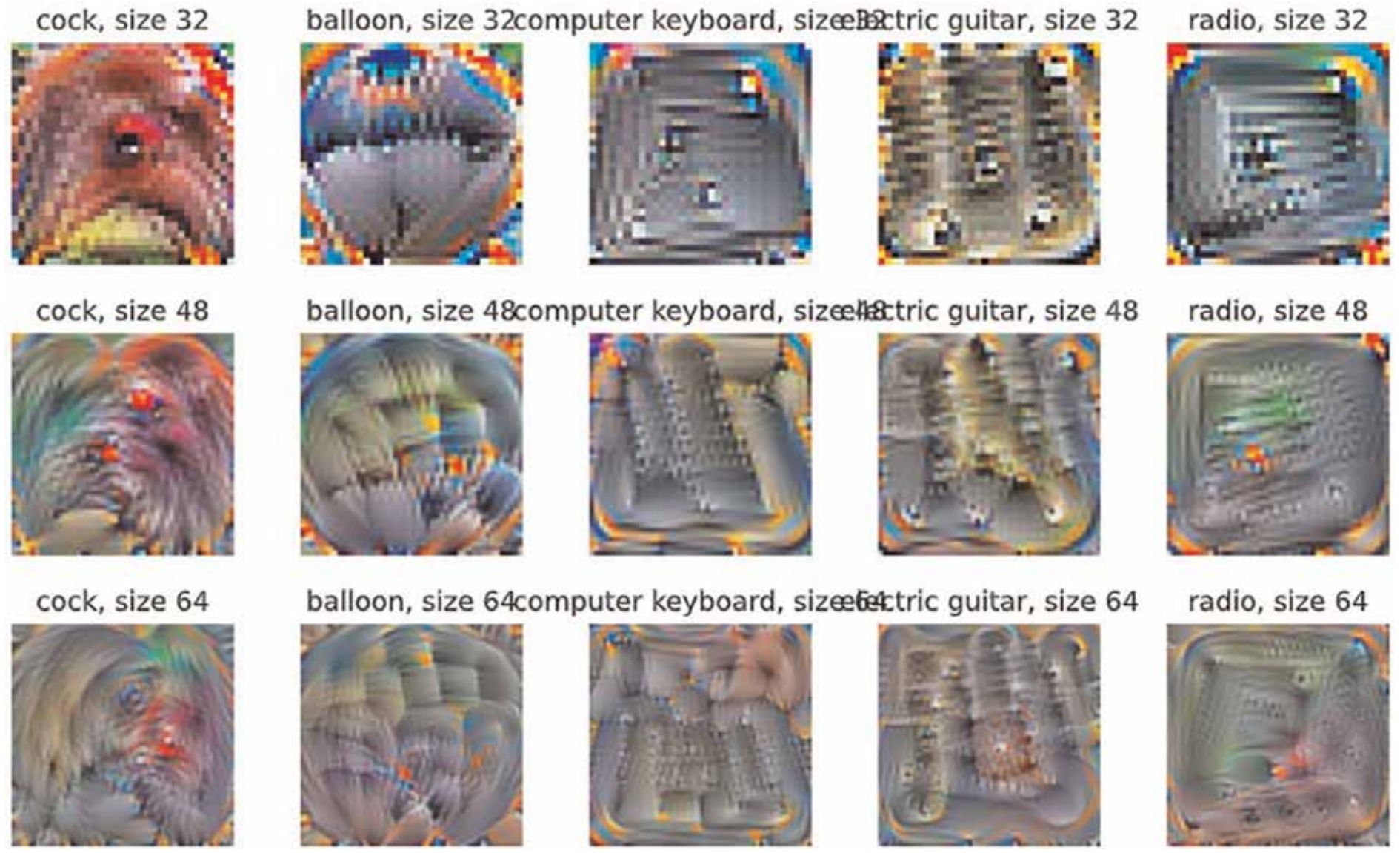

**Figure 7.**
*Five images used as patches: cock, balloon, computer keyboard, electric guitar, and radio. The sizes for the patch images: 32*32, 48*48, and 64*64.*

| Patch image | Size of the patch image | | |
|---|---|---|---|
| | **32*32** | **48*48** | **64*64** |
| cock | 78.75 | 92.01 | **97.82** |
| balloon | **81.17** | **92.35** | 97.44 |
| computer keyboard | 0.04 | 68.46 | 92.97 |
| electric guitar | 54.47 | 87.08 | 95.93 |
| radio | 22.35 | 75.49 | 94.03 |

**Table 11.**
*Top-1 accuracy (%) of ResNet34 model for different patch sizes.*

| Patch image | Size of the patch image | | |
|---|---|---|---|
| | **32*32** | **48*48** | **64*64** |
| cock | 93.48 | 98.59 | **99.84** |
| balloon | **93.73** | **98.88** | 99.83 |
| computer keyboard | 1.22 | 91.63 | 99.45 |
| electric guitar | 77.43 | 97.15 | 99.64 |
| radio | 62.08 | 93.67 | 99.37 |

**Table 12.**
*Top-5 accuracy (%) of ResNet34 model for different patch sizes.*

| Patch image | Size of the patch image | | |
|---|---|---|---|
| | **32*32** | **48*48** | **64*64** |
| cock | 0.00 | 0.27 | 0.94 |
| balloon | **85.06** | **95.67** | **98.76** |
| computer keyboard | 17.17 | 80.81 | 97.22 |
| electric guitar | 70.14 | 93.64 | **98.76** |
| radio | 7.73 | 81.34 | 95.59 |

**Table 13.**
*Top-1 accuracy (%) of GoogleNet model for different patch sizes.*

| Patch image | Size of the patch image | | |
|---|---|---|---|
| | **32*32** | **48*48** | **64*64** |
| cock | 0.09 | 8.69 | 32.63 |
| balloon | **96.76** | **99.80** | **99.99** |
| computer keyboard | 66.72 | 96.81 | 99.94 |
| electric guitar | 90.45 | 99.60 | 99.98 |
| radio | 65.15 | 97.21 | 99.84 |

**Table 14.**
*Top-5 accuracy (%) of GoogleNet model for different patch sizes.*

| Patch image | Size of the patch image | | |
|---|---|---|---|
| | **32*32** | **48*48** | **64*64** |
| cock | 0.00 | 0.00 | 0.01 |
| balloon | **14.70** | **35.41** | 40.25 |
| computer keyboard | 0.02 | 0.08 | **47.46** |
| electric guitar | 1.91 | 5.97 | 46.74 |
| radio | 0.53 | 8.88 | 43.44 |

**Table 15.**
*Top-1 accuracy (%) of DenseNet-161 model for different patch sizes.*

| Patch image | Size of the patch image | | |
|---|---|---|---|
| | **32*32** | **48*48** | **64*64** |
| cock | 0.09 | 0.08 | 0.16 |
| balloon | **41.63** | **69.44** | 69.75 |
| computer keyboard | 0.76 | 1.46 | **79.86** |
| electric guitar | 13.10 | 22.67 | 75.43 |
| radio | 10.36 | 46.63 | 74.84 |

**Table 16.**
*Top-5 accuracy (%) of DenseNet161 model for different patch sizes.*

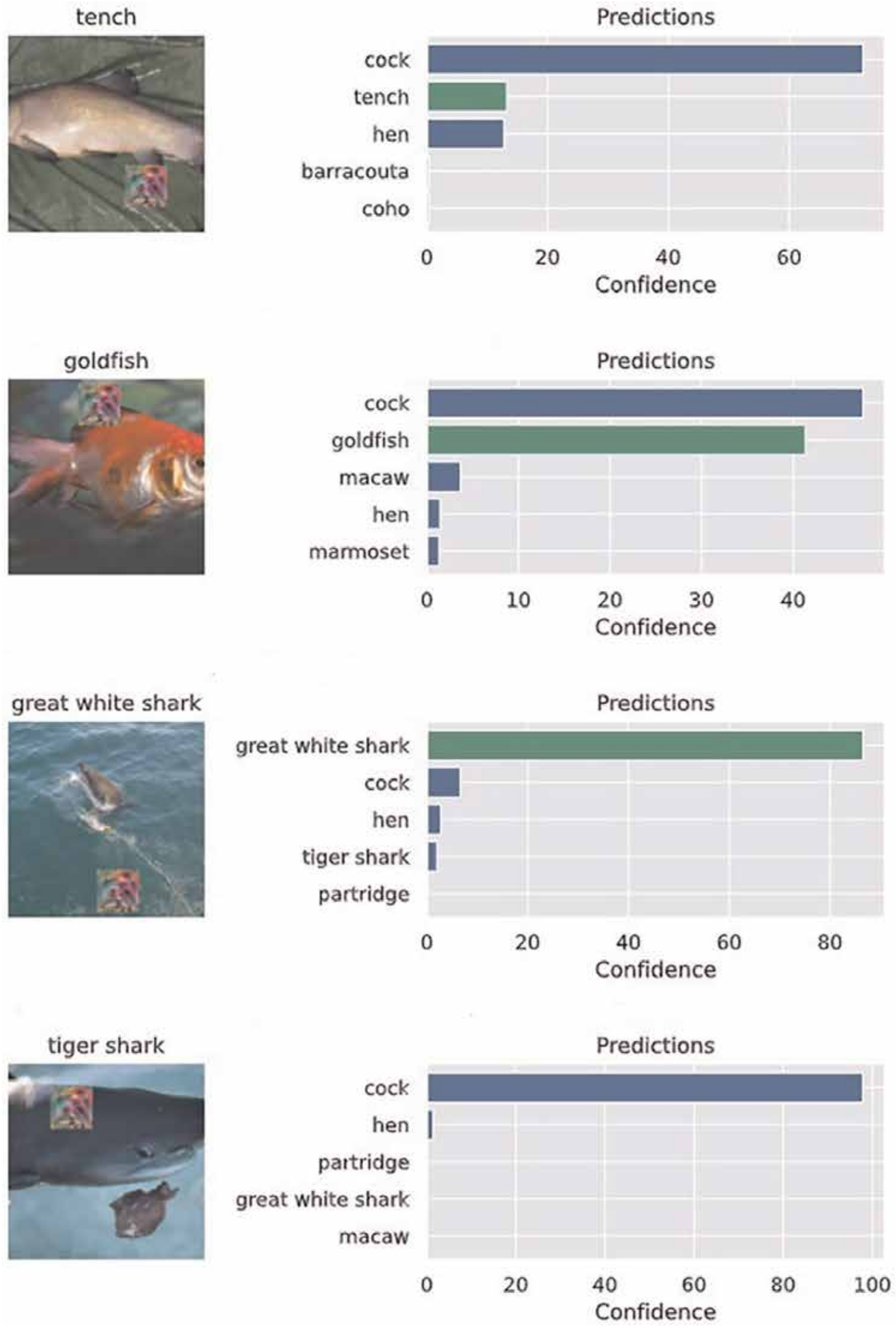

**Figure 8.**
*The classification results of the ResNet-34 model in the presence of a patch image of a balloon with size 64*64.*

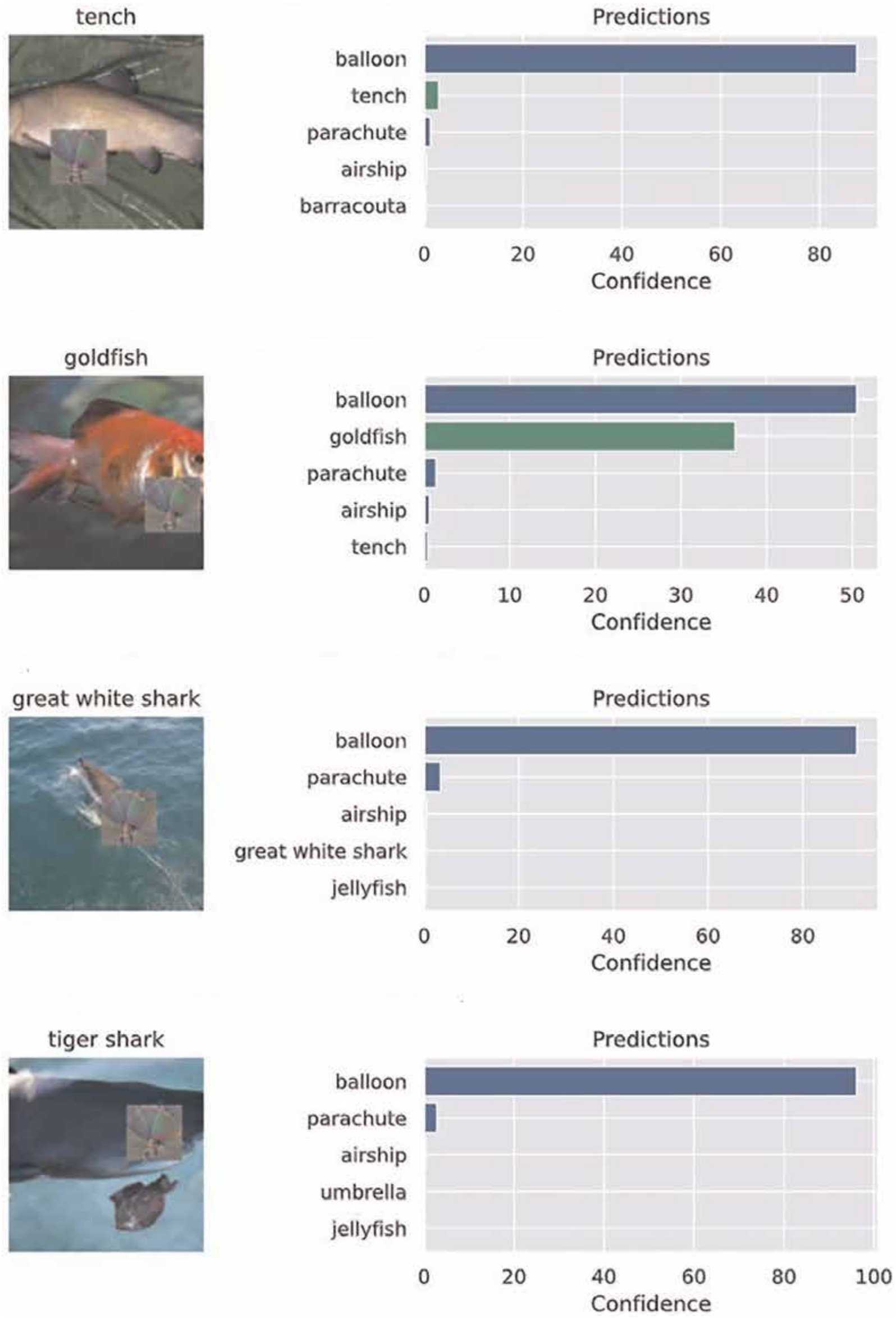

**Figure 9.**
*The classification results of the GoogleNet model in the presence of a patch image of a balloon with size 64*64.*

## 5. Conclusion

In this chapter, some adversarial attacks on CNN-based image classification models are discussed. In particular, two attacks, e.g., the FGSM attack and adversarial patch attack are presented in detail. The former attack involves changing the pixels of an image in the direction of their maximum gradients so that the value of the loss function is maximized. While the resultant adversarial image is impossible to distinguish from the original image by human eyes, the highly trained classification models will most likely classify the adversarial image into a class that is different from its ground truth. For the adversarial patch attack, an image patch of a different class is inserted in the original image in such a way that the trained models will be deceived and forced to incorrectly classify the original image into the class of the patch image. It is observed in the study that with the increase in the amount of perturbation created in the original image by the FGSM attack, the error in the classification increases till a threshold level is reached at which the attack saturates. No further increase in perturbation usually leads to a further decrease in the classification accuracy of the models. For the adversarial patch attack, the attack effectiveness increases with the increase in the patch size.

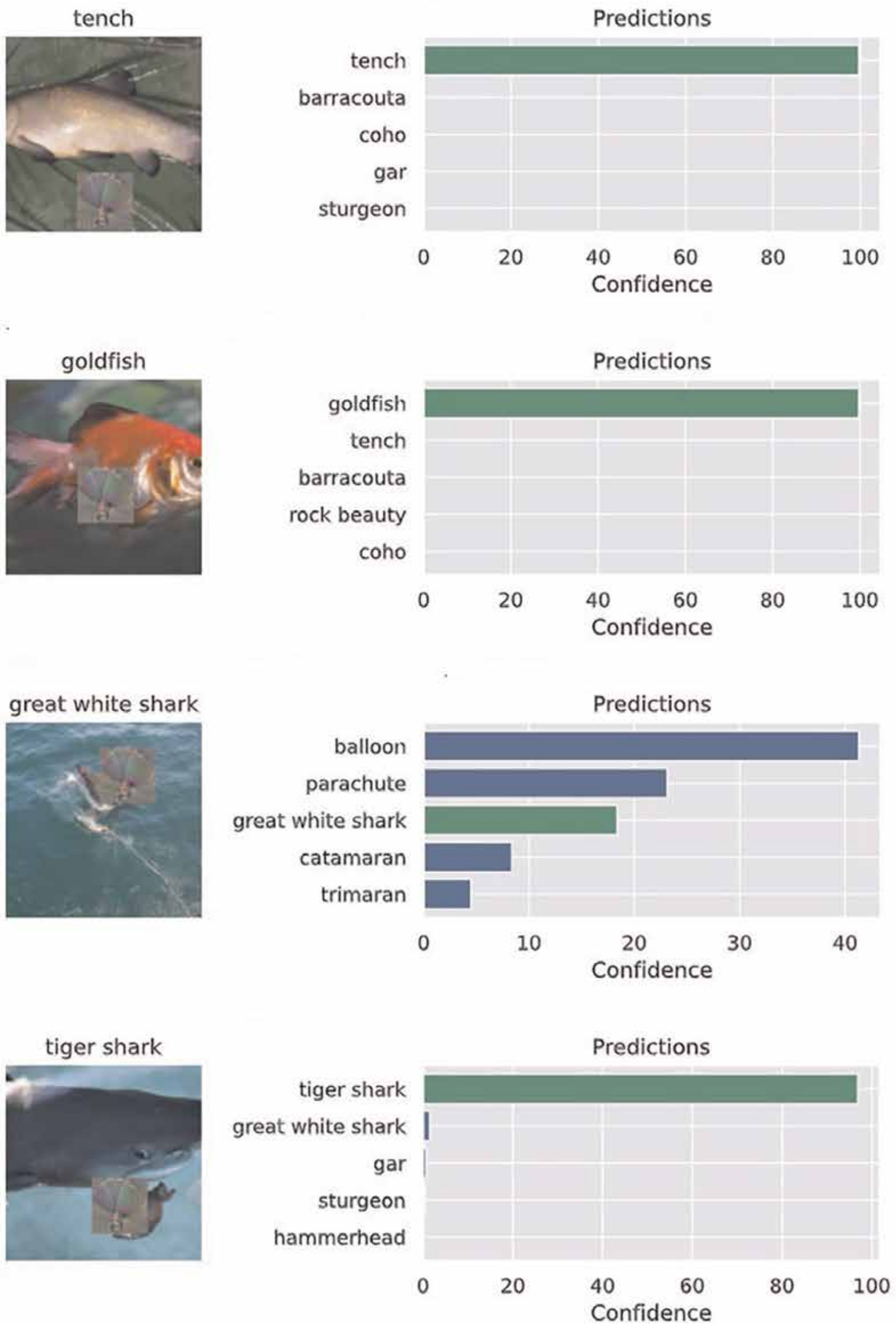

**Figure 10.**
*The classification results of the DenseNet-161 model in the presence of a patch image of a balloon with size 64*64.*

## Author details

Jaydip Sen* and Subhasis Dasgupta
Department of Data Science, Praxis Business School, Kolkata, India

*Address all correspondence to: jaydip.sen@acm.org

**Chapter 3**

# Recent Results on Some Word Oriented Stream Ciphers: SNOW 1.0, SNOW 2.0 and SNOW 3G

*Subrata Nandi, Srinivasan Krishnaswamy and Pinaki Mitra*

## Abstract

In this chapter, we have studied three word-oriented stream ciphers SNOW 1.0, SNOW 2.0 and SNOW 3G in a detailed way. The detailed description includes the working principles of each cipher, security vulnerabilities and implementation issues. It also helps us to study the challenges in each cipher. As SNOW 3G is used as a confidentiality and integrity component in 3G, 4G and 5G communications, the after study of this article may instigate the reader to find the fixes from different cryptanalysis and also find a new suitable design in Mobile telephony security.

## 1. Introduction

In modern era of communication, mobile devices, tablets, computers are developed with huge processing power, memory and storage. When one mobile device communicates with a remote server or another mobile device, the communication always takes place in a secret way. For any bank transaction or any online purchase through PC, we always require communication link between the source and the server which ensures the authentication, confidentiality and integrity of the channel. Before the year 1999, Block ciphers was the only way to provide confidentiality in any kind of communication in word oriented environment. AES, DES, 3-DES, BlowFish, Serpent, Twofish are some of the common used block cipher algorithms. The problems associated with block ciphers are mainly processing power, throughput in comparison to stream cipher. Stream cipher works very efficiently in hardware due to its simple design and good statistical properties. But it lacks in software based applications. But is it possible to make a word-oriented cipher which will be faster than Block cipher, at least gives security of AES [1] (Advance Encryption Standard) and suits in software as well as hardware? This initiate the design and analysis of word-oriented stream cipher. The basic building block of word oriented stream cipher is designed by LFSR (Linear Feedback Shift Register) with multi input multi output (MIMO) delay blocks and a Nonlinear function. In this kind of design, LFSR plays the role of generating sequence with uniform distribution. It generates $m-$sequence. As, sequence from LFSR can be easily cryptanalyzed by Barleycamp Massey Algorithm [2], Nonlinear maps are used along with LFSR to increase the linear complexity as well as nonlinearity of the output sequence. Generally, S-box, Addition modulo $2^n$ ($\boxplus$), Subtraction modulo $2^n$ ($\boxminus$) are used as Nonlinear function. Word-based Cipher acts as a pseudorandom number generator (PRNG) which produces vector in each clock cycle as keystreams. Plaintext block and keystream block are encrypted with bitwise EXOR operator and it creates cipher text block. In the receiving end, cipher text block and the same keystream generator produces the plaintext block using the same bitwise exor operator. In the next subsection, we discuss about some existing word-based LFSR.

### 1.1 Related work

The research on word-based stream cipher was coined by Bart Preneel in FSE 1994. In NESSIE (New European Schemes for Signature, Integrity, and Encryption) competition (2000–2003), six stream ciphers (BMGL, Leviathan, LILI-128, SNOW [3], SOBER-t16 [4] and SOBER-t32 [5]) were submitted. Among which SNOW 1.0, SOBER-t16 and SOBER t-32 were found as word oriented stream ciphers. In 2002, SNOW, SOBER-t16 and SOBER t-32 were found with security flaws with certain cryptographic attacks (Distinguishing attack, Guess and Determine attack and Linear cryptanalysis). In 2002, SNOW 2 [6] was proposed by Ekdahl and Johansson, the same author published SNOW 1.0. But two cryptographic attacks, Algebraic attack and Linear Distinguishing attack made SNOW 2.0 vulnerable. After SNOW 2.0, in 2006 SNOW 3G and in 2008 Sosemanuk [7] (as a Estream finalist) came into the literature. SNOW 3G was selected as 3GPP Confidentiality and Integrity Algorithms UEA2 and UIA2. It was also analyzed by Fault Analysis attack [8] in 2009. Sosemanuk cipher is an modified of version of SNOW 2.0. There are some attacks on Sosemanuk like Linear masking method [9], byte based guess and determine attack [10]. Ekdahl et al. recently proposed SNOW-V [11] stream cipher with the feature of 256-bit security and huge throughput in 5G environment. Still, we find fast correlation attack [12] with $2^{251.93}$ complexity and improved guess and determining attack [13] with $2^{406}$ complexity on SNOW-V.

In this literature, we are trying to present detailed study of SNOW 1.0, SNOW 2.0 and SNOW 3G and discuss the basic problems related to it.

## 2. Symbols used and their meaning

The following mathematical symbols will be used in this article (**Table 1**).

## 3. Preliminaries

In this section, we present some definitions and related concepts useful to understand the context of next sections.

**Primitive polynomial:** A polynomial that generates all elements in an extension field over a base field is called Primitive Polynomial. It is also irreducible polynomial. There are $\frac{\phi(q^n-1)}{n}$ primitive polynomials of degree $n$ in $\mathcal{GF}(q)[X]$, where $\phi$ is the Euler phi function.

**Example 1.** $x^4+x+1$ and $x^4+x^3+1$ are two primitive polynomials of $\mathcal{GF}(2^4)$.

**Linear feedback shift register (LFSR):** LFSR is an important source of PRNG in stream cipher design. It is very fast and easy to implement in hardware. It consists of some D flip flops and a feedback polynomial. If $f$ is the primitive polynomial of degree $n$ and $\{x_1, x_2, \cdots, x_n\}$ where each $x_i \in \mathbb{F}_2$, is the state of the LFSR, the state update function of the LFSR $L$ is defined:

| Symbol | Meaning |
|---|---|
| $\mathbb{F}_{b^n}$ | Finite field of cardinality $p^n$, where b is a prime number |
| $\mathbb{F}_p^n$ | $n$-dimensional vector space over $\mathbb{F}_p$ |
| $GF(P)$ | Galois field with elements $\in \{0, 1, \cdots, p-1\}$ |
| $v^T$ | A vector $v$ with transposed form $v = (v_1, v_2, \cdots, v_n)$ |
| $\oplus$ | XOR |
| $\boxplus$ | Addition modulo $2^{32}$ |
| $M^{n \times n}$ | Matrix M with $n$ rows and $n$ columns |
| $x \lll$ | Cyclic shift of x to 7 step left |
| $M_m(F_2)$ | Matrix Ring of $m \times m$ matrices over Finite Field $F_2$ |

**Table 1.**
*Symbol and their meaning.*

$$L : \{x_1, x_2, \cdots, x_n\} \rightarrow \{f(x_1, x_2, \cdots, x_n), x_2, \cdots, x_{n-1}\} \tag{1}$$

If the feedback polynomial of a LFSR is primitive, it can generate all the nonzero states in its period. But LFSR based PRNG is vulnerable to Barleycamp Massey attack. It finds the initial state and the feedback polynomial of the LFSR if $2 \times n$ keystream can be accessed from the LFSR state. So, various forms of nonlinear feedback shift register (NLFSR) like Nonlinear combiner generator, Nonlinear feedforward generator, Clock control generator are used as a keystream generator to resist BMA attack. But LFSRs are slow in smartphone, PC, embedded system applications with respect to word oriented operation. So word oriented PRNG's like RC4, SOBER, SNOW, SNOW 2.0 etc. came to the market to serve the purpose of PRNG. The important factor in word oriented LFSR is primitive polynomial over extension field. These papers [14–16] are a good source of materials to study primitive polynomials over extension field.

Let $b$ be the number of $m$ input output delay blocks ($D_0, D_1, \cdots, D_b$ where each $D_i \in \mathbb{F}_2^m$) and gain matrices $B_0, B_1, \cdots, B_{b-1} \in \mathbb{F}_2^{m \times m}$ of a multi-input multi-output LFSR (MIMO LFSR). Initial state of the MIMO LFSR is of $mb$ bits. The state update function of a $\sigma-$LFSR, $A_{mb}$ is defined as:

$$A_{mb} = \begin{bmatrix} 0 & I & 0 & \cdots & 0 \\ 0 & 0 & I & \cdots & 0 \\ \vdots & \vdots & \vdots & \cdots & \vdots \\ 0 & 0 & 0 & \cdots & I \\ B_0 & B_1 & B_2 & \cdots & B_{b-1} \end{bmatrix} \in \mathbb{F}_2^{mb \times mb}$$

where $0, I \in \mathbb{F}_2^{m \times m}$ are all zero and identity matrix respectively. The characteristic polynomial of $A_{mb}$,

$$f(x) = x^n + B_{b-1}x^{n-1} + B_{b-2}x^{n-2} + \cdots + B_0 \tag{2}$$

is called a primitive polynomial over $\mathbb{F}_{2^m}$ if periodicity of the polynomial is $2^{mb} - 1$. Primitive MIMO LFSR is a good PRNG as the keystreams generated from it satisfy balancedness, span-n, 2-level autocorrelation property according to Golomb's randomness criterion. But only LFSR cannot be used as a PRNG due to small linear complexity. Linear complexity is the length of the smallest linear feedback shift register which can generate the sequence. To increase the linear complexity, nonlinear functions are used in PRNG along with LFSR (**Figure 1**).

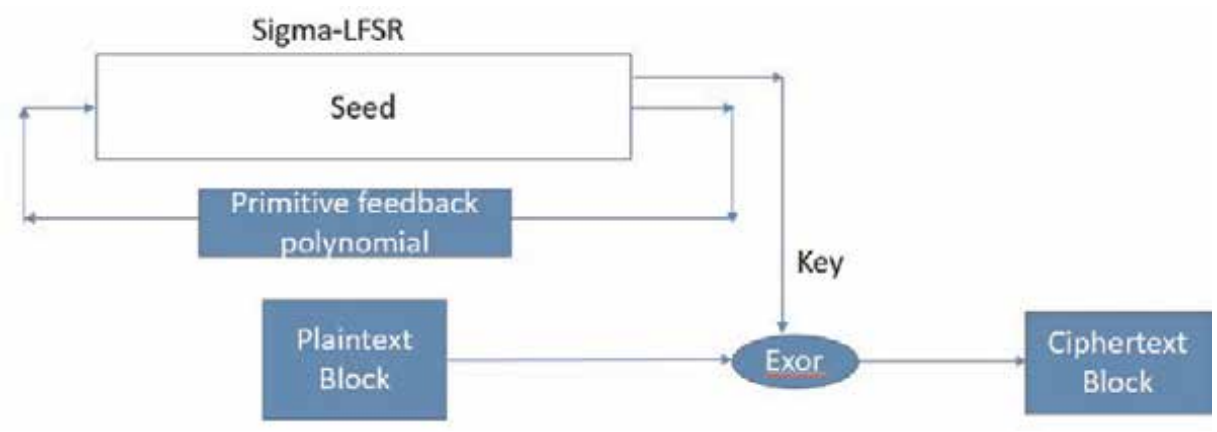

**Figure 1.**
*Word oriented LFSR based Encryption.*

**Substituation Box (S-Box):** An S-Box or substitution box $f$ is a vectorial Boolean function [17] which is defined as follows:

$$f : \mathbb{F}_{2^n} \rightarrow \mathbb{F}_{2^n}$$

It is nothing but the permutation of $n$ elements from one set to another. We can represent $f$ as $(f_1, f_2, \cdots, f_n)$ where each $f_i$ is the component Boolean function [18] of the S-box.

$$f_i : V_n \rightarrow \mathbb{F}_2$$

There are $n!$ S-box'es for a set of $n$ elements. We can categorize S-boxes into two section.

1. **Affine S-box:** If all the component functions are affine functions.

2. **Non Affine S-box:** If at least one component function is nonlinear function.

In cryptology, researchers are interested on Non affine S-boxes whose all component functions are nonlinear. S-box should have good cryptographic characteristics such as balancedness, good nonlinearity, resiliency, optimal algebraic immunity, good differential uniformity [19].

**Example 2.** One of the S-boxes used in DES(Data Encryption Standard) is:

| **0** | **1** | **2** | **3** | **4** | **5** | **6** | **7** | **8** | **9** | **10** | **11** | **12** | **13** | **14** | **15** |
|---|---|---|---|---|---|---|---|---|---|---|---|---|---|---|---|
| | 4 | 13 | 1 | 2 | 15 | 11 | 8 | 3 | 10 | 6 | 12 | 5 | 9 | 0 | 7 |

*There are four boolean function with respect to this S-box. Their Algebraic Normal Form(ANF) are:*

1. $f_1 : y0^*y1^*y3 + y0^*y2 + y0^*y3 + y1 + y3$

2. $f_2 : y0^*y1 + y0^*y2^*y3 + y0^*y2 + y0^*y3 + y0 + y1^*y2^*y3 + y1^*y2 + y1^*y3 + y1 + y2^*y3 + 1$

3. $f_3 : y0^*y1^*y3 + y0^*y1 + y0^*y2^*y3 + y0 + y1^*y2^*y3 + y1^*y2 + y1^*y3 + y2^*y3 + y3 + 1$

4. $f_4 : y0^*y1^*y3 + y0^*y1 + y0^*y2 + y1^*y3 + y2 + y3 + 1$

## 4. SNOW 1.0 KSG

In this section, we demonstrate SNOW 1.0 Keystream generator and various attacks possible on it (**Figure 2**).

SNOW 1.0 consists of two parts as LFSR part and FSM part. The LFSR of SNOW 1.0 has 16 delay blocks $St_i$, each can store 32 values. It means $St_i \in \mathbb{F}_{2^{32}}$. The LFSR has a primitive feedback polynomial over $\mathbb{F}_{2^{32}}$ which is

$$p(y) = y^{16} + y^{13} + y^7 + \alpha \tag{3}$$

where $\alpha$ is the generating element of $\mathbb{F}_{2^{32}}$. The irreducible polynomial $f(y)$ used to generate $\mathbb{F}_{2^{32}}$ as an ideal is $f(y) = y^{32} + y^{29} + y^{20} + y^{15} + y^{10} + y + 1$ such that $f(\alpha) = 0$.

$$\mathbb{F}_{2^{32}} = \mathbb{F}_2[y] / f(y)$$

The FSM (Finite State Machine) part comprised of registers $Reg1$, $Reg2 \in \mathbb{F}_{2^{32}}$ and substitution box $S$,

$$\begin{aligned} Fm &: \{0,1\}^{32} \times \{0,1\}^{32} \to \{0,1\}^{32} \\ Fm_t &= (St_{t+i} \boxplus Reg1_t) \oplus Reg2_t \end{aligned} \tag{4}$$

Here, $\boxplus$ operator is integer addition modulo $2^{32}$ such as $x \boxplus y = (x + y)(\text{mod} 2^{32})$. In the FSM the registers are updated as follows.

$$Reg1_{t+1} = ((Fm \boxplus Reg2_t) \lll) \oplus Reg1_t \tag{5}$$

$$Reg2_{t+1} = S(Reg1_t) \tag{6}$$

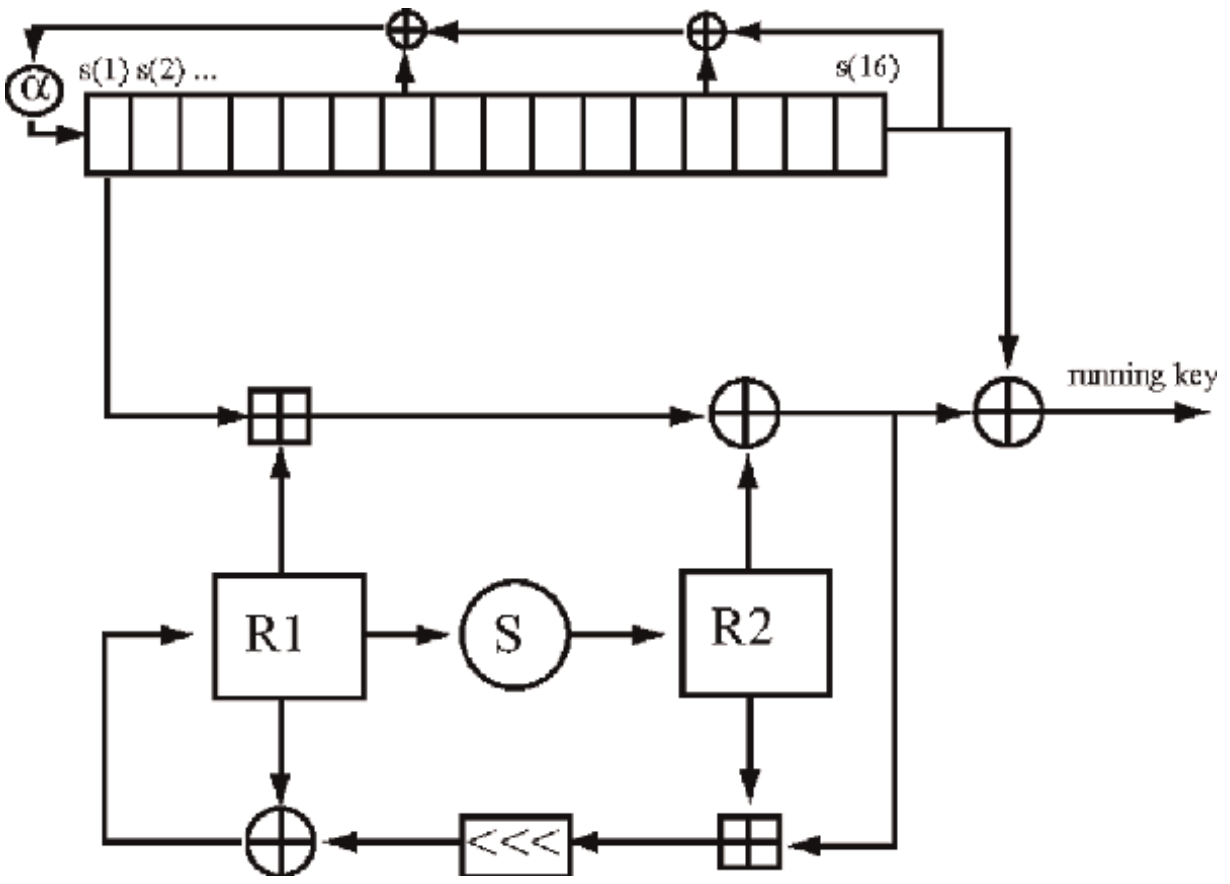

**Figure 2.**
*Block Diagram of SNOW 1.0.*

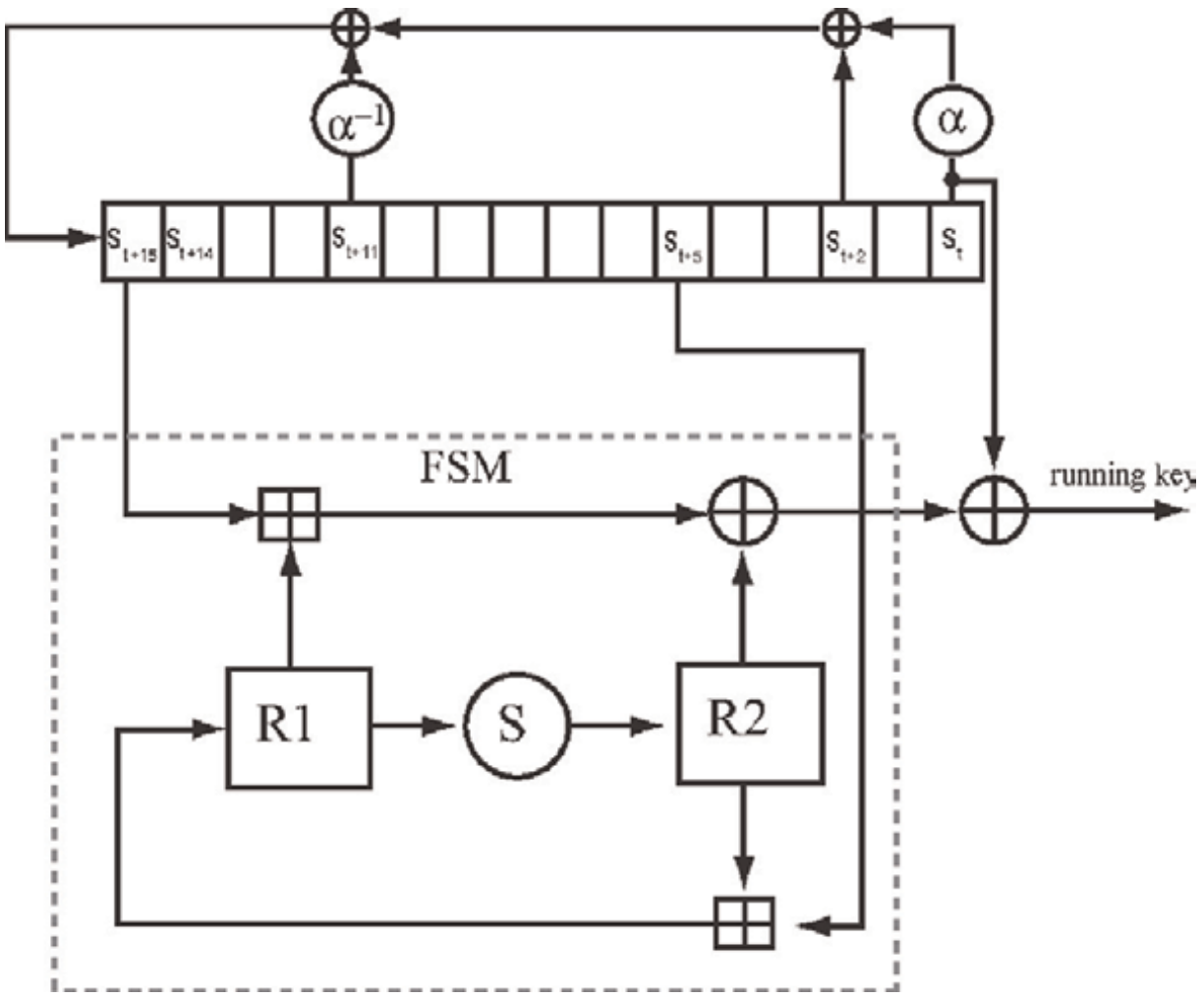

**Figure 3.**
*Block Diagram of SNOW 2.0.*

The S-box $S$ operation acts as follows (**Figure 3**). The input y is broken into 4 bytes. Each of the bytes are changed to another byte by a nonlinear mapping from 8 to 8 bits. The nonlinear mapping is

$$x = Y^7 \oplus \beta^2 \oplus \beta \oplus 1 \tag{7}$$

where $x$ is output of nonlinear map, $Y$ is the input and both considered to be representing elements $\in \mathbb{F}_{2^8}$ with the polynomial basis $\{\beta^7 \cdots \beta, 1\}$. $\beta$ is the root of the irreducible polynomial $h(y) = y^8 + y^5 + y^3 + y + 1$ such that $h(\beta) = 0$. The nonlinear mapping followed by a permutation of the bits in the output word.

### 4.1 SNOW 1.0 algorithm

In this section, we demonstrate the working nature of SNOW 1.0. It starts with Initialization, LFSRupdate, FSMupdate and finally ends with SNOW 1.0 algorithm.

---

**Algorithm 1:** Initialization().

---

**Input:** Key $K = (k_1, \cdots, k_8) \in \mathbb{F}_{2^{256}}$ where each $k_i \in \mathbb{F}_{2^{32}}$ and Initialization vector $IV = (IV_2, IV_1) \in \mathbb{F}_{2^{64}}$ where each $IV_2, IV_1 \in \mathbb{F}_{2^{32}}$.

1: $(St_t, \cdots, St_{t+15}) \leftarrow (k_1 \oplus IV_1, k_2, k_3, k_4 \oplus IV_2, k_5, k_6, k_7, k_8, k_1 \oplus 1, k_2 \oplus 1, k_3 \oplus 1, k_4 \oplus 1, k_5 \oplus 1, k_6 \oplus 1, k_7 \oplus 1, k_8 \oplus 1)$
2: $(Reg1_t,\ Reg2_t) \leftarrow (0, 0)$
3: **for all** $t \leftarrow 1$ to 32 **do**
4: $Fm_t \leftarrow (Reg1_t \boxplus St_t) \oplus Reg2_t$
5: $FSMupdate()$
6: $LFSRupdate()$
7: $(St_t, \cdots, St_{t+15}) \leftarrow (St_t, \cdots, St_{t+15}) \oplus Fm_t$
8: **end for**

---

---

**Algorithm 2:** FSMupdate().

---

**Input:** $St_t, Fm_t$
**Output:** Output of the FSM $Fm_t$ at time $t$. [19]
$Fm_t \leftarrow (Reg1_t \boxplus St_t) \oplus Reg2_t$
$Reg1_{t+1} = ((Fm_t \boxplus Reg2_t) \lll) \oplus Reg1_t$
$Reg2_{t+1} = S(Reg1_t)$
$Reg1_t = Reg1_{t+1}$
$Reg2_t = Reg2_{t+1}$

---

**Algorithm 3:** LFSRupdate()

---

$temp = \alpha(St_t \oplus St_{t+3} \oplus St_{t+9})$
$(St_t, \cdots, St_{t+15}) \leftarrow (temp, St_t, \cdots, St_{t+14})$
**Algorithm 4** SNOW 1.0().
**Input:** $K, IV$
**Output:** Keystream $z_t$ at time t.
$Initialization(K, IV)$
$t \leftarrow 1$
**while** $t \leq GivenNumber$ **do**
$z_t = Fm_t \oplus St_t$
$FSMupdate()$
$LFSRUpdate()$
Output $z_t$
$t \leftarrow t + 1$
**end while**

---

### 4.2 Weaknesses in SNOW 1.0

1. **Guess and determine attack**: It is one type of key recovery attack. It [20] utilizes the relationship between internal values (recurrence relation in a shift register) and the relationship used to establish the key-stream values from the registers values. In this attack, value of some registers are guessed and then the relationships are utilized to find other internal values.

   The problem in SNOW 1.0 is the recurrence relation

$$St_{t+16} = \alpha(St_t \oplus St_{t+3} \oplus St_{t+9}) \tag{8}$$

   If we square Eq. (1),

$$St_{t+32} = \alpha^2(St_t \oplus St_{t+6} \oplus St_{t+18}) \tag{9}$$

   We can find out the distance of three words between $St_t$ and $St_{t+3}$, and the distance of 6 words between $St_{t+3}$ and $St_{t+9}$. So the attacker can use $(St_{t+i} \oplus St_{t+6+i})$ as a single input to both the equation. Another aspect the use of $\lll$ operator (Circular shift operator) helps in finding relation between FSM and Reg2 (**Table 2**).

| | Data complexity | Process complexity |
|---|---|---|
| Guess-and-determine attack (method 1) | $\mathcal{O}(2^{64})$ | $\mathcal{O}(2^{256})$ |
| Guess-and-determine attack (method 2) | $\mathcal{O}(2^{95})$ | $\mathcal{O}(2^{100})$ |

**Table 2.**
*GD attack complexity*

2. **Distinguishing attack**: In this kind of attack linear approximation of the nonlinear part is done first and combined with the linear part. Coppersmith et al. [21] observed that only $\alpha$ present which in $\mathbb{F}_{2^{32}}$. Using Frobenious automorphism ($\phi : y \to y^{2^{32}}$) they eliminated $\alpha$ and gave a new linear relation over $GF(2)$.

$$p(y)^{2^{32}} + p(y) = y^{16\times 2^{32}} + y^{13\times 2^{32}} + y^{7\times 2^{32}} + y^{16} + y^{13} + y^{7} \tag{10}$$

The best linear approximation of the two consecutive round input outputs of the FSM from the following.

$$\delta = (St_t)_{15} \oplus (St_t)_{16} \oplus (St_{t+1})_{22} \oplus (St_{t+1})_{23} \oplus (Fm_t)_{15} \oplus (Fm_{t+1})_{23} \tag{11}$$

where $(St_t)_k$ signify k th bit of $St_0$ state of the LFSR at time $t$. The bias of the linear approximation evaluated was at least $2^{-9.3}$. And the author calculated $2^{101.6}$ rounds keystream requirement for distinguishing the output sequence from SNOW 1.0 and the sequences from true random bit generator.

## 5. SNOW 2.0 KSG

This section discusses all about SNOW 2.0 Keystream generator. We also mention about some cryptographic attacks on SNOW 2.0.

SNOW 2.0 is the updated KSG over SNOW 1.0. Here, the primitive polynomial over $GF(2^{32})$ is chosen by studying the weakness of the primitive polynomial in SNOW 1.0. Let $\delta$ be the generating element of the primitive polynomial $f(y) = y^8 + y^7 + y^5 + y^3 + 1$, such that $f(\delta) = 0$ and $\alpha$ be the generator of the primitive polynomial $g(y) = y^4 + \delta^{23}y^3 + \delta^{245}y^2 + \delta^{48}y + \delta^{239}$ such that $g(\alpha) = 0$. We can represent each element in $\mathbb{F}_{2^{32}}$ with the help of the basis $\{\alpha^3, \alpha^2, \alpha, 1\}$. Using the above 2 extension fields the generator polynomial of SNOW 2.0

$$H(y) = \alpha y^{16} + y^{14} + \alpha^{-1}y^5 + 1 \in \mathbb{F}_{2^{32}}[Y] \tag{12}$$

is calculated and the recurrence relation of $H(y)$ is as follows:

$$St_{t+15} = \alpha^{-1}St_{t+11} + St_{t+2} + \alpha St_t \tag{13}$$

where $St_t \in \mathbb{F}_{2^{32}}$ is the state of the first delay block in clock time $t$.

The FSM part of SNOW 2.0 is same as SNOW 1.0, except $St_{t+5}$ is used as a input to the FSM. It makes more dependency of state vectors to the FSM. We can evaluate the FSM $Fm_t$ as:

$$Fm_t = (St_{t+15} \boxplus Reg1_t) \oplus Reg2_t \tag{14}$$

and the keystream $z_t$ is given by

$$z_t = Fm_t \oplus St_t \tag{15}$$

The updation of registers $Reg1_{t+1}$, $Reg2_{t+1}$ from $Reg1_t$, $R2_t$ are related as follows:

$$Reg1_{t+1} = St_{t+4} \boxplus Reg2_t \tag{16}$$

$$Reg2_{t+1} = S(Reg1_t) \tag{17}$$

Here S is the S-box which takes 4 bytes $(b_0, b_1, b_2, b_3)$ as input and uses AES S-box followed by mixcolumn operation to output 4 bytes.

$$\begin{bmatrix} b_0^{t+1} \\ b_1^{t+1} \\ b_2^{t+1} \\ b_3^{t+1} \end{bmatrix} = \begin{bmatrix} X & X+1 & 1 & 1 \\ 1 & X & X+1 & 1 \\ 1 & 1 & X & X+1 \\ X+1 & X & 1 & 1 \end{bmatrix} \begin{bmatrix} S(b_0^t) \\ S(b_1^t) \\ S(b_2^t) \\ S(b_3^t) \end{bmatrix} \tag{18}$$

In the above equation, the matrix used is for Mixcolumn operation where the value in $X \in F_{2^8}$ and the S-box $S : \mathbb{F}_{2^8} \to \mathbb{F}_{2^8}$ is a permutation function used in SubByte step defined as:

$$S(y) = \begin{cases} 0, & \textbf{if} \, y = 0 \\ y^{-1}, & \forall y \in \mathbb{F}_2^8 - \{0\} \end{cases}$$

### 5.1 Key initialization

In SNOW 2.0128 bits or 256 bits key (K) and a initialization vector IV (public) is used. The $IV \in \{0, 1, \cdots, 2^{128} - 1\}$ and the two memory registers are set to 0. The cipher is then clocked 32 times where no keystream is produced and the FSM output is feeded as following:

$$St_{t+15} = \alpha^{-1} St_{t+11} \oplus St_{t+2} \oplus \alpha St_t \oplus Fm_t \tag{19}$$

The cipher is then switched into the normal mode, but the first output of the keystream is discarded. After $2^{50}$ keystream the cipher's key $K$ is changed to a new value for resisting from cryptanalysis.

### 5.2 SNOW 2.0 algorithm

In this section, we describe the working principle of SNOW 2.0 algorithm which consists of Initialization, LFSRupdate, FSMupdate.

---

**Algorithm 5:** Initialization().

---

**Input:** Key $K = (k_0, \cdots, k_7) \in \mathbb{F}_{2^{256}}$ where each $k_i \in \mathbb{F}_{2^{32}}$ and Initialization vector $IV = (IV_3, IV_2, IV_1, IV_0) \in \mathbb{F}_{2^{128}}$ where each $IV_i \in \mathbb{F}_{2^{32}}$.

$(St_t, \cdots, St_{t+15}) \leftarrow (k_0 \oplus 1, k_1 \oplus 1, k_2 \oplus 1, k_3 \oplus 1, k_4 \oplus 1, k_5 \oplus 1, k_6 \oplus 1, k_7 \oplus 1, k_0, k_1 \oplus IV_3, k_2 \oplus IV_2, k_3, k_4 \oplus IV_1, k_5, k_6, k_7 \oplus IV_0)$
$(Reg1_t, Reg2_t) \leftarrow (0, 0)$
**for all** $t \leftarrow 1$ to 32 **do**
$Fm_t \leftarrow (Reg1_t \boxplus St_{t+15}) \oplus Reg2_t$
$FSMupdate()$
$LFSRupdate()$
$(St_t, \cdots, St_{t+15}) \leftarrow (St_t, \cdots, St_{t+15}) \oplus Fm_t$
**end for**

---

**Algorithm 6:** FSMupdate()

---

**(Input)** $St_{t+15}, St_{t+5}$
**Output:** Output of the FSM $Fm_t$ at time $t$.
$Fm_t \leftarrow (Reg1_t \boxplus St_{t+15}) \oplus Reg2_t$
$Reg1_{t+1} = St_{t+5} \boxplus Reg2_t$
$Reg2_{t+1} = St(Reg1_t)$
$Reg1_t = Reg1_{t+1}$
$Reg2_t = Reg2_{t+1}$

---

**Algorithm 7:** LFSRupdate()

---

$temp = \alpha^{-1} St_{t+11} \oplus St_{t+2} \oplus \alpha St_t$
$(St_{t+15}, \cdots, St_t) \leftarrow (temp, St_{t+15}, \cdots, St_{t+1})$

---

**Algorithm 8:** SNOW 2.0()

---

**Input:** $K, IV$
**Output:** Keystream $z_t$ at time t.
$Initialization(K, IV)$
$t \leftarrow 1$
**while** $t \leq 2^{50}$ **do**
$z_t = F_t \oplus S_t$
$FSMupdate()$
$LFSRUpdate()$
Output $z_t$
$t \leftarrow t + 1$
**end while**

---

## 5.3 Cryptographic attack on SNOW 2.0

### *5.3.1 Distinguishing attack*

In this kind of attack a distinguisher algorithm is constructed to distinguish the output keystream from a PRNG and same length output from a true random number generator. If the distinguishing algorithm complexity is less than the brute force search algorithm, this is called an attack on the cipher.

1. Watanabe et al. [21] 2003 used linear masking method to distinguish the output of SNOW 2.0 from a TRNG. Basically it tries to find out linear relation between the output of the keystream with FSM and LFSR. SO to serve this purpose we need to find a mask $\mathcal{T} \in \mathbb{F}_{2^{32}}$ with high bias such that

$$\mathcal{T}St_{t+16} \oplus \left(\mathcal{T}.\alpha^{-1}\right).St_{t+11} \oplus \mathcal{T}St_{t+2} \oplus (\mathcal{T}\alpha).St_t = 0 \tag{20}$$

holds. The 2 rounds approximation of FSM

$$\mathcal{T}_0St_t \oplus \mathcal{T}_1St_{t+1}\mathcal{T}_5St_{t+5} \oplus \mathcal{T}_{15}St_{t+15} \oplus \mathcal{T}_{16}St_{t+16} = \mathcal{T}_0z_t \oplus \mathcal{T}_1z_{t+1} \tag{21}$$

with assumption that all the masks values are same, becomes possible of two nonlinear approximation such as S-box and the three $\boxplus$ operator,

$$\mathcal{T}S(X) = \mathcal{T}X \tag{22}$$

$$\mathcal{T}(X \boxplus y) = \mathcal{T}X \oplus \mathcal{T}y \tag{23}$$

The bias of the total approximation can be found with complexity $\mathcal{O}\left(2^{-112.25}\right)$. So we need about $2^{225}$ words to distinguish SNOW 2.0 from true random bit sequence.

2. In 2006 FSE, Nyberg et al. [22] improved this attack by approximating FSM (Finite state machine) and output of the cipher with different linear mask $(\mathcal{T}, \lambda \in \mathbb{F}_2^{32})$

$$\mathcal{T}F(x) = \lambda x \tag{24}$$

$$\mathcal{T}(z_{t+16} \oplus z_{t+2}) \oplus \mathcal{T}\alpha.z_t \oplus \mathcal{T}\alpha^{-1}.z_{t+11} \oplus \lambda(z_{t+17} \oplus z_{t+3}) \oplus \lambda\alpha.z_{t+1} \oplus \lambda\alpha^{-1}.z_{t+12} = 0 \tag{25}$$

measured the bias of the above relation with correlation $(\mathcal{T}, \lambda)$ which is defined as:

$$correlation(\mathcal{T}, \lambda) = \left(\#\left\{x \in \mathbb{F}_2^{32} : \mathcal{T}F(x) = \lambda x\right\} - \#\left\{x \in \mathbb{F}_2^{32} : \mathcal{T}F(x)! = \lambda x\right\}\right) \tag{26}$$

They also investigated the diffusion property of Mixcolumn, improved the search complexity of linear distinguishing attack.

### *5.3.2 Correlation attack*

In correlation attack [23] over extension field, the correlation of output keystream with the LFSR output is calculated for a particular $N$ (# available words). If the correlation or bias value is far greater than $\frac{1}{2^n}$, we find linear relation between input and output and also find out the initial state of the LFSR. It is also one kind of key recovery attack. Another kind of correlation attack is Fast Correlation attack [24] where the each output of a keystream $z_i$ is written as:

$$z_i = u_i + e_i \tag{27}$$

$u_i$ is the output of the LFSR and $e_i$ is considered as error in the discrete memoryless channel which is the nonlinear function attached with LFSR. So, finding initial state of the LFSR is equivalent of solving the decoding problem in error correcting code. We consider LFSR as $(N, l)$ linear code, where $l$ is the size of the LFSR.

1. In 2008, Lee et al. [9], in Asiacrypt 2008 presented more improved result with second LFSR derivation technique to mount correlation attack with time complexity $\mathcal{O}(2^{212.38})$, space complexity $\mathcal{O}(2^{202.83})$ bits and data complexity $\mathcal{O}(2^{197.77})$ bits to find out the initial state of SNOW 2.0 LFSR.

2. In 2015, Bin Zhang et al. [25] published paper on Fast correlation attack on SNOW 2.0 over extension field to find out the initial state of the LFSR with data complexity $\mathcal{O}(2^{163.59})$, time complexity $\mathcal{O}(2^{164.15})$ which is $2^{49}$ times faster than the previous one.

3. In 2018, Funabiki et al. [26] updated the FCA attack on SNOW 2.0 by using a MILP aided search for the linear mask very efficiently. It also uses the k-tree algorithm in [25] to find the key of SNOW 2.0 in data complexity $\mathcal{O}(2^{162.34})$ and time complexity $\mathcal{O}(2^{162.92})$.

4. In 2020, Gond et al. [27] published their work on FCA on SNOW 2.0 by slightly modifying the idea of k-tree algorithm. It finds the key of SNOW 2.0 in data complexity $\mathcal{O}(2^{159.62})$ and time complexity $\mathcal{O}(2^{162.86})$.

**Theorem 6.1** *The carry bit $c_i$ in the addition $X \boxplus Y = Z$ is equal to zero with probability* $\frac{1}{2} + \frac{1}{2^{i+1}}$.

*5.3.3 Algebraic attack*

Any stream cipher can be expressed with respect to algebraic equations where the variables of the equations are nothing but the initial state of the LFSR. We know that the challenge is to solving system of nonlinear equations with respect to the keystreams available to the us. It is well known to us that Solving such system of nonlinear equations over finite field is NP-Hard problem. But there are some approaches in the literature to mount algebraic attack like linearization, Grobner basis, Finding low degree annihilators [28] of a Boolean function etc.

In 2005, [29] Olivier Billet et. el cryptanalyzed SNOW 2.0 with algebraic attack like following:

1. Assuming $\boxplus$ in the cipher as $\oplus$ operation, the equations at time stamp $t$ help in mounting algebraic attack:

$$Reg2_t = Reg2_0 + \sum_{i=0}^{t} z^i + \sum_{i=0}^{t} (St_{4+i} + St_{15+i} + St_i) \tag{28}$$

where only known information is output keystreams$(z)$.

$$Reg2_{t+1} = S(Reg1_t) \tag{29}$$

$$= S(Reg2_t + z^t + St_{15+t} + St_t) \tag{30}$$

From each S-boxes we can find 39 linearly independent equations, So total $39 \times 4 = 156$ quadratic equations can be found from equation (11) for one keystream. The authors took linearization as a tool to make $\sum_{i=0}^{2} \binom{544}{i} \approx 2^{17}$ many unknown variables. The equations can be solved using Graobner basis with the help of about less than 17 keystreams. It results to find the initial state of the cipher with time complexity $\mathcal{O}(2^{51})$.

#### *5.3.4 Guess and determining attack*

In this kind of attack, the attacker first assumes the value of some registers and determine the value of the rest registers following the guesses. Later, keystream is generated from the cipher. If the keystream is equal to the keystream found by known keystream, the guess is a valid one. The terminology for the minimum guessed values of the cipher is called guessed basis. First systematic algorithm was proposed by Ahmadi et al. [30] which was a Viterbi like algorithm. Guessed basis for this algorithm was 8 and time complexity of the algorithm is $\mathcal{O}(2^{265})$. The next updated result is found from the article [31] which uses two auxiliary equations. Moreover, the guessed basis for this result is 6 and time complexity of the algorithm is $\mathcal{O}(2^{192})$.

### 5.4 KDFC SNOW

To resist from Algebraic attack, another version of SNOW 2.0 called Key dependent feedback configuration (KDFC) SNOW is proposed in [32]. It replaces the LFSR over $\mathcal{F}_{2^{32}}$ by $\sigma-$LFSR over $M_{32}(F_2)$. Moreover, it also implements the idea of different feedback matrices [33] for the $\sigma-$LFSR based on the key of the cipher. The whole process of changing the existing feedback matrix is done on the initialization phase of the cipher (**Figure 4**).

KDFC Scheme hides the feedback matrix of the $\sigma-$LFSR which helps the SNOW 2.0 to resist from some known plaintext attacks like Algebraic attack, Distinguishing Attack, Some Fast correlation attacks, Guess and Determining Attacks.

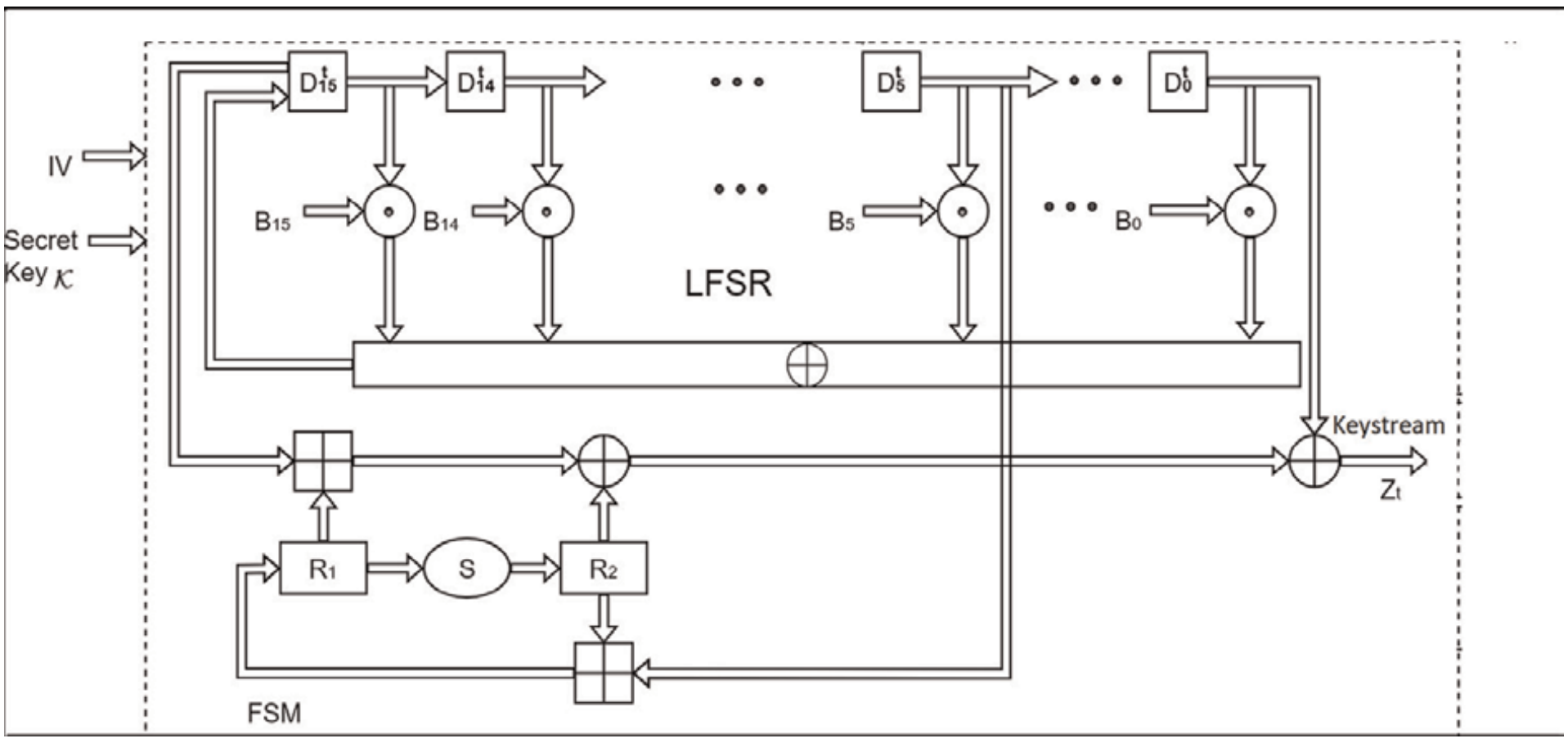

**Figure 4.**
*Block diagram of KDFC-SNOW.*

## 6. SNOW 3G KSG

SNOW 3G is an updated version of SNOW 2G. To get resistance from Algebraic Attack on SNOW 2.0, this cipher is proposed. It is used in mobile telephony for 4G and 5G communication as authenticated encrypted word oriented stream cipher in UEA2 and UIA2. The basic building block of SNOW 3G is as follows (**Figure 5**):

The LFSR configuration of SNOW 3G is as same as SNOW 2.0. The only changes are in the FSM configuration. One extra register R3 and extra Sbox is proposed by the designer. We discuss only the FSM construction of SNOW 3G in the following paragraph.

**FSM:** The FSM of SNOW 3G consists of two input words $St_{15}$ and $St_5$. It generates a 32-bit output word FW.

$$Fw = (St_{15} \boxplus R1) \oplus St_5 \tag{31}$$

On the next step, the registers are updated. To do so, we calculate one value $r$ like following:

$$r = R2 \boxplus (R3 \oplus St_5) \tag{32}$$

Next, we update the registers

$$R3 = SBox2(R2) \tag{33}$$

$$R2 = SBox1(R1) \tag{34}$$

$$R1 = r \tag{35}$$

**SBox1:** This Sbox is same as the Sbox described in SNOW 2.0.

**SBox2:** The SBox2 (S2 in Eq. (23)) is constructed using the Dickson polynomial. It is defined as an element of $GF(2^8)$ which is generated by $x^8 + x^6 + x^5 + x^3 + 1$.

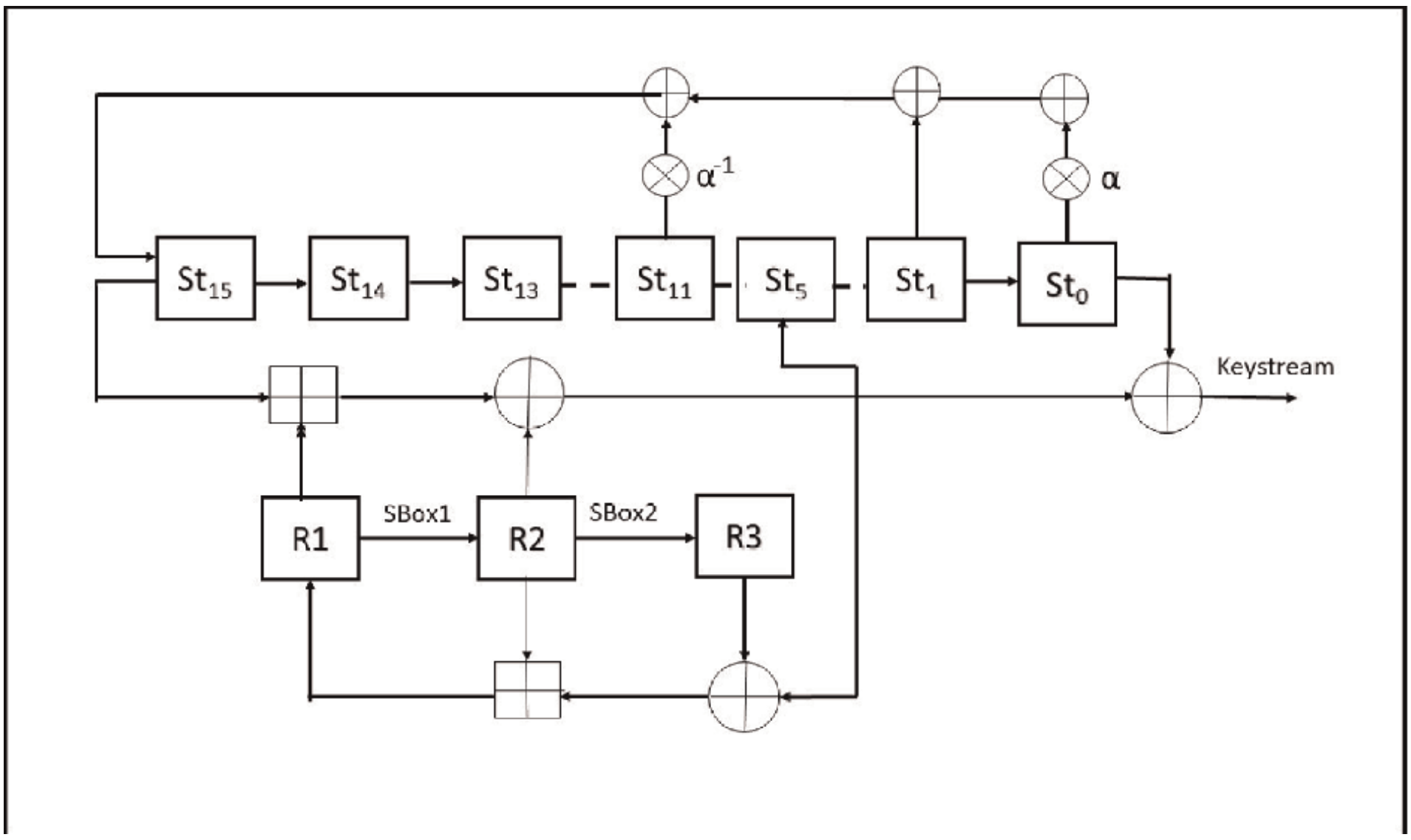

**Figure 5.**
*Block diagram of SNOW 3G.*

Suppose $b_i \in GF(2^8)$, using that we find an 32−bits inputs $b = b_1||b_2||b_3||b_4$ which is given as a input to the SBox2. It works as the following:

$$\begin{bmatrix} b_0^{t+1} \\ b_1^{t+1} \\ b_2^{t+1} \\ b_3^{t+1} \end{bmatrix} = \begin{bmatrix} Y & Y+1 & 1 & 1 \\ 1 & Y & Y+1 & 1 \\ 1 & 1 & Y & Y+1 \\ Y+1 & Y & 1 & 1 \end{bmatrix} \begin{bmatrix} S2(b_0^t) \\ S2(b_1^t) \\ S2(b_2^t) \\ S2(b_3^t) \end{bmatrix} \tag{36}$$

Here, $Y \in GF(2^8)$ and the matrix of Y and 1 is used for mixcolumn transformation.

## 6.1 Cryptographic attack on SNOW 3G

### *6.1.1 Fault attack on SNOW 3G*

In this attack [8, 34] the attacker has a access of the physical device and it can cause transient fault attack on the device. In this kind of fault attack, the attacker can reset the device to its original state and apply a fault to the same device to get new keystream. The attacker can find the secret key of the cipher by verifying the faulty keystream and the actual keystream.

In SNOW 3G KSG, fault is injected between two computations of the following registers:

- Register $R1^t$.
- Register $R2^t$.
- LFSR state $St_0^t$.

We need to keep track that we consider the memory locations of $R1^{t+1}$, $R2^{t+1}$, $St_{15}^{t+1}$ after computation of a new word. Next we can find out $X_t \oplus X_{t'}$ which can give value either 1 or 0 for a certain fault injected positions, where $X_t$ denotes the value the above mentioned registers in RAM and $X_{t'}$ is the faulty valued of the registers. We also verify the $Z_t \oplus Z_{t'}$ value to get the confirmation of the fault occurrence. Finally, the state of the LFSR of SNOW 3G can be found by at least 24 fault injections and solving 512 linear equation by Gaussian elimination.

**Note:** Besides, all the attacks on SNOW 2.0 written before except algebraic attacks are also performed on SNOW 3G. It results SNOW 3G as a 128-bit secure cipher.

In **Table 3**, we enlist the attacks possible on SNOW 1.0, SNOW 2.0 and SNOW 3G.

| | Distinguishing attack | Algebraic attack | Fast correlation attack | Guess and determining attack | Fault attack |
|---|---|---|---|---|---|
| SNOW 1.0 | ✓ | X | X | ✓ | X |
| SNOW 2.0 | ✓ | ✓ | ✓ | ✓ | ✓ |
| SNOW 3G | ✓ | X | ✓ | ✓ | ✓ |

**Table 3.**
*Various attacks on SNOW 1.0, SNOW 2.0, SNOW 3G.*

## 7. Implementation issues

### 7.1 Vector-vector multiplication over $\mathcal{F}_{2^{32}}$

Whole SNOW family of ciphers use LFSR with $y * \beta$ operation over $\mathcal{F}_{2^{32}}$ where $\beta$ is a primitive element of $\mathcal{GF}(2^{32})$. SNOW 2.0 and SNOW 3G uses the following approach:

$$y.\beta = (y \ll 8) \oplus Table(y \gg 24) \tag{37}$$

Here, Table(x) function takes 8-bit vector as input and return 32-bit vector as output. It can be described as follows:

$$Table(x) = F_1(x, 23, 0xA9)||F_1(x, 245, 0xA9)||F_1(x, 48, 0xA9)||F_1(x, 239, 0xA9) \tag{38}$$

where $0xA9$ is the hexadecimal representation of the primitive polynomial $x^8 + x^7 + x^5 + x^3 + 1$ over $\mathcal{F}_2$, $F_1$ is a function which takes 16 bit inputs and positive integer $i$ to 8-bit vector. $F_1(W, i, x) = W$ if $i == 0$ and $F_1(W, i, x) = F_2(F_1(W, i-1, x), x)$, otherwise, where $W, x \in \mathcal{F}_2^8$. And $F_2(W, x) = (W \ll_8 1) \oplus x$ if leftmostbit(W)==1, else $F_2(W, x) = (W \ll_8 1)$. $F_2$ is a function which takes two 8− bit inputs and gives 8- bit output. This whole procedure reduces the time complexity of vector-vector multiplication over finite field to $\mathcal{O}(1)$ complexity.

### 7.2 Addition modulo $2^{32}$ ($\boxplus$)

Let us take two vectors $P, Q \in \mathcal{F}_2^{32}$ and addition modulo $2^{32}$ is defined as follows:

$$P \boxplus Q = (P + Q)(\text{mod}2^{32}) = (P + Q)\&(0xFFFFFFFF) \tag{39}$$

Here, $+$ is normal addition and & is bitwise AND-operator.

## 8. Conclusion

In this article, we can observe that SNOW 1.0 has problems in its design regarding the $\lll_7$ shift operator. The interaction between the $\lll_7$ shift operator and the S-box is one of the reasons for the large correlation in the FSM. Besides, the S-box used in SNOW 1.0 is not up to the mark. So, changing the S-box with Rijndael improves its power [35] against Guess and Distinguishing attack. In this context, SNOW 2.0 is better than SNOW 1.0 with respect to the design issues and cryptographic attacks. Still, it is susceptible to several attacks. Among all the attacks, the best-known attack is Algebraic Attack [29] which is taken care in the upgraded version SNOW 3G. A new S-Box and another Register are introduced in SNOW 3G to circumvent the problems in SNOW 2.0. Besides, the only practical attack possible for SNOW 2.0 and SNOW 3G is Fault Attack [8, 34] which is possible if someone access the memory location of the SNOW algorithm. This results SNOW 3G to be used in mobile telephony with 128 bit security. Though, KDFC-SNOW is another approach which is also potential candidate to resist some known plaintext attacks on SNOW 2.0, it requires some research on the implementation issue in the initialization phase. Besides, KDFC-SNOW cannot resist the recent fast correlation attack by Gond et al. [27] on SNOW 2.0. A possible research towards the improvement of SNOW family may be the finding an word LFSR with 64- or 128-bit block size with 256 bit key security which would be beneficial in 5G communication. Also, resistance of Fast Correlation Attack may be another kind of research which can be taken as future study.

## Author details

Subrata Nandi[1], Srinivasan Krishnaswamy[2] and Pinaki Mitra[1]*

1 Department of Computer Science and Engineering, IIT Guwahati, Assam, India

2 Department of Electronics and Electrical Engineering, IIT Guwahati, Assam, India

*Address all correspondence to: pinaki@iitg.ac.in

Chapter 4

# Role of Access Control in Information Security: A Security Analysis Approach

*Mahendra Pratap Singh*

## Abstract

Information plays a vital role in decision-making and driving the world further in the ever-growing digital world. Authorization, which comes immediately after authentication, is essential in restricting access to information in the digital world. Various access control models have been proposed to ensure authorization by specifying access control policies. Security analysis of access control policies is a highly challenging task. Additionally, the security analysis of decentralized access control policies is complex because decentralization simplifies policy administration but raises security concerns. Therefore, an efficient security analysis approach is required to ensure the correctness of access control policies. This chapter presents a propositional rule-based machine learning approach for analyzing the Role-Based Access Control (RBAC) policies. Specifically, the proposed method maps RBAC policies into propositional rules to analyze security policies. Extensive experiments on various datasets containing RBAC policies demonstrate that the machine learning-based approach can offer valuable insight into analyzing RBAC policies.

## 1. Introduction

Access control [1] ensures secure access to resources, devices, and data through policies. It regulates who can access which computing environment and its components. There are various access control models, such as Discretionary Access Control (DAC), Mandatory Access Control (MAC), Role-Based Access Control (RBAC), etc., that can specify and enforce access control policies. Among these, RBAC [2] is a widely adopted access control model that groups job functions into roles to simplify the administration. In RBAC, permissions are actions on objects assigned to roles instead of users. Therefore, a user can get specific permission only if the user is a member of the role to which the permission is assigned. RBAC can be described as a 6-tuple access control model, and its components are as follows.

- U, P, R, and Sessions represent a set of users, permissions, roles, and sessions, respectively. Sessions are not considered in the proposed approach because they do not influence security analysis.

- PA denotes permissions to role assignments and is represented as $PA \subseteq P \times R$.
- UA denotes users to role assignments and is represented as $UA \subseteq U \times R$.
- Each session is mapped to a single user and is represented as $S \rightarrow U$.
- Each session can have one or more roles and is represented as Session $\rightarrow 2^R$.
- The role hierarchy is defined as a partial order relation on roles and is represented as $RH \subseteq R \times R$.

The RBAC components (UA, PA, RH) determine whether users can access a particular resource, system, or data. Therefore, any change in these components would take a system to a new state. Hence, verifying whether it is safe is necessary before moving the system into a new state.

Security analysis aims to answer critical questions, such as whether a state is reachable to at least one user, whether all the reachable states satisfy security property, etc. An undesired state would be one in which an authorized user does not get access despite being entitled to it or an unauthorized user gains access.

One should consider various security properties before deploying a system. In this paper, we focus on safety and reachability properties that are described as follows:

**Safety Property:**

- Whether a user $u$ can access permission $p$.
- Whether a user $u$ can perform an access right $r$ on an object $o$.

**Reachability Property:**

- Whether a role $R$ is reachable to a user u.
- Whether a permission p is reachable to a role $R$.

If the evaluation outcome of a safety query mentioned above is no, then the system is safe. In contrast, if the evaluation outcome of a reachability query is yes, then the system is reachable.

The rest of the chapter is organized as follows. Previous works related to this research are presented in Section 2. Section 3 explains the proposed model, whereas Section 4 describes the security analysis of RBAC policies. Section 5 presents the result analysis and parameters used in the model, while Section 6 concludes the work and provides future research directions.

## 2. Related work

This section reviews the literature on recent applications, administration, and security analysis of RBAC policies.

Over the last few years, various access control models have been proposed. Among these, RBAC [2] is one of the well-adapted access control models. In RBAC, permissions are available to users according to their membership in specific roles. A large group of users can be grouped into roles to access resources according to permission assigned to the roles. Therefore, users should have a specific role to gain the required permissions for a task. The role-role relation enables delegation of authority and separation of authority. The main advantage of RBAC is the ease of policy administration.

### 2.1 Applications of RBAC

Recently, Kim et al. [3] have demonstrated RBAC usage in video surveillance using smart contracts, whereas Shaobo et al. [4] used RBAC to ensure fine-grained access to electronic health records. Additionally, Gurucharansing et al. [5] demonstrated the use of RBAC in specifying large-scale application access control policies. A Blockchain-based RBAC model with separation of duty presented by Ok-Chol et al. [6].

### 2.2 Administration of RBAC

Despite the advantages of RBAC, the administration of RBAC is complex and crucial for its proper management. Sandhu et al. [7] have presented the formal definition, intuition, and motivation of a role-based model for the administration of RBAC, ARBAC97 (administrative RBAC). The primary basis for the model is the simplification of administration along with scalability and administrative convenience. The components of the ARBAC97 model are user-role assignment (URA97), permission-role assignment (PRA97), and role-role assignment (RRA97).

### 2.3 Security analysis of RBAC

Apart from the administration, security analysis of the RBAC policies needs to be considered seriously. Alpern et al. [8] have formally defined safety and liveness security properties. Additionally, a topological characterization of both properties is also given. Their work captures all the main distinctions of the security properties. Koch et al. [9] have proposed safety state change rules where the RBAC states are posed as graph formalism in the RBAC model. Safety is defined as if a provided graph can become a subgraph of another graph. They have demonstrated that safety is decidable because a state change rule cannot simultaneously add and remove components to a graph. The proposed notion of safety captures the general notion but needs to cover bounded safety. Phillips et al. [10] proposed an access control model for servers, databases, inter-operating legacy, etc. Their work presents several theorems and lemmas to validate integrity and security. The combination of security and integrity ensured the proposed approach's liveness and safety. This approach does not consider the constraints of the RBAC model.

Li et al. [11] have proposed a security analysis approach using role-based trust management language for RBAC. They defined the problems related to security analysis and presented a way to represent and capture several security properties in a complex RBAC system. Specifically, two problem classes, namely AAR (Assignment and Revocation) and AATU (Assignment and Trusted Users), are discussed in the paper. The approach is based on reducing the two problem classes into another similar role-based trust-management language. This way, a relationship between the RT framework and RBAC is established. The approach produces efficient algorithms to solve significant queries. They demonstrated that several problems in security analysis need to be more concrete and intractable.

Jha et al. [12] performed the security analysis of the URA97 component of the ARBAC97 model using model-checking and logic programming approaches. Their work results demonstrate that the logic programming approach is better for many roles than the model-checking approach. Rakkay et al. [13] performed the security analysis by modeling and analyzing RBAC policies with the help of CPN tools and Colored Petri Nets (CP-Nets). The approach elaborates on the CP-Net model, which explains a generic access control structure based on RBAC policies. The significant advantage of CPN tools and CP-Nets is to provide an analytical framework and a graphical representation, which the security administrators use to understand why some permissions are denied or granted. Also, the framework and model are used to verify the security constraints.

Ferrera et al. [14] have proposed an approach to verify RBAC security policies using an abstraction-based tool. The proposed method converts data into imperative programs and performs security analysis. VAC tool was used to convert policies into crucial programs. An interval-based static analysis was carried out on the critical programs to show the correctness of policies. Martin et al. [15] have proposed a data-mining methodology to infer access control policies. The proposed approach is based on a tool developed for automatically generating requests, evaluating the requests to obtain the responses, and finally, using machine learning on the response-request pair to infer policy properties. The tool assists a user in identifying those requests, which can identify mistakes in the policy specification.

Most approaches mentioned above need to consider the overhead of translation of access control policies from one format (say XACML) to a specific format to perform security analysis. To address this, Singh et al. [16] have presented a framework that enables the specification and enforcement of heterogeneous access control policies, such as RBAC and ABAC, as data in the Database. Additionally, Singh et al. [17] have also presented a novel methodology for analyzing the security properties of heterogeneous access control policies. The proposed methodology models policies as facts using Datalog and analyses them through the $\mu$z tool in the presence of the administrative model. In addition, an approach to analyzing unified access control policies is also presented in [18] that captures policies as data in the Database.

It can be observed from the above literature survey that it is the first attempt to analyze access control policies using a machine learning-based model. The following section presents the proposed machine learning-based approach for analyzing RBAC policies.

## 3. Proposed security analysis approach

This section presents the approach for analyzing the security properties of RBAC policies. The proposed approach uses a rule-based machine learning algorithm to map RBAC policies into propositional rules. **Figure 1** shows the proposed model, and the description of its components is as follows.

### 3.1 Extraction of RBAC policy data from the unified database Schema

Generally, RBAC policies are specified using XACML, but we captured them as data in a unified database schema presented in [19]. There can be various combinations of RBAC policy data, but we have considered the following.

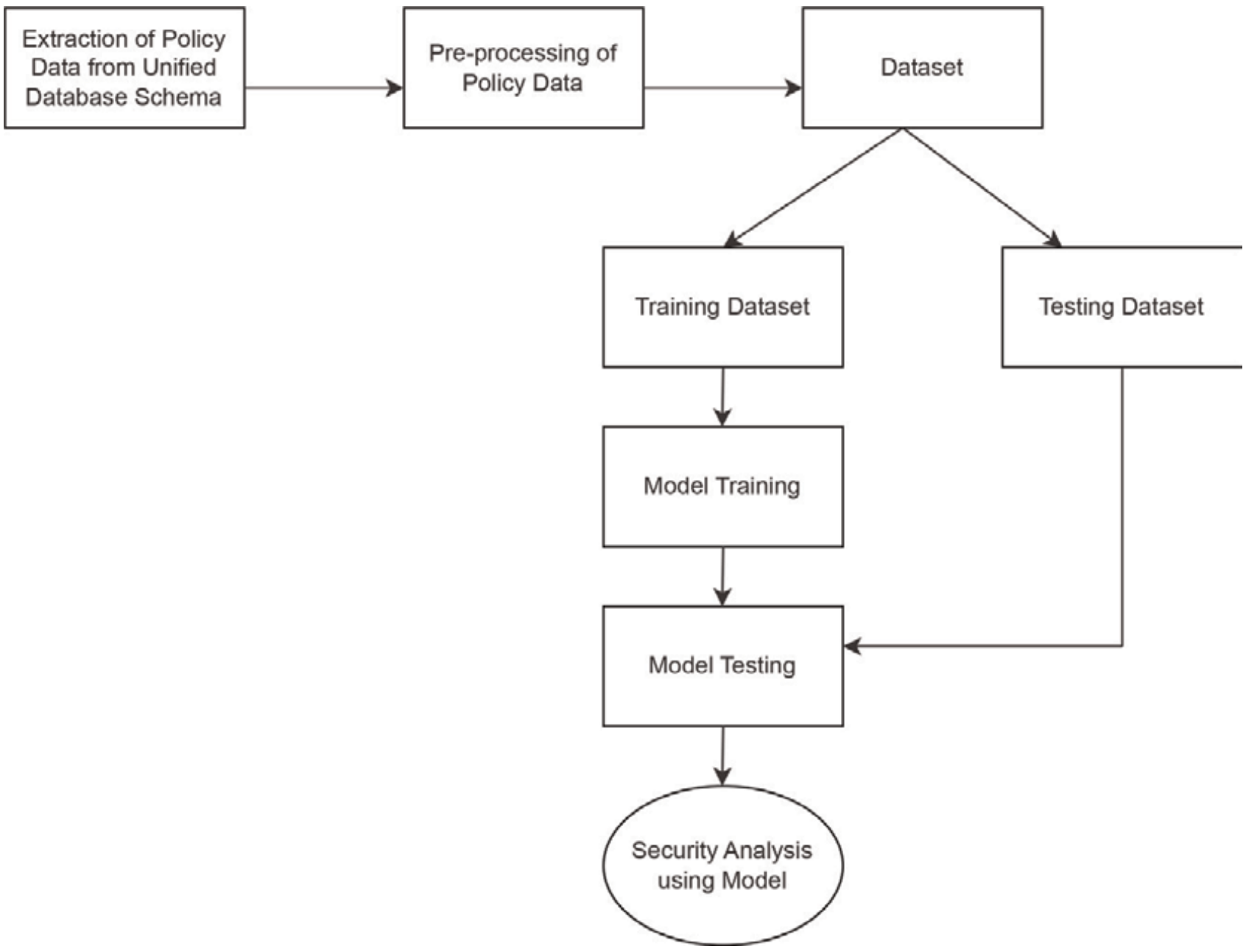

**Figure 1.**
*Proposed model.*

- Roles-Users: It captures the direct and indirect (through role hierarchy) association of roles with users.
- Permissions-Roles: It captures the direct and indirect (through role hierarchy) associations of permissions with roles.
- Users, Roles, and Permissions: It captures direct and indirect (through role hierarchy) associations of permissions with users through the role(s).
- Users, Roles, Rights, and Objects: It captures direct and indirect (through role hierarchy) associations of objects and rights with users through the role(s).

In the above combinations, attributes such as role, user, permission, object, and right act as features. Moreover, an extra feature, named label, has also been added. The above policy data combinations are fed to the following subsection for further processing.

### 3.2 Pre-processing of policy data to generate a dataset

In pre-processing, the policy data obtained in Subsection 3.1 was passed as input to AWK [20] for text manipulation and labeling. AWK is a Linux/Unix text manipulation utility that searches and substitutes text. Its output is a file containing only valid entries. Each valid entry represents authorized access and is marked as a safe state.

For security analysis, it is essential to consider valid and invalid entries. Therefore, another file containing both valid and invalid entries was created using the cartesian product of the attributes. Unlike valid entries, invalid entries represent unsafe states that indicate unauthorized access. Both valid and invalid entries were combined to form a broad system with safe and unsafe states. The valid and invalid entries were labeled 'Permit' and 'Deny', respectively.

The step-by-step process to create the dataset is as follows:

i. First, it takes a policy data file, say File 1, containing valid entries.

ii. Then generates another file, say File 2, through the Cartesian product of attributes.

iii. Next removes the common entries between File 1 and File 2 from File 2 and labels each entry as Deny.

iv. Labels all the entries of File 1 as Permit.

v. Combines the two files to produce a file containing both 'Permit' and 'Deny' labeled entries.

vi. Then, converts the file obtained above to a CSV file.

vii. Finally, the CSV file is converted to an ARFF file.

The above process is used to create the testing and following training datasets.

- Test dataset with all 'Permit' entries.
- Test dataset with all 'Deny' entries.
- Test dataset with a combination of 'Permit' and 'Deny' entries.

The datasets created in this subsection are used in the following subsections for training and testing the proposed model.

### 3.3 Proposed model and its training

This subsection describes the proposed model and its training using the dataset created in Subsection 3.2. We use a rule-based machine learning algorithm to create the model that takes the labeled dataset as input and maps it into propositional rules.

In the rule-based algorithm, a machine learning process identifies, evolves, or learns 'rules' to apply, manipulate or store.

The structure of a propositional rule is as follows:

$$\text{If}(\text{attribute1} < \text{a}) \text{ and } (\text{attribute2} > \text{b}) \ldots \ldots \text{and } (\text{attributen} < \phi\,) = \, > \text{class label}$$

Where a, b, …, and $\phi$ are the values the algorithm identifies from the policy attribute1, attribute2, …, and attributen, respectively, to generate the rule. In the above rule, the left side denotes the pre-condition or antecedent, a combination of attributes, whereas the right side shows the rule's consequent/class label.

We develop the model using **JRipper** [21] algorithm that implements a propositional rule learner known as RIPPER. This rule learner was proposed by William W.

Cohen and is based on a very effective and common technique of REP found in the decision tree algorithms.

The rule learner divides the training dataset into pruning and growing. First, the growing set is used to form an initial rule set with the help of some heuristic approach. Later, the large rule set is simplified repetitively using different pruning operators. The pruning operator with the most error reduction is selected at every simplification stage. The terminating point of simplification is when none of the operators reduces errors. It provides propositional rules for the training dataset used for classification.

The description of steps involved in the JRipper algorithm are as follows:

- **Initialization Stage:** In this stage, the rule set (RS) is initialized as RS={} for every class from most frequent to least frequent one.

- **Growing and Pruning Phase:** This phase repeats several steps until the error rate is below 50% or there is no positive example. In the growing phase, terms are added to a greedy rule to make the rule perfect. The rule is pruned incrementally in the pruning phase. The formula used to measure the pruning value is $2p / (p + n) - 1$, where p is the positive example covered by the rule, and n is the negative example covered by the rule.

- **Optimization Phase:** After identifying the rule set, two variants are created for every rule using random data, and then pruned. For the first variant, an empty rule is used. The addition of antecedents makes the second variant. The most petite description length is calculated for the original rule and its variants. If any residual positives are observed, they are used to identify more rules.

- **Delete Phase:** If the description length for any rule exceeds the limit, the rule gets deleted. The remaining rules are appended to form the final rule set.

We evaluate the model's performance using testing datasets in the following subsection.

### 3.4 Model testing

After training the model, the next step is to test the model. Testing datasets created in Subsection 3.2 are used to predicate the model's accuracy. Like the training dataset, the testing dataset contained policy data in ARFF file format. Every policy data is mapped to a predicate rule, then compared with the model's predicate rules. The outcome of the evaluation and accuracy is reported in Section 5.

In the following section, we use an example to demonstrate the security analysis of RBAC policies through the proposed model.

## 4. Analysis of security property

This section shows how the proposed model can be used to verify the safety and reachability properties of RBAC policies.

To understand the security analysis effectively, we consider an RABC system that consists of 20 users, 3 roles, and 30 permissions and the following UA, PA, and RH assignments:

**Role**={1,2,3}
**User**={1,2,3....20}
**Permission**={1,2,3,4....30}
RH=$\{3 \preccurlyeq 1\}$
**UA**={({1,2..7},1), ({8,9..13},2), ({14,15..20},3)}
**PA**={({1,2..10},1), ({11,12..20},2), ({21,22..30},3)}

We create a model using the proposed approach for the above specification that contains propositional rules. The propositional rules correspond to some of the policies specified above are as follows:

- **Rule 1**: (User ID<=7) and (Role ID=1)=>Permit
- **Rule 2**: (User ID<=7) and (Role ID=2)=>Deny
- **Rule 3**: (User ID<=7) and (Role ID=3)=>Deny
- **Rule 4**: (User ID>=8) and (User ID<=13) and (Role ID=1)=>Deny
- **Rule 5**: (User ID>=8) and (User ID<=13) and (Role ID=2)=>Permit
- **Rule 6**: (User ID>=8) and (User ID<=13) and (Role ID=3)=>Deny
- **Rule 7**: (User ID>=14) and (User ID<=20) and (Role ID=1)=>Deny
- **Rule 8**: (User ID>=14) and (User ID<=20) and (Role ID=2)=>Deny
- **Rule 9**: (User ID>=14) and (User ID<=20) and (Role ID=3)=>Permit
- **Rule 10**: (Permission ID<=10) and (Role ID=1)=>Permit
- **Rule 11**: (Permission ID<=10) and (Role ID=2)=>Deny
- **Rule 12**: (Permission ID<=10) and (Role ID=3)=>Deny
- **Rule 13**: (Permission ID>=11) and (Permission ID<=20) and (Role ID=1) =>Deny
- **Rule 14**: (Permission ID>=11) and (Permission ID<=20) and (Role ID=2) =>Permit
- **Rule 15**: (Permission ID>=11) and (Permission ID<=20) and (Role ID=3) =>Deny
- **Rule 16**: (Permission ID>=21) and (Permission ID<=30) and (Role ID=1)=>Permit
- **Rule 17**: (Permission ID>=21) and (Permission ID<=30) and (Role ID=2)=>Deny
- **Rule 18**: (Permission ID>=21) and (Permission ID<=30) and (Role ID=3) =>Permit

Similarly, the propositional rules are specified for the remaining policies.

To perform the security analysis, we consider safety and reachability security properties defined in Section 1, and their analysis is as follows.

- **Security Analysis of Safety Property:** Test cases in the following form are passed to the model to analyze the safety property.
  - (User ID=5) and (Role ID=1) and (Permission ID=17)
  - (User ID=23) and (Role ID=2) and (Permission ID=27)
  - (User ID=9) and (Role ID=3) and (Permission ID=15)

The model classified all the test cases mentioned above as 'Deny' for the following reasons.

  - User with USER ID=5 is not authorized to access Permission with Permission ID=17 through role with Role ID=1.
  - User with USER ID=23 is not authorized to access Permission with Permission ID=27 through role with Role ID=2.
  - User with USER ID=9 is not authorized to access Permission with Permission ID=15 through role with Role ID=3.

It can be noticed from the above analysis that the Role ID in each test case cannot provide permission to the user according to the rules present in the model. Thus, safety property satisfies.

- **Security Analysis of Reachability Property:** The test cases in the following form are considered to analyze the reachability property.
  - (User ID=5) and (Role ID=1)
  - (User ID=14) and (Role ID=3)
  - (User ID=9) and (Role ID=2)

The model classified all the test cases mentioned above as 'Permit' for the following reasons.

  - User with USER ID=5 has role with Role ID=1.
  - User with USER ID=14 has role with Role ID=3. For
  - User with USER ID=9 has a role with Role ID=2.

It can be seen from the above analysis the Role ID in each test case is available to the user according to the rules present in the model. Thus, the reachability property holds.

The experimental analysis of the proposed model is demonstrated in the following section.

## 5. Experimental results and analysis

Several experiments were performed on the system having 64 GB RAM and an Intel Core i7 processor to observe the impact of various components of RBAC policies.

To evaluate the performance of the proposed model, we created three synthetic RBAC policy datasets shown in **Table 1** using Oracle 12c that capture policies as data. To reflect the real-world scenario, it can be observed from **Table 1** that the number of users and the number of objects were increased 100 times, whereas the number of rights was increased only two times. The number of permissions and roles were increased 200 times and four times, respectively. Additionally, the number of permission-role assignments was increased to 100, while the number of permissions and user-role assignments were increased to 200 times.

The following parameters were used to measure the performance of the proposed model:

- True Positive(TP) rate is the ratio of instances correctly classified for a class to the total number of instances.

$$\text{TP rate} = \text{TP}/(\text{TP} + \text{FN})$$

- False Positive(FP) rate is the ratio of negative events wrongly categorized as positive to the total number of actual negative events.

$$\text{FP rate} = \text{FP}/(\text{FP} + \text{TN})$$

- Precision is the ratio of instances that belong to that class to the total number of instances classified as that class.

$$\text{Precision} = \text{TP}/(\text{TP} + \text{FP})$$

- Recall is the ratio of instances classified as a given class to the actual number of instances that belong to that class.

| Relation | Dataset 1 | Dataset 2 | Dataset 3 |
|---|---|---|---|
| Users | 75 | 750 | 7500 |
| Objects | 100 | 1000 | 10,000 |
| Rights | 5 | 5 | 10 |
| Permissions | 100 | 1000 | 20,000 |
| Roles | 10 | 20 | 40 |
| Permission Role Assignment | 100 | 1000 | 20,000 |
| User Role Assignment | 150 | 1500 | 15,000 |
| Permission Object Assignment | 100 | 1000 | 20,000 |

**Table 1.**
*Number of entries for relations in the datasets.*

$$\text{Recall} = \text{TP}/(\text{TP} + \text{FN})$$

- Confusion Matrix: It shows how many instances of a particular class are correctly or incorrectly classified.

- F-measure depends upon precision and recall.

$$\text{F-measure} = (2^{*}\,\text{precision}^{*}\,\text{recall})/(\text{precision} + \text{recall})$$

- k-fold cross validation: The training dataset is divided into k-sets. Of the k sets, the k-1 is used for training, and the remaining one is used for testing. It is repeated k times, and a different set is used for testing each time. After k iterations, the average error across k-trials is measured.

The following four models were trained and tested to classify instances as Permit or Deny using Datasets 1, 2, and 3. The value of k was kept at 10 for Dataset 1, while 2 for Datasets 2 and 3. The description of the models is as follows:

- URA is composed of users and roles.

- PRA consists of permissions and roles.

- URP is made up of users, permissions, and roles.

- UROR comprises users, objects, rights, and roles.

The following parameter values were obtained for the above models.

- **Accuracy of Model:** It shows how many instances of labeled data were correctly identified by the model in percentage. In other words, it shows a model's reliability in reflecting the RBAC policies.

- **Time to Build the Model:** It shows the time the algorithm takes to construct a rule set for the classifier from the labeled data set.

- **Accuracy of Test Results:** It shows how many entries of the test dataset were correctly classified by the classifier.

**Table 2** shows the model's accuracy, the time to build a model, and the accuracy of test results. It can be observed from the table that model and test result accuracy are near about 100% for all three datasets. Additionally, there is no significant increase in time to build models for Datasets 1, 2, and 3.

| Dataset | Accuracy of Model (%) | | | | Time to Build the Model (Sec) | | | | Accuracy of Test Results (%) | | | |
|---|---|---|---|---|---|---|---|---|---|---|---|---|
| | URA | PRA | URP | UROR | URA | PRA | URP | UROR | URA | PRA | URP | UROR |
| 1 | 96.67 | 87.80 | 99.99 | 99.07 | 0.03 | 0.18 | 10.67 | 197.79 | 96.13 | 100 | 100 | 100 |
| 2 | 99.22 | 99.33 | 99.99 | 99.98 | 4.96 | 9.71 | 132.72 | 814.48 | 100 | 100 | 100 | 100 |
| 3 | 99.85 | 99.94 | 99.99 | 99.99 | 1315.62 | 56.97 | 446.73 | 672.63 | 100 | 100 | 100 | 100 |

**Table 2.**
*Performance of models.*

| Dataset | Access | Predicted: Deny | | | | Predicted: Permit | | | |
|---|---|---|---|---|---|---|---|---|---|
| | | URA | PRA | URP | UROR | URA | PRA | URP | UROR |
| 1 | Actual: Deny | 302 | 691 | 70,650 | 370,650 | 14 | 19 | 0 | 0 |
| | Actual: Permit | 11 | 103 | 1 | 35 | 423 | 187 | 4349 | 4315 |
| 2 | Actual: Deny | 10,633 | 17,070 | 218,000 | 224,991 | 27 | 30 | 0 | 9 |
| | Actual: Permit | 101 | 104 | 35 | 78 | 4249 | 2796 | 217,465 | 224,922 |
| 3 | Actual: Deny | 224,976 | 719,965 | 8,399,996 | 6,999,998 | 24 | 35 | 4 | 2 |
| | Actual: Permit | 429 | 480 | 0 | 1 | 74,571 | 79,520 | 8,710,000 | 7,059,999 |

**Table 3.**
*Confusion matrix of models.*

The confusion matrix of models is shown in **Table 3**. From the table, it can be seen that most of the models predicted instances accurately. Therefore, it can be concluded that the proposed model is scalable and can be a viable option.

## 6. Conclusion

The security analysis of the RBAC policies has been performed through the Machine Learning-based model that uses the JRipper algorithm. The proposed model could map most policies correctly to rule sets for each classifier. The results show that the proposed model is highly reliable and efficient for the security analysis of RBAC policies. Additionally, it can also be observed that the model's efficiency has improved significantly due to an increase in the RBAC policy datasets. Thus, the proposed approach can be considered a viable solution for performing the security analysis of large policy sets.

In the future, the proposed model can be extended to analyze the other security properties of RBAC with and without an administrative model. Moreover, it can also be used to analyze RBAC extensions (such as TRBAC, ESTRBAC, etc.) security properties in the presence and absence of an administrative model.

## Abbreviations

| | |
|---|---|
| DAC | Discretionary Access Control |
| MAC | Mandatory Access Control |
| RBAC | Role-Based Access Control |
| ARBAC97 | Administrative RBAC |
| URA97 | User-Role Assignment |
| PRA97 | Permission-Role Assignment |
| RRA97 | Role-Role Assignment |
| ABAC | Attribute-Based Access Control |
| TRBAC | Temporal Role-Based Access Control |
| ESTARBAC | Extended Spatio-Temporal Role-based Access Control |
| AAR | Assignment and Revocation |
| AATU | Assignment and Trusted Users |
| XACML | Extensible Access Control Markup Language |

## Author details

Mahendra Pratap Singh
Department of Computer Science and Engineering, National Institute of Technology Karnataka, Surathkal, Mangaluru, India

*Address all correspondence to: mahoo15@gmail.com

Chapter 5

# Enhanced Hybrid Privacy Preserving Data Mining Technique

*Naga Prasanthi Kundeti, Chandra Sekhara Rao MVP, Sudha Sree Chekuri and Seshu Babu Pallapothu*

## Abstract

At present, almost every domain is handling large volumes of data even as storage device capacities increase. Amidst humongous data volumes, Data mining applications help find useful patterns that can be used to drive business growth, improved services, better health care facilities etc. The accumulated data can be exploted for identity theft, fake credit/debit card transactions, etc. In such scenarios, data mining techniques that provide privacy are helpful. Though privacy-preserving data mining techniques like randomization, perturbation, anonymization etc., provide privacy, but when applied separately, they fail to be effective. Hence, this chapter suggests an Enhanced Hybrid Privacy Preserving Data Mining (EHPPDM) technique by combining them. The proposed technique provides more privacy of data than existing techniques while providing better classification accuracy as well as evidenced by our experimental results.

## 1. Introduction

Modern machine learning models are applied on large volumes of data accumulated over time. The data used for training or building models may contain personal data. Data owners may not want to share their personal data. To safeguard privacy of personal data, this paper seeks to perform data analysis without revealing the sensitive personal information of users.

Privacy has often been defined in many different ways. Westin (1968) defined privacy as “the assertion of individuals, groups or institutions to specify when, how and to what extent their information can be shared to others”. Bertino et al. [1] defined privacy as “the security of data about an individual contained in an electronic repository from unauthorized disclosure”.

Privacy threats can be categorized into three types, namely (a) Membership Disclosure, (b) Attribute Disclosure and (c) Identity Disclosure.

*Membership Disclosure:* Such threats occur when an attacker manages to check the presence of specific user data in a data set and infers certain meta-information thereof.

*Attribute Disclosure:* In this type of attack, some sensitive user data can be anecdoted by the attacker by connecting data entries with some data from other sources.

*Identity Disclosure:* Here, an attacker can identify all sensitive data about a person by making a particular data admission in a data set, thereby revealing his identity and threatening his safety.

Privacy preservation methods protect data from data leakage by altering the original data, minimizing exposure as specified in literature [2, 3]. Prominent techniques include randomization, perturbation, suppression, generalization etc. In order to preserve useful data after altering the data, various data utility metrics such as discernability metric, KL-divergence, entropy-based information loss etc. are applied as mentioned in literature.

The data shown in tabular form is processed with each row representing an entity in the real world. The attributes of a data table can be categorized into four types viz., Identifier Attributes (Ids), Quasi-identifier Attributes (QIDs), non-Sensitive Attributes (NSAs) and Sensitive Attributes (SAs). The attributes that help identify a person from a given data are called identifier attributes. For ex: SSN, Aadhar id etc. Generally, such attributes are removed from data before sharing the data for data analysis to protect personal identity. Sensitive attributes contain delicate personal information health condition, financial status etc. Such attributes do not share or remove sensitive personal data to avoid bad results. So, the sensitive data is maintained but personal identity also needs to be protected. Quasi identifiers are the attributes purportedly used by attackers to disclose identity of the individual when combined with some background knowledge. Hence, such quasi identifiers need to be modified to prevent identity disclosures by attackers. The last attributes i.e., non-sensitive attributes do not disclose any information about individuals by retained them intact while sharing data for analysis.

So, while sharing data for analysis, several privacy preservation methods are proposed like randomization, perturbation etc. to protect privacy [4, 5]. Though data transformations are applied to provide privacy of data, yet it may lead to inaccurate data mining results, thereby reducing its utility. Hence, to balance both privacy preservation and accuracy in data mining results, Privacy Preserving Data Mining (PPDM) techniques are applied. In the process, divergence of data is minimized and actual data is validated from the analysts' perspective through some metrics that evaluate the privacy level and data utility of different PPDM techniques [1, 6, 7].

## 2. Review of PPDM techniques

Data present in various data sources can be privacy enabled with application of different privacy preserving techniques. Some of them are Generalization, Suppression, Anatomization and Perturbation.

- Generalization: In this method, a data field is swapped with a more generalized data filed value. In case of numerical attributes, the data field value is exchanged with a range of values. In case of categorical attributes, generalization is performed based on value generalization hierarchy.

- Suppression: This method, averts information disclosure by abolishing some values of attribute. Original data field values are replaced with("*").

- Anatomization [7]: This method works by separating quasi-identifiers and sensitive attributes. They are located in two different tables so that connecting QIDs to sensitive attributes become very tedious.

- Perturbation: Here. original data field values are substituted by artificial values with the similar statistical information.

Samarati and Sweeny [8, 9] proposed a popular privacy model i.e., k-anonymization. Further, k-anonymity for a table is defined as follows [10]:

"Let T(A1,…,An) be a table.

Let QI be the set of quasi-identifiers corresponding to table T.

T fulfills k-anonymity property with respect to QI if and only if each sequence of values in T[QI] appears at least with k occurrences in T[QI]".

Generalization and suppression techniques are applied on Quasi Identifiers (QIDs) as part of k-anonymization. All the QIDs in a group of size 'k' have similar values on ensures that the confidential data about individual users is not revealed when data is shared for analysis purpose. So, K-anonymized data provides privacy of data. An attacker can still infer sensitive information about individuals using a K-anonymized table and some background knowledge, if the value of sensitive attribute is same for all individuals in a given k-group. Let us consider the k-anonymized table shown below in **Table 1**.

While *k*-anonymity is a promising approach in group based anonymization due to its ease of use and the varied array of algorithms that perform it, yet it is vulnerable to many attacks. When attackers access background information, they can cause massive damage to sensitive data, including the following:

- **Homogeneity Attack**: These attacks leverage cases in which the sensitive attribute values for a sensitive attribute within a set of *k* records are alike. In these cases, in spite of k-anonymization, sensitive attribute value for the group of *k* records may be precisely foreseen.

- **Background Knowledge Attack**: Such attacks leverage an association among one or more sensitive attribute with QID (quasi-**i**dentifier) attributes to decrease the set of possible values for accessing the delicate attribute. Machanavajjhala et al. [11] showed that knowledge of heart attacks occuring at a condensed rate among Japanese patients could be utilized to slim the range of values for a delicate attribute of a patient's illness.

An attacker who has access to this 3-anonymous table can use background knowledge from other data sources and identify all patients in Mumbai having disease 'Flu'.

| QI: Age | QI: city | Sensitive attribute: Disease |
|---|---|---|
| 20–30 | mumbai | Flu |
| 20–30 | mumbai | Flu |
| 20–30 | mumbai | Flu |
| 30–40 | Delhi | Cancer |
| 30–40 | Delhi | Cancer |
| 30–40 | Delhi | Cancer |

**Table 1.**
*3-anonymized table.*

So, sensitive information about an individual residing in Mumbai is revealed. To overcome this security breach, l-diversity principle is applied on sensitive attribute.

Agarwal et al. [4] defines *l*-diversity as, "Let a q*-block be a set of tuples such that its non-sensitive values generalize to q*. A q*-block is *l*-diverse if it contains *l* 'well represented' values for the sensitive attribute S. A table is *l*-diverse, if every q*-block in it is *l*-diverse."

Li et al. [12] defined *l*-diversity as "an equivalence class is said to have *l*-diversity if there are at least *l* "well-represented" values for the sensitive attribute. A table is said to have *l*-diversity if every equivalence class of the table has *l*-diversity".

Aggarwal and Yu [4] showed the likelihood of more than one sensitive field when the *l*-diversity problem becomes more difficult due to added dimensionalities.

## 3. Methodology

Kundeti et al. [13] had introduced a hybrid privacy preserving data mining (HPPDM) technique that provided privacy and lesser attacks, which, however can be extended to create more privacy by applying the l-diversity principle. In fact, L-diversity provides more privacy against different background attacks.

Algorithm: - Enhanced Hybrid Privacy Preserving Data Mining (EHPPDM).

Input: - Adult Dataset D.

Output: - Privacy enabled Adult Data setD'.

Step1: Categorize attributes of Adult Data set into Identifiers, Quasi Identifiers, Sensitive and Non-Sensitive Attributes.

Step2: Consider the Quasi Identifiers and create value generalization hierarchies for quasi identifiers.

Step3: Apply geometric perturbation technique in numerical quasi identifiers to obtain perturbed numerical quasi identifier.

Step4: Create generalization hierarchies in categorical quasi identifiers and choose different levels in generalization hierarchy based on k-value chosen for anonymization.

Step5: apply l-diversity for sensitive attributes based on number of different values for classpresent.

Step 6: Obtain the privacy preserved Adult data set D′.

## 4. Implementation

An Enhanced Hybrid Privacy Preserving Data Mining (EHPPDM) technique is implemented by using R language. The ARX anonymization tool is used for performing K-Anonymization.

An adult Dataset from the UCI machine learning repository is used for evaluating the EHPPDM technique. The dataset consists of 15 attributes, including the Class attribute. The attributes are age (numerical), work-class (categorical), fnlwgt (numerical), education (categorical), education-num (numerical), marital-status (categorical), occupation (categorical), relationship (categorical), race (categorical), sex (categorical), capital-gain (numerical), capital-loss (numerical), hours-per-week (numerical), native-country (categorical) and class variable. These attributes can be divided into quasi-identifiers, sensitive attributes and Insensitive attributes. The quasi identifiers in this data set are age, work class, education and nativity.

Class attribute is sensitive attribute, while the remaining attributes are classified as Insensitive attributes.

Among the quasi identifiers, age is the numerical attribute. The Geometric data perturbation technique [14] is applied on numerical quasi identifier i.e. age. Value generalization hierarchies are created for categorical quasi identifier attributes. K-anonymization algorithm is applied to these categorical quasi identifiers. For different values of K, different anonymization levels are obtained, which provide privacy at different levels. The k-values considered are 50, 100, 150, 200, 250, 300, 350, 400, 450, 500. After anonymization, the anonymized data sets are applied with classification algorithms like naive bayes, J48 and decision tree. The accuracies of classification are noted down.

To further enhance the privacy of data, l-diversity is applied on sensitive attribute i.e. Class attribute. L-diversity is applied to reduce background attacks and linkage attacks. As l-diversity ensures that the class attribute value in a given anonymized group does not have single value, the attacker cannot identify an individual's sensitive attribute value. The anonymized and l-diversity-applied dataset is obtained. After applying classification algorithms on the anonymized data, their accuracies are tabulated. Later, the risk analysis for various types of attacks is represented in the various figures.

**Figure 1** depicts the classification accuracies for Adult data set when applied with k-anonymization. K-anonymization for different values of k is applied. **Figure 2** displays the classification accuracies for the adult data set wherein *l*-diversity is applied to decrease background attacks. After detecting the increase in privacy, *l*-diversity principle is applied.

The classification accuracies of Hybrid Privacy Preserving Data Mining (HPPDM) [13] technique for adult data set are shown in **Figure 3**. The experimental results depict better classification accuracies with HPPDM technique as compared to k-anonymization.

**Figure 4** illustrates the classification accuracies for adult data set when Enhanced Hybrid Privacy Preserving Data Mining (EHPPDM) is applied.

**Figures 5-8** illustrate the risk analysis for adult data set.

**Figure 5** demonstrates the risk analysis against various types of attacks when k-anonymization is applied on adult data set. **Figure 6** displays the risk analysis against various types of attacks when k-anonymization and l-diversity are applied

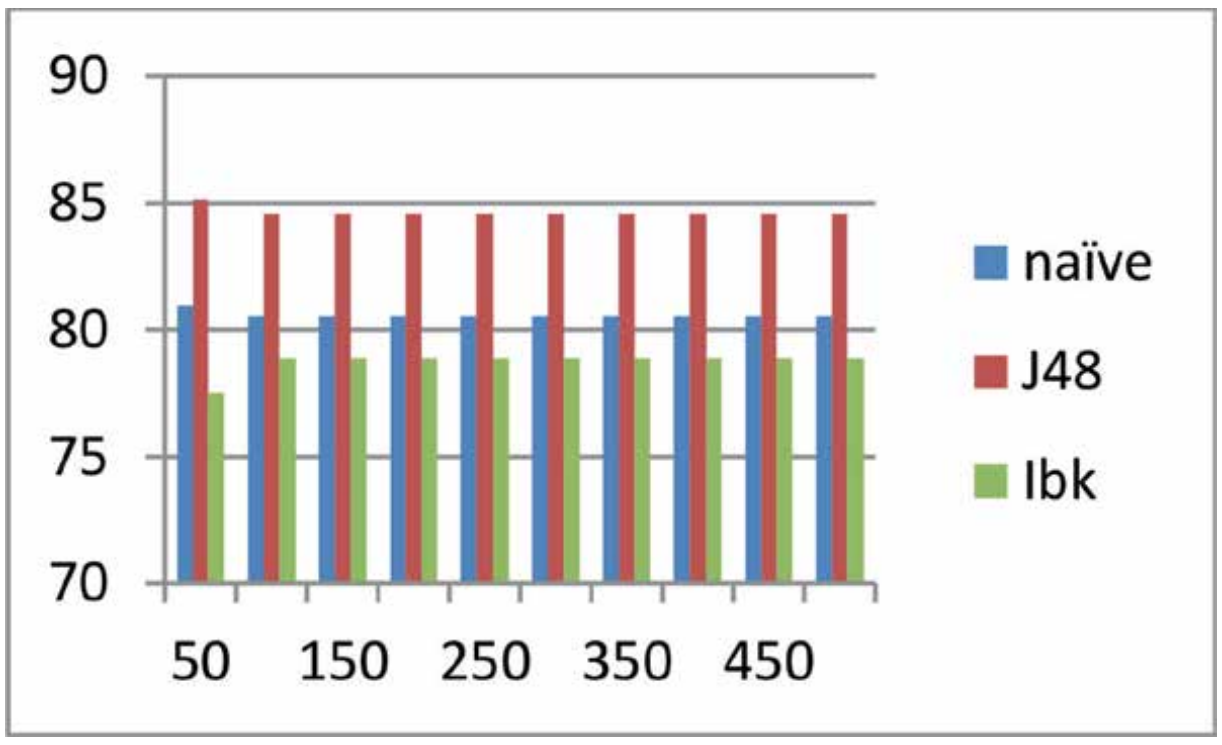

**Figure 1.**
*Classification accuracies for adult K-anonymized data for different k-values.*

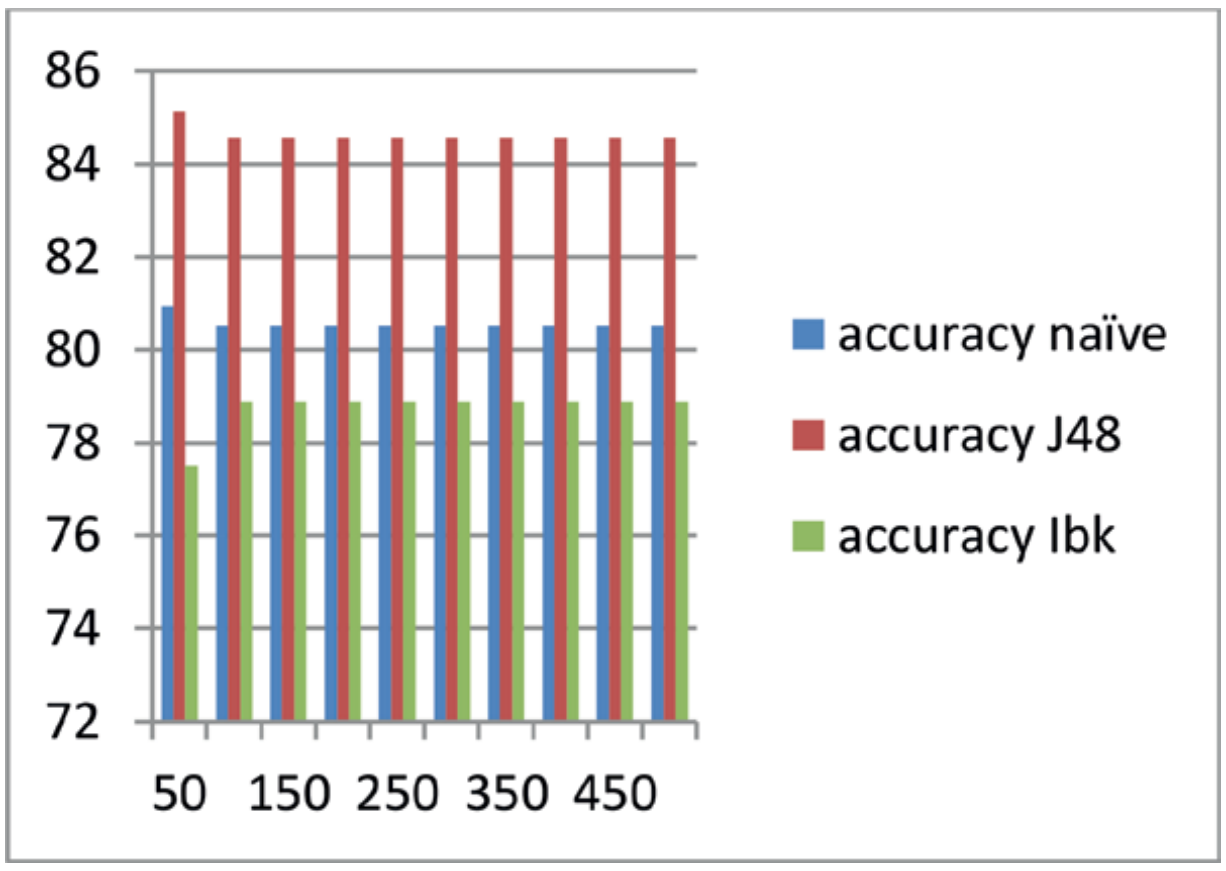

**Figure 2.**
*Classification accuracies for adult K-anonymized and l-diversity (l-value = 2) applied.*

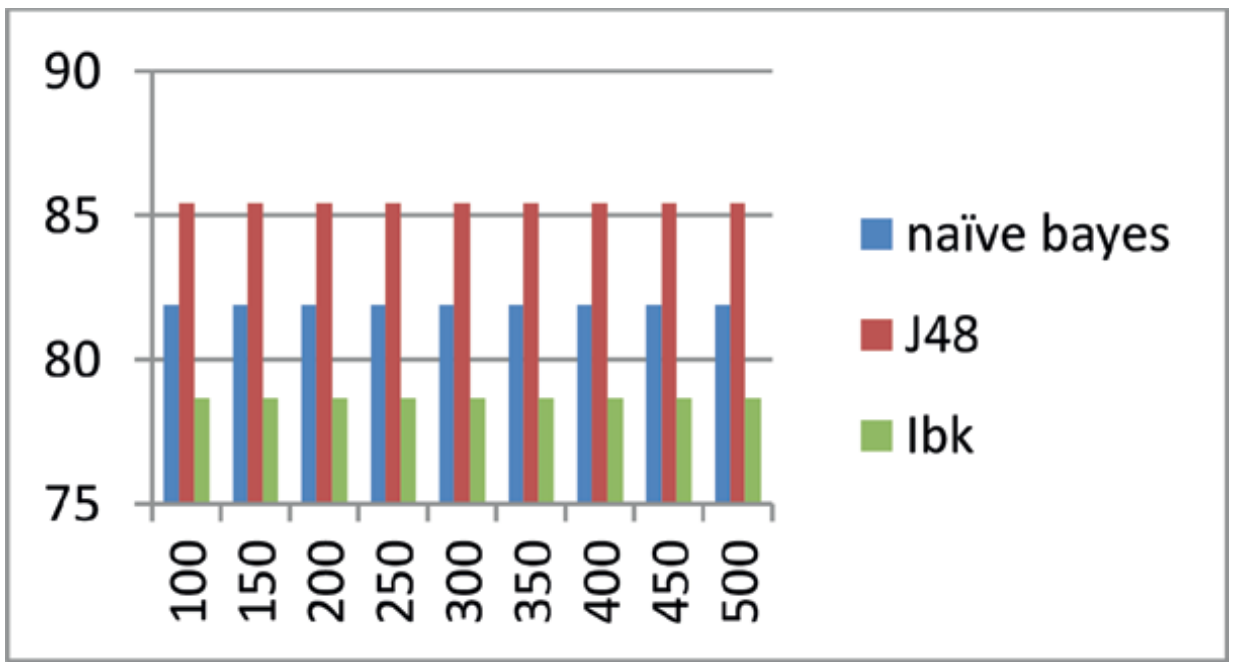

**Figure 3.**
*Classification accuracies for adult after applying hybrid privacy preserving data mining technique.*

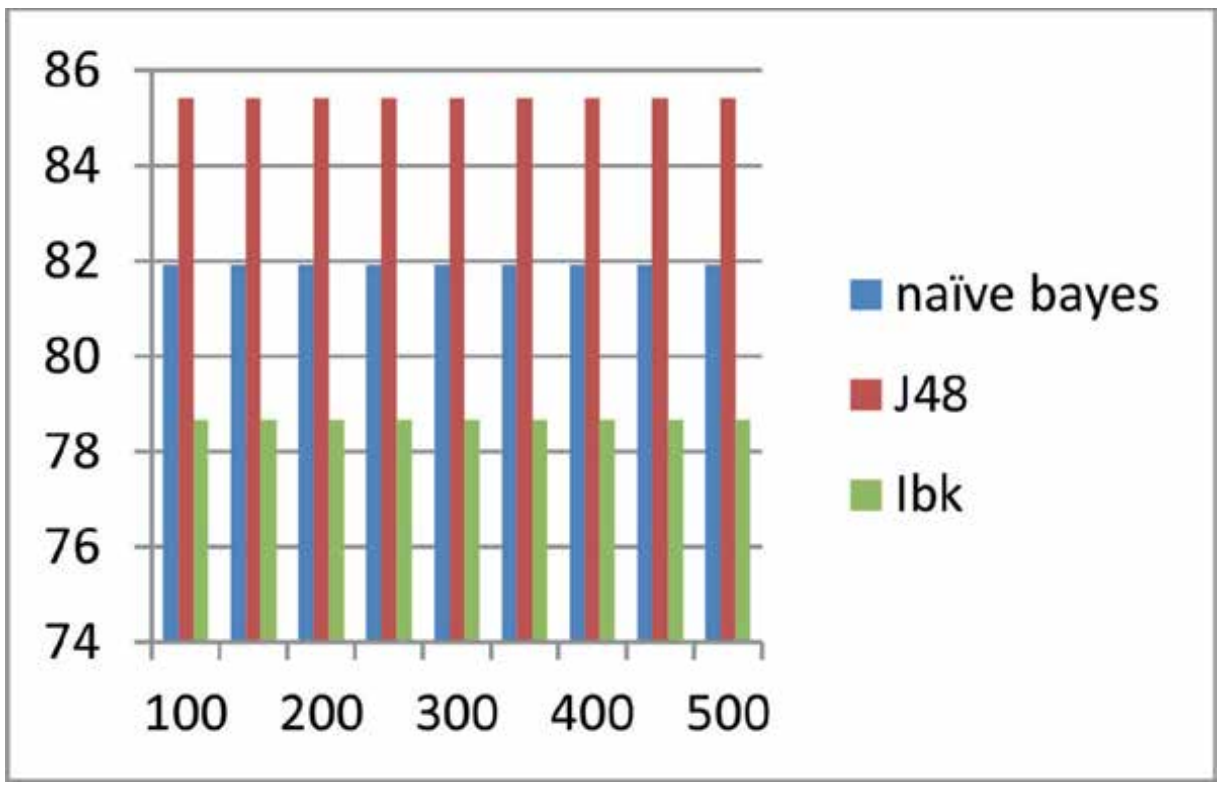

**Figure 4.**
*Classification accuracies for enhanced hybrid privacy preserving data mining technique.*

on adult data set. **Figure 7** illustrates the risk analysis against various types of attacks when Hybrid Privacy Preserving Data Mining (HPPDM) technique [13] is applied on adult data set. **Figure 8** depicts the risk analysis against various types of attacks when

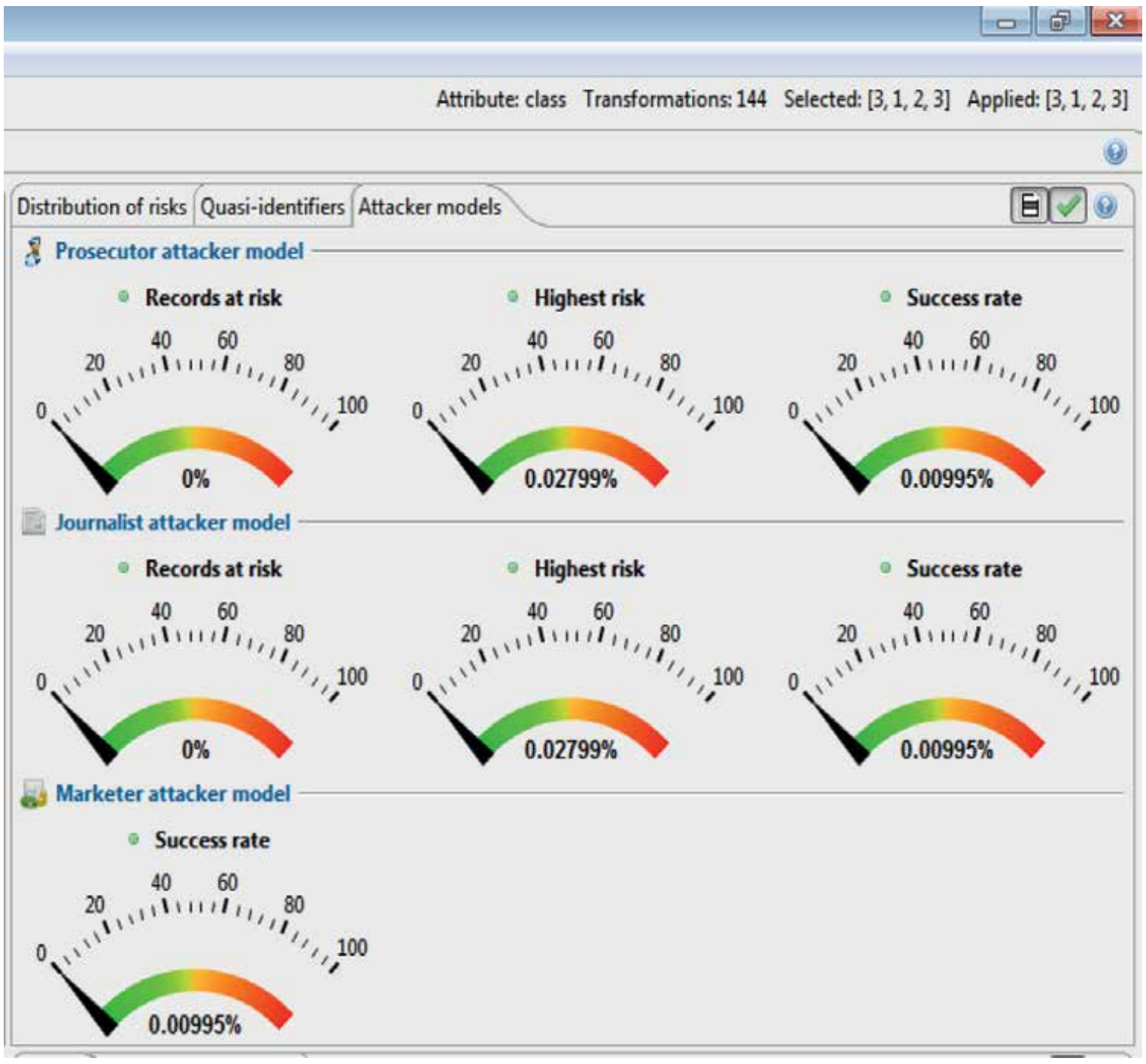

**Figure 5.**
*Risk analysis for various types of attacks after applying k-anonymization (k-value = 100).*

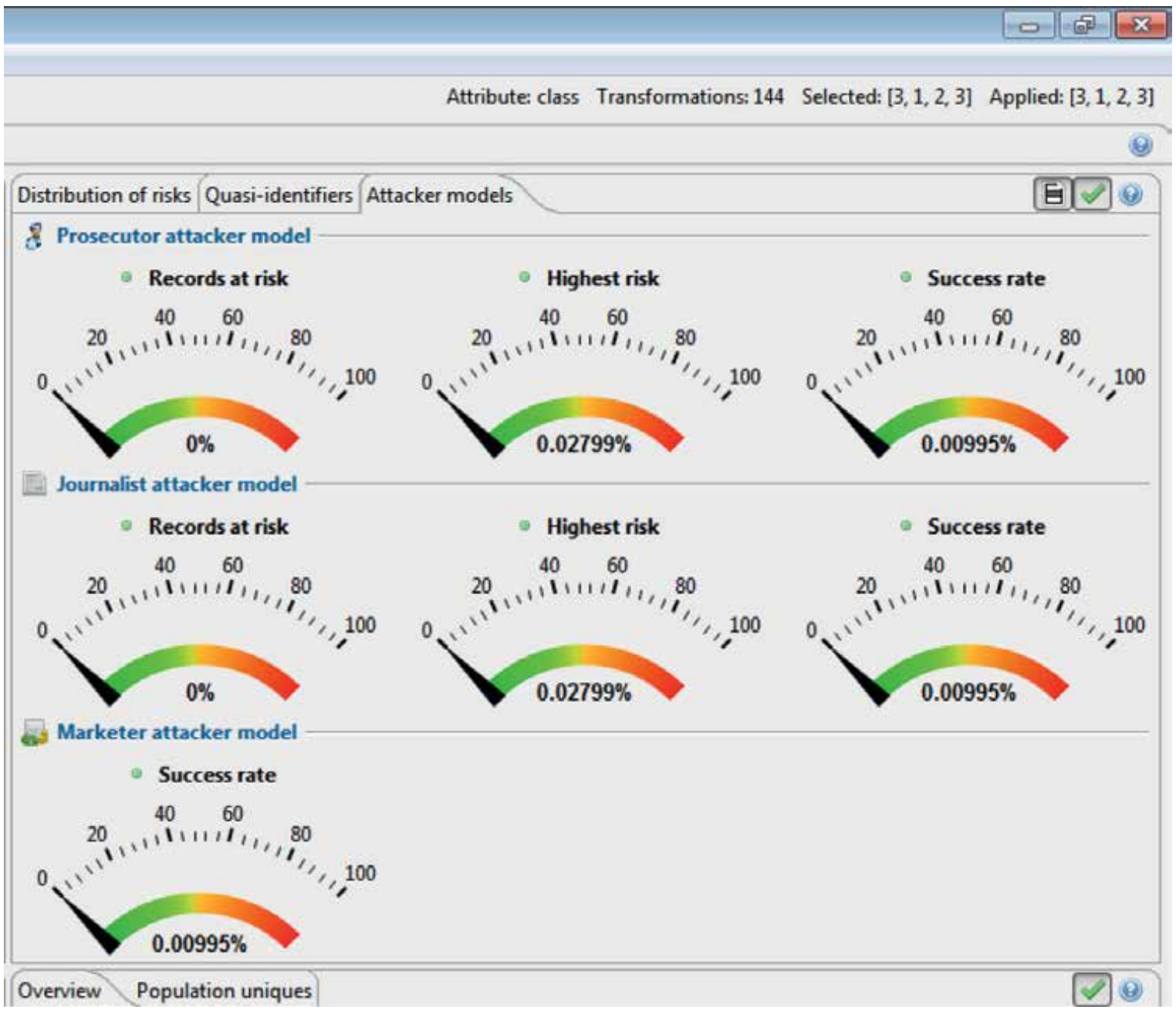

**Figure 6.**
*Risk analysis for various types of attacks after applying k-anonymization (k-value = 100) and l-diversity (l-value = 2).*

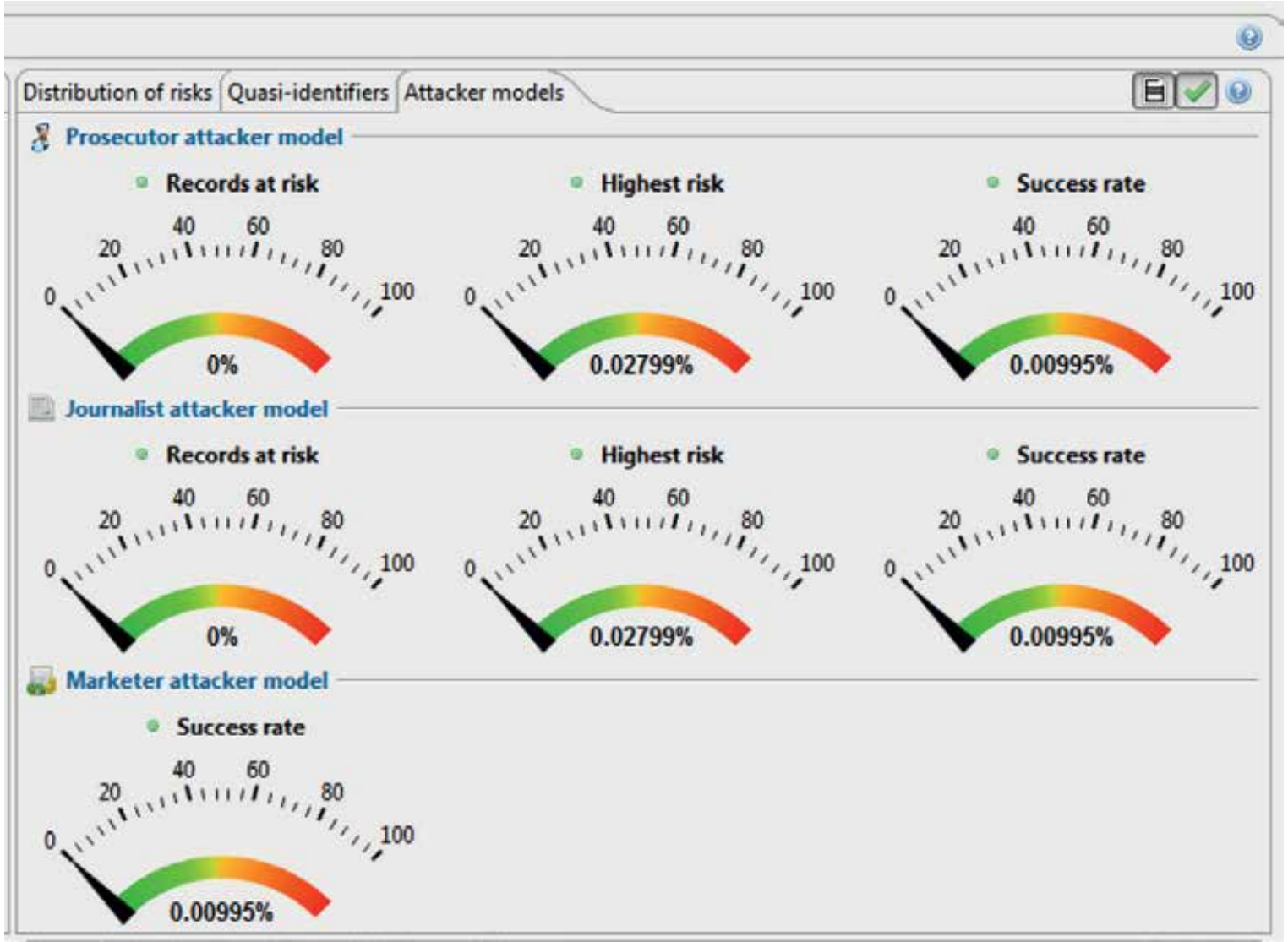

**Figure 7.**
*Risk analysis for various types of attacks after applying hybrid privacy preserving data mining (HPPDM) technique for kvalue = 100.*

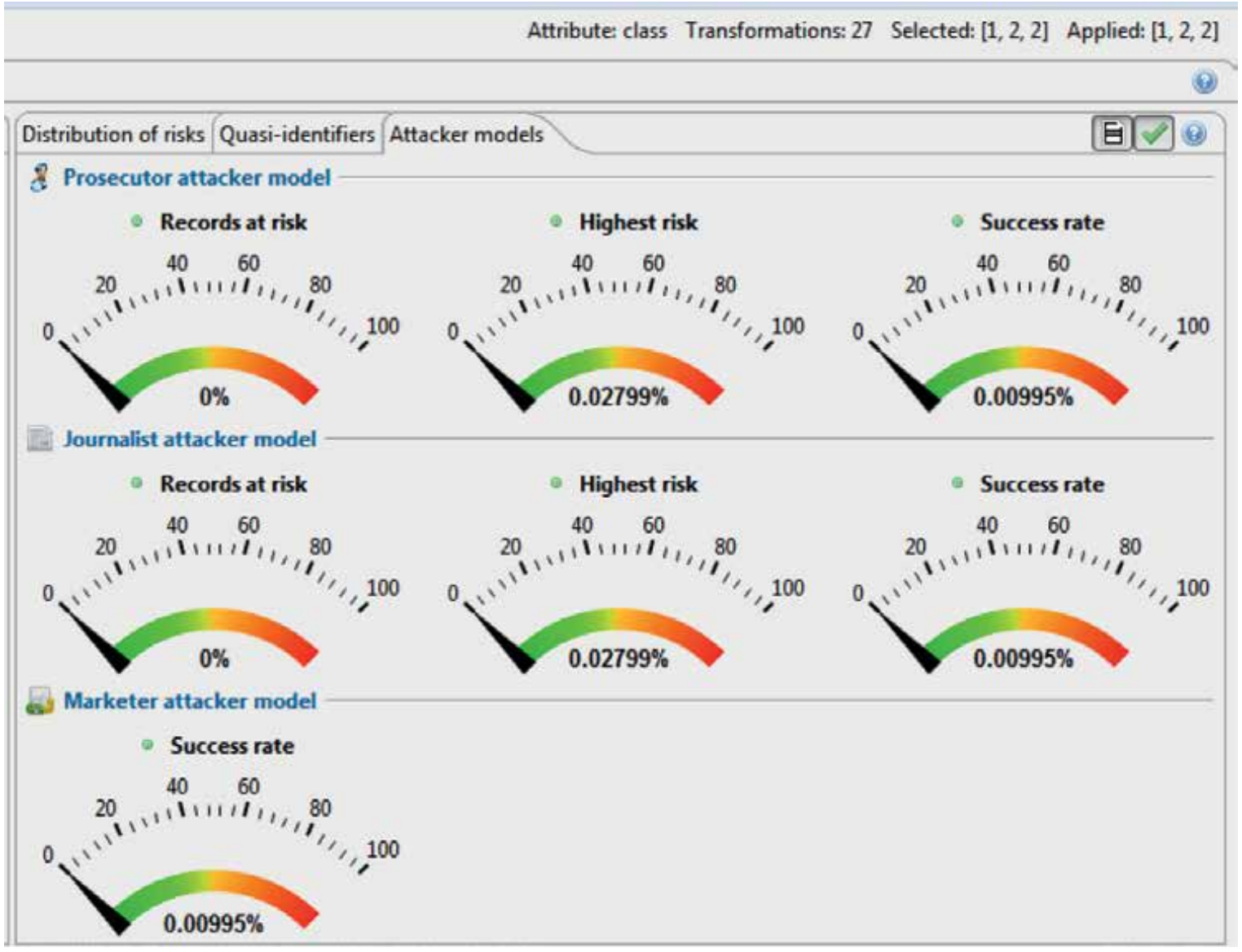

**Figure 8.**
*Risk analysis for various types of attacks after applying enhanced hybrid privacy preserving data mining (EHPPDM) technique for kvalue = 100,l-2 diversity.*

Enhanced Hybrid Privacy Preserving Data Mining (EHPPDM) technique is applied on adult data set. The Experimental results confirm reduction of risks to negligible levels when HPPDM and EHPPDM techniques are applied.

## 5. Conclusion

The proposed Enhanced Hybrid Privacy Preserving Data Mining (EHPPDM) technique is applied on adult datasets from UCI machine learning repository. In fact, EHPPDM technique combines two privacy preservation techniques viz., perturbation and k-anonymization. The numerical quasi identifiers are applied with geometric data perturbation, whereas categorical quasi identifiers are applied with k-anonymization technique. To enhance privacy and reduce attacks, l-diversity (lvalue = 2) is applied on sensitive attributes. The experimental results showed that classification accuracy considerably increased by applying the proposed EHPPDM technique. Moreover, the EHPPDM technique can be extended by including t-closeness property in future works.

## Author details

Naga Prasanthi Kundeti[1*], Chandra Sekhara Rao MVP[2], Sudha Sree Chekuri[2] and Seshu Babu Pallapothu[3]

1 Department of CSE, Lakireddy Balireddy College of Engineering, Mylavaram, Affiliated to JNTU, Kakinada, India

2 Department of CSE, RVR & JC college of Engineering, Guntur, India

3 Department of Mathematics and statistics, K.B.N. College, Vijayawada, India

*Address all correspondence to: prasanthi.kundeti@gmail.com

Section 2

# Watermarking Methods

Chapter 6

# Review on Watermarking Techniques Aiming Authentication of Digital Image Artistic Works Minted as NFTs into Blockchains

*Joceli Mayer*

## Abstract

The recent creation of Non Fungible Tokens (NFTs) has enabled a multibillionaire market for digital artistic works including images or sequence of images, videos, and animated gifs. With this new trend issues regarding fraud, stolen works, authenticity, and copyright came along. The goal of this chapter is to provide an overview of the watermarking techniques that can be employed to mitigate those issues. We will discuss transparency, robustness, and payload of watermarking techniques aiming to educate the artists, researchers, and developers about the many approaches that watermarking techniques provide and the resulting trade-offs. We focus on fragile watermarking techniques due to their high transparency for embedding into artistic works. We discuss the spread spectrum and Least Significant Bit techniques. We describe the usual process of NFT minting into a blockchain and propose a more secure certification protocol with watermarking which employs the same usual NFT minting offered by current marketplaces. The proposed certification protocol mints a checksum string into a blockchain, ensuring the validity of the watermark and the information embedded into this watermark. This proposed protocol validates the date of creation and author identification which are transparently embedded in the artistic work, thus, increasing the security and confidence of markets for artistic works transactions.

## 1. Introduction

The world of digital art has found an innovative way to trade and/or advertise their artistic image works after the recent creation of Non Fungible Tokens (NFTs) associated with a blockchain and some service to sell and buy the images or sequence of images, videos, and animated gifs. The main innovation conferred by NFTs is that the ownership of the digital artistic work is verifiable after the digital asset or a link to the asset with a URL (Universal Resource Locator) is minted into a blockchain.

After the first NFT work was created in 2014, named "quantum", a multibillionaire business has grown around NFTs and blockchains. The market cap of trading NFTs totaled over 23 billion dollars last year. Along with this surge of lucrative trading digital art through NFT, markets came also another black market with players that trade unauthorized copies of digital art disposed of at the markets. As a result, many artists started to include visible and invisible watermarks in their works in the hope that it would prevent stealing or provide additional legal evidence about the authorship to be disputed in a court of law. Moreover, protocols including the watermarked NFTs and the embedded data in the watermarks are being designed to provide the buyers some extra confidence that the work is actually original and created or owned by the seller, avoiding or mitigating a problem created in the market of NFTs: unauthorized copies sold to unwise buyers.

A complication issue is that the artistic work needs to be shown in the markets and is easily copied and re-sold as another NFT. The illegal trader just copies the advertised digital art in the market and mints the digital work (or minting its URL as is the usual practice) as his or her own using the same NFT technology and one of the many available blockchains and storage/displaying servers and services. This illegal trading is particularly damaging to low-cost artistic works which would not be worthwhile to start a legal prosecution against the illegal trader. The costs and difficulties to prove ownership in a court of law are prohibitive. Moreover, many artists that do not even create an NFT for their work are being stolen as illegal traders create the NFT before the actual owner.

The goal of this chapter is to provide an overview of the watermarking techniques that can be employed to mitigate the problem of authentication in this multibillionaire market of NFTs. We will discuss transparency, robustness, and payload of watermarking techniques divided into three categories: transparent with low impact in the artistic work, very robust with high impact in the artistic work and transparent and reversible watermarks. The discussion aims to educate the artists, researchers, and developers about the many approaches that watermarking techniques provide and the trade-offs that each watermarking technique imposes. As the technology of digital art trading evolves, these watermarking technologies and trading protocols will take place to provide a safer and more lucrative environment to the sellers and buyers in this innovative market.

## 2. Security of authorship for minted NFTs

The process of registering digital data (coin, NFT, image, video, etc) into a blockchain, due to a somewhat complex and secure cryptographic protocol employed, provides a very high probability that the digital data can be securely assigned to a owner along with some extra data such as a URL, date and other information about the transaction. The process of registering the data into a blockchain is named minting, due to its similarity to printing (minting) fiduciary money. This process is considered very secure, publicly accessed, and it is verifiable in a noncentralized way by many participants in the process. The decentralized finance (DeFi) approach is based on blockchain to assure proper secure transactions (digital coins or smart contracts) without the need for a unique institutional agent, such as a bank or government [1].

### 2.1 Minting Process for NFTs

Regardless of the blockchain chosen for minting, there is a cost associated with the computing energy spent to process and validate the transactions in the blockchain, usually referred to as "gas" fees. For this reason, the minting of NFTs usually requires an associated storage server to upload the actual image, video, or animated GIF image. Otherwise, the "gas" fees become prohibitive high due to a large amount of data (bytes) required to be verified by the computing servers. Therefore, due to registration costs, in practice, only some data related to the NFT (URL, author, date, or some small information) is actually minted into the blockchain. This raises some issues regarding the security of the digital asset since it is stored in external servers and not registered into the blockchain. Currently, some services are provided for that external storage, however, they do not use blockchain technology and the security is left to the service provider's considerations. Recently, it has been reported that US$ 1,7 million of NFTs was stolen by a hacker from a very popular NFT service name OpenSea.

Therefore, additional technologies need to be provided to digital art creators in order to enable more confidence in the transactions. Besides cryptographic protocols, watermarking techniques are being employed by artists aiming protection for copyright, authentication, and mitigation of frauds in the NFT market.

## 3. Watermarking techniques applied to NFTs

There are a variety of watermarking techniques such as visible, fragile, semi-fragile, strong, and reversible watermark. These techniques may be used to achieve different goals of authentication, copyright protection, tracking, or fraud detection. Some techniques properties, namely, robustness, transparency, and payload are required depending on the desired goal.

### 3.1 Watermarking properties and tradeoffs

#### *3.1.1 Robustness*

Robustness is a desirable property for a technique in the sense that the watermark, which may contain copyright or authentication information, is able to survive a given attack. There are two types of attacks: malicious and nonmalicious attacks. Nonmalicious attacks refer to normal transformations that one digital work may suffer during transmission or processing as a change of image format, from a JPEG to PNG for instance, or a mild filtering or histogram equalization. On the other hand, malicious attacks are designed to either remove the watermark and/or to substitute it with another watermark for fraudulent purposes. Some malicious attacks may include geometric (shearing, horizontal flipping, collage) and volumetric transformations (noise addition, color map modification, filtering, JPEG compression) [2].

#### *3.1.2 Transparency*

Transparency is a very desired property in the context of artistic digital works. The watermarking technique should be as invisible as possible in order to not affect the image quality since the work is presented by a given site or application for potential buyers. However, many artists use available software to insert very visible watermarks over the original work. This approach intends to provide a sample for the digital work either for advertising the author's artistic qualities and/or for indicating that a watermark-free can be purchased after by contacting the author. The approach aims to mitigate possible stealing of the work and re-selling under other authors' names. As a result, the watermark location is visible, and usually damages the image quality and presentation to a certain degree and additionally, operations of collage with image processing can be used to remove the visible watermarks creating a watermark-free similar version in order to re-sell the stolen art in the same site or another similar service for NFT trading.

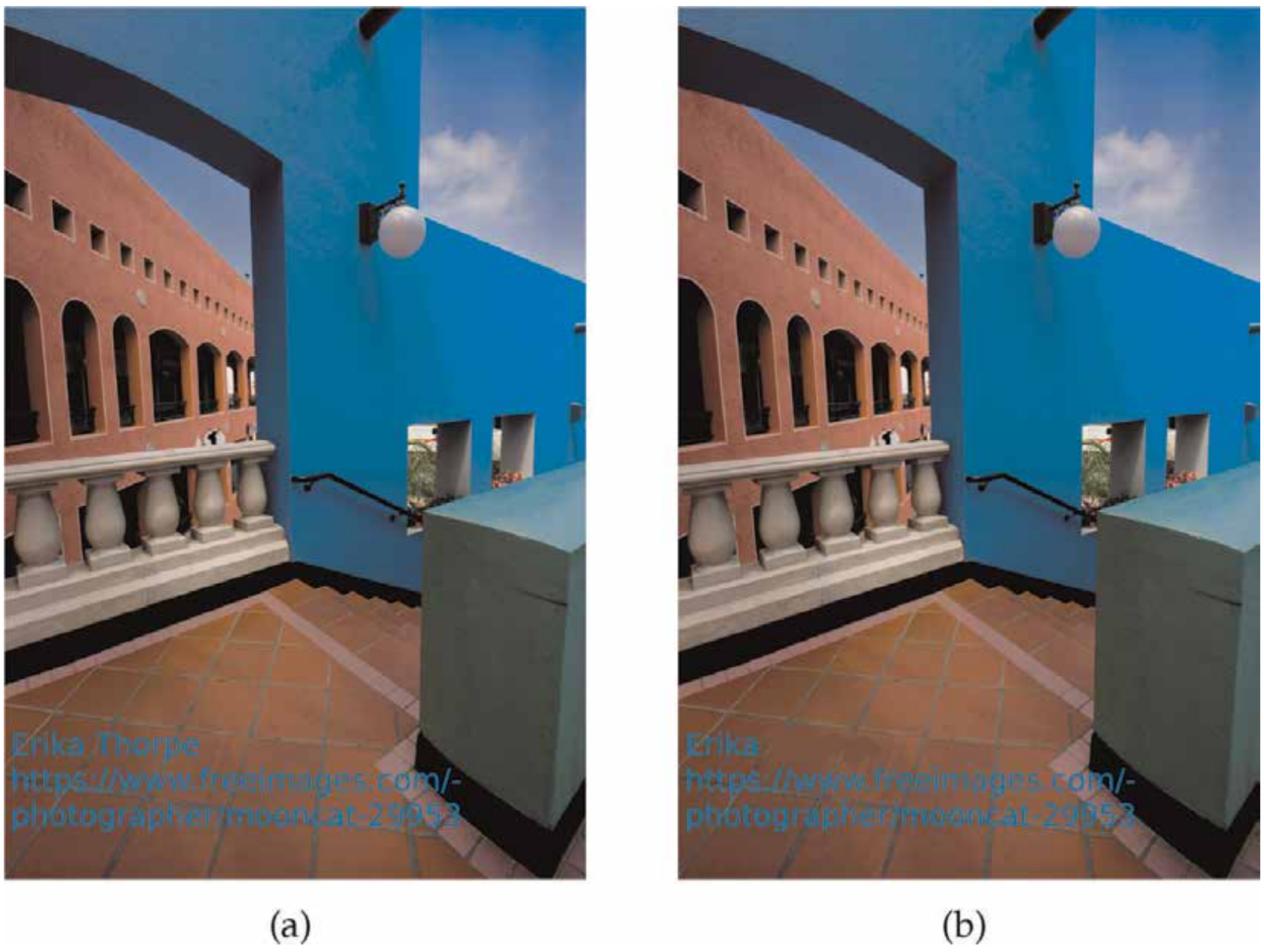

**Figure 1.**
*(a) A copyright-free image from [3] had the author's name and URL included at the bottom. (b) By using image tools, the surname of the author has been removed.*

In order to illustrate the point, in **Figure 1a**, a copyright-free image from [3] had the author name (found in the metadata image information) and the original URL drawn over the image bottom, and in **Figure 1b**, using image tools, the surname of the author has been removed, indicating how easily visible watermarks can be tampered with.

Moreover, in this scenario, a buyer has no guarantee that the received digital work is indeed original as it was created by the seller or it is a stolen edited copy. Therefore, for this NFT trading scenario, very transparent watermarks can be employed to convey authentication information along with some certification protocol provided by a trusted organization. Although invisible watermarks are more complicated and there is no available standard protocol for the artists, the need for a more secure market has been recognized and some corporations are building a trusting environment and friendly applications using watermarking techniques and blockchains that enable certification for the authors along with their digital works.

### *3.1.3 Payload*

Payload is the amount of information measured in bytes that a watermarking technique is able to carry into the artistic image work. The required amount depends on the security protocol used and, on the need to convey some particular information such as author ID, URL, date of minting, and so on. For each given watermarking technique there is a trade-off among robustness, transparency, and payload. A technique with high robustness usually provides a relatively low transparency and a small payload. Conversely, a very transparent technique usually has low robustness and low payload. However, low robustness might be desired in some authentication applications. In such applications, the goal is to keep the digital work authenticated only if it has not been tampered with, thus very transparent and low robustness are proper for the NFT scenario where scarcity and authenticity are essential.

## 4. Semi-fragile and reversible watermarks

Robust watermarking techniques usually produce very low transparency and as stated before, transparency is a very important property when dealing with artistic works, therefore robust watermarking may not be a good choice for NFT authentication. On the other hand, very transparent watermarking can be achieved with fragile, semi-fragile, and also reversible techniques. Among these techniques, some are based on the spacial domain approach using Spread Spectrum (SS) [4] or Least Significant Bits (LSB) techniques. Others techniques rely on the transform domain approach using either Discrete Cosine Transform (DCT) as it is a basis for JPEG compression or Discrete Wavelet Transform (DWT) [2].

The semi-fragile approach allows a small degree of distortion imposed by nonmalicious attach such as image format transcoding, i.e., converting the digital image work from JPEG format to PNG format. However, large distortions usually meant for fraudulent purposes such as image horizontal flipping will result in losing the watermark and the digital work will not be authenticated anymore. In the next section, we describe an authentication protocol aiming to help to provide a more secure market for NFTs.

Reversible watermarking is designed to be able to remove the watermark with a proper secret key in order to restore the original artistic work. This approach is quite interesting for the NFT scenario where image quality is highly desirable. We describe how this feature can be achieved in the next sections.

### 4.1 Spatial domain techniques

Spread spectrum and LSB-based techniques are widely used and are able to provide a transparent watermarking for authentication purposes. These techniques can be designed as region based in order to locate which regions have been tampered with.

#### *4.1.1 Spread spectrum techniques*

The spread spectrum approach [5] is an additive operation on the spatial domain resulting in the watermarked image:

$$Im_W = Im + \alpha b W, \tag{1}$$

where $Im$ is the original image (or frame of a video), $\alpha$ is the scaling parameter designed according to a desired robustness and transparency, $b$ is an antipodal bit $\in \{-1, +1\}$ and $W$ is a watermarking image of same size as $Im$. Usually, this watermarking image is built as white noise, generating a signal with a large spread spectrum in the frequency domain. The antipodal bit $b$ is used to convey one bit of information along with the watermark signal authentication, in some cases it can be discarded, remaining only the weighted watermark signal, i.e., $Im_W = Im + \alpha W$.

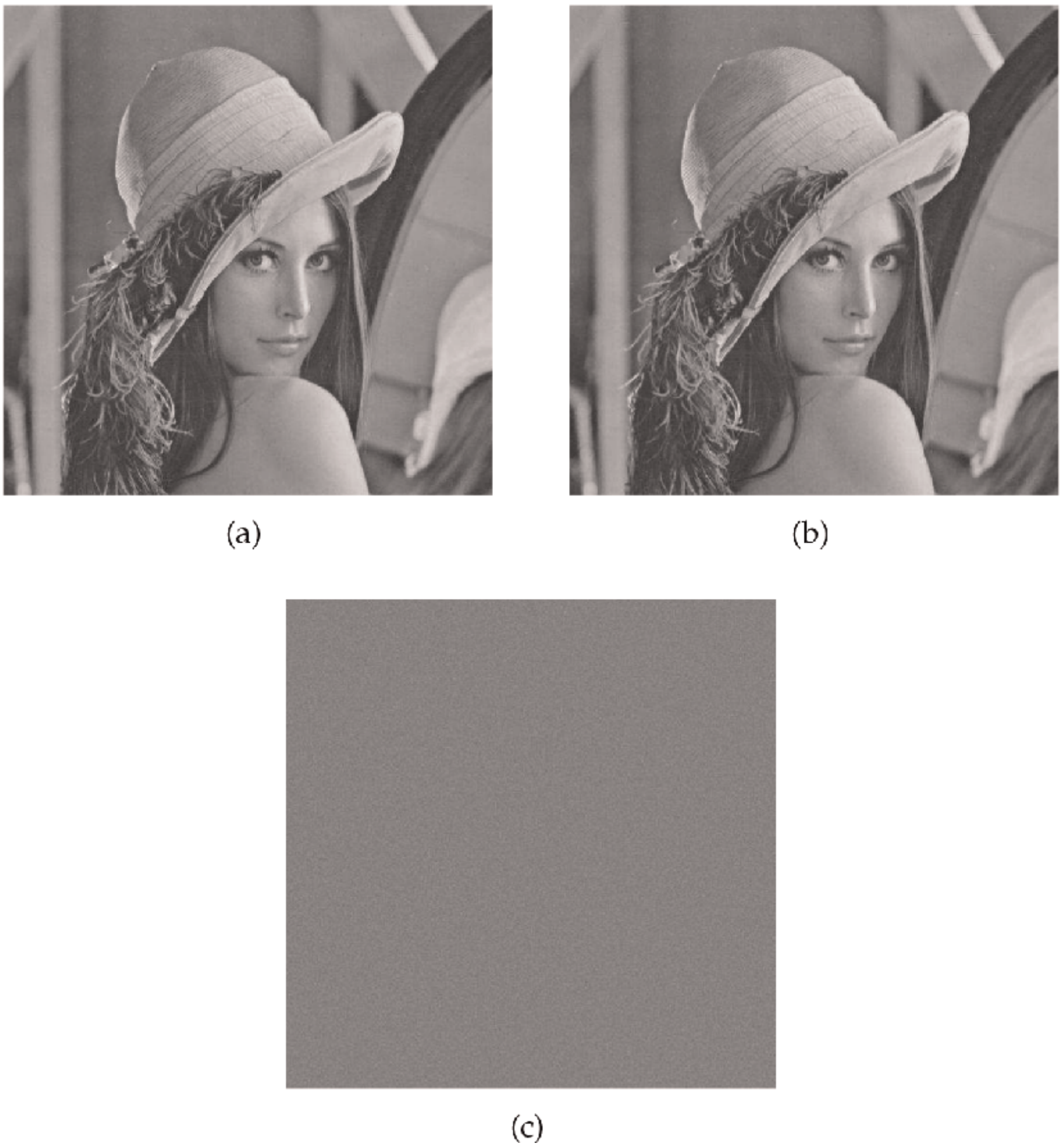

**Figure 2.**
*(a) Original Lenna Image. (b) Watermarked Lenna Image embedded with 10 bits presents very low perceptual impact using the multibit technique in [4]. (c) Difference among the images scaled for visibility.*

In other cases, more bits can be inserted with a more complex watermark signal composed of a weighted sum of pseudorandom sequences of the same dimension as the original image. These pseudorandom sequences can be further optimized for their orthogonality [6]. In either case, the weight $\alpha$ can be computed to satisfy a given tradeoff between robustness and transparency as defined in [4]. **Figure 2** illustrates the very high transparency achieved using an elaborated multibit spread spectrum technique.

#### *4.1.2 LSB techniques and reversible approach*

The watermarking embedding is performed by changing the last K least significant bits of image pixels. The resulting impact for the last 2 bits, for instance, is usually very small and results in a very transparent embedding. Moreover, the approach of LSB embedding can also be reversible using a property of the binary operation XOR (exclusive OR). To understand how it works for the case of LSB embedding in the last bit of each pixel, assume a secret key, named $Im_K$ as one 1-bit image of the same dimension as the original image, $Im$. Using an XOR ($\oplus$) operation for each last bit $i$, the embedding watermark is given as:

$$W(i) = Im_K(i) \oplus Im(i) \qquad (2)$$

Next, the last bit of each pixel of the image, $Im(i)$, is replaced by the watermark $W(i)$, resulting in the watermarked image $Im_W$. The process allows the authentication of the digital image using a given protocol. Moreover, given the secret key $Im_K$, the last bits, changed previously, of the original image can be restored:

$$Im(i) = W(i) \oplus Im_K(i) \qquad (3)$$

By replacing the last bit of each pixel of the watermarked image, $Im_W(i)$ by $Im(i)$, all bits of the original image are properly restored. This property can be used to improve the security of the NFT market. The LSB can be applied to more than the last bit, decreasing the transparency and increasing the payload. Notice that the LSB approach is a very fragile technique where any image modification will damage the watermark. This fragility is acceptable for authentication purposes within a certification protocol and services associated in order to improve the NFT market security and acceptance (**Figure 3**).

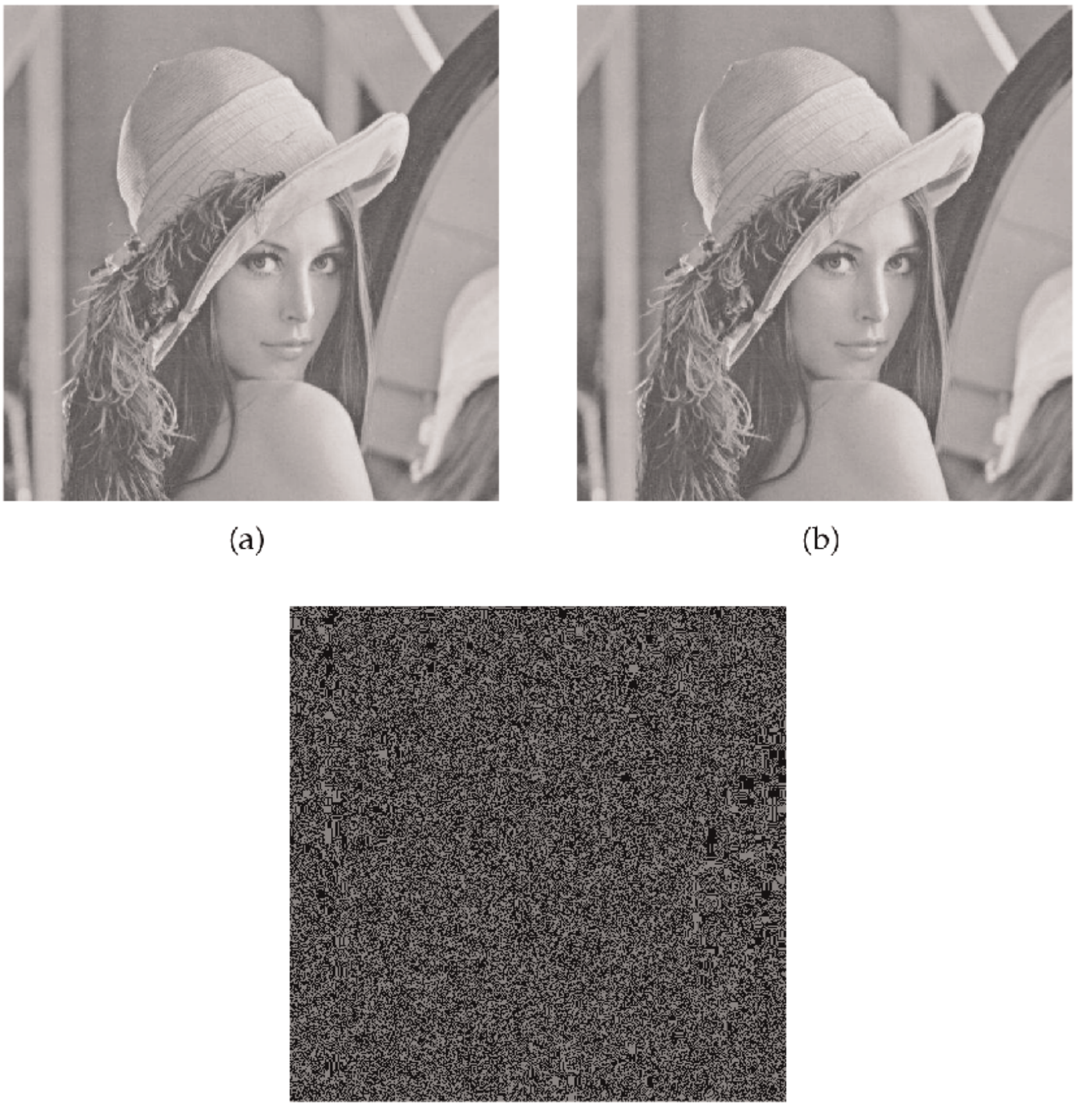

**Figure 3.**
*(a) Original Lenna Image. (b) Watermarked Lenna Image embedded with 1 bits (all zeros in this example) presenting very low perceptual impact using the LSB. (c) Difference among the images scaled by 100 times for visibility.*

### 4.2 Frequency domain techniques

In the frequency domain, it is possible to embed a waterkmark considering some model of human perception in relation to the frequency. In this way the technique can be properly adjusted to transparency according to such a human perception model and, as consequence, to reduce the visual impact of the embedding as compared to spatial domain techniques. The most used transforms for this purpose are the DCT and the DWT. There are a variety of approaches that modify some coefficients in the frequency domain in order to embed a watermark, a review on these advanced techniques is found in [2].

## 5. Certification protocol for watermarking NFTs

The process of minting an NFT into a blockchain is complex and a detailed example of the minting process for an NFT into the Ethereum blockchain is found in [7]. However, services provided by NFT marketplaces can mitigate the complexity for the users. Some of the top NFT marketplaces include OpenSea, Axie Marketplace, Larva Labs/CryptoPunks, NBA Top Shot Marketplace, Rarible, SuperRare, Foundation, Nifty Gateway, Mintable, and ThetaDrop. Using these services the process is simplified to a minimal number of steps which are explained in [8]. Moreover, the Rarible NFT marketplace offers a feature called "Lazy Minting": all fees are charged to the buyer, only after buying the Work is actually minted, the seller receives the Work price amount minus the fees, including the minting "gas". This feature is very interesting to incentive artists and creators [9].

As explained before due to the costs of mining, usually mentioned as "gas" fees, only the URL to the artistic work is actually minted into the blockchain. Usually, the data (representing the image, video, or another Work format) is stored in an Interplanetary File System (IPFS) which is a decentralized protocol and peer-to-peer network for storing and sharing data in a distributed file system. For example, the Pinata [10] system provide a convenient IPFS API and toolkit, to store NFT asset and metadata to ensure that the NFT is truly decentralized.

The minting process validates in a blockchain the transaction associated with the URL of the data (image, video, music, work) stored in an IPFS. Notice that the data itself is not minted into the blockchain. Watermarking is another verification layer of the authentication process along with procedures and evaluations provided by the marketplaces to verify for frauds of many types. As stated before, many artists are employing visible and invisible watermarking to reduce the number of frauds or even to help to detect when an artistic Work is stolen. Other approaches, out of the scope of this work, can be used to help to detect frauds, such as techniques to investigate image similarities and image forensics [11].

The third entity for certification purposes of the transaction can be implemented to help to validate the watermarking process using an Rivest–Shamir–Adleman (RSA) cryptographic protocol. The RSA is a public-key cryptosystem that is widely used for secure data transmission. The acronym "RSA" comes from the surnames of Ron Rivest, Adi Shamir and Leonard Adleman, who publicly described the algorithm in 1977 [12]. Both the work owner and the certification entity can use their private and public keys to improve the authenticity of the Work, the creator (authorship), and the certification entity by using the extra signature (watermark) embedded into the artistic Work. The checksum of this extra validation signature (watermark) can be also minted into a blockchain to register the transaction for extra security, keeping the decentralized approach for NFTs and digital coins.

Using the RSA cryptography process one can generate a watermark $W$ to embed into the image (Work) in order to certificate the creation date, $DATECREA$, the owner identification, $USERID$, and other information. Assume an RSA symmetric cryptography using an encryption process named $PUB(., Key_{PUB})$ and a decryption process named $PRIV(., Key_{PRIV})$ with corresponding public and private keys $Key_{PUB}$ and $Key_{PRIV}$ such that

$$W = PRIV(Key_{PRIV}, PUB(Key_{PUB},\ W)). \tag{4}$$

These processes need these public and private keys in order to properly encrypt and decrypt messages. The public key used for encryption may be distributed publicly without compromising the security while the private key should be only known to the message sender or the Work creator/owner. In the following, we present a certification protocol that validates the authenticity of the Work and the ownership of the creator.

### 5.1 Proposed watermarking certification protocol

Let's assume a certification entity is used for giving better credibility to the artists by showing and dealing with the artistic works, registering the transactions into a blockchain for public auditing as well as for validation of the embedded watermark. This entity can be one of the current marketplaces that register the URL of the Work along with other information into the blockchain, which is usually the Ethereum blockchain. Other related approaches can be found in [13–15].

For a given image, $Im$, a watermark, $W$, can be embedded using one of the many watermarking techniques, including the spread spectrum and LSB techniques explained above. The Work owner (buyer or creator) can use the services of the marketplace to create private and public keys, $Key_{PRIVUSER}$ and $Key_{PUBUSER}$, the private key is kept secure under the user personal and digital wallet. The marketplace also creates those keys, $Key_{PRIVMKT}$ and $Key_{PUBMKT}$ for this transaction. The private keys should be kept secret from the owner and the marketplace. On the other hand, the process of embedding and extracting the watermark is public. The creation of the watermark is based on the user identity given by the marketplace when the account of owner is created, $USERID$, the date of the creation of work, $DATECREA$ and date of transaction (or minting into the blockchain), $DATEMI$, which are properly combined by concatenation, | operator. The owner encrypts his part of the watermark, $W_1$ using the public key of the marketplace and the marketplace encrypts its part of the watermark, $W_2$, using the public key of the owner, such that the final watermark, $W$, is composed of XOR operation, $\oplus$, from both parts:

$$W = PUB(Key_{PUBMKT}, USERID|DATECREA) \oplus PUB(Key_{PUBUSER}, USERID|DATEMI) \tag{5}$$

The watermark $W = W_1 \oplus W_2$ is then embedded into the work before storing the watermarked Work in an IPFS server. Notice that when necessary, the part $W_1$ can be generated by the entity marketplace that knows the part $W_2$ and the extracted watermark $W$ by using the reversible property of the XOR operation explained above in the LSB technique section. The part $W_2$ can be generated by the owner in the same way. Moreover, a checksum, $CHKSUM$, of the watermark can be generated by one of the available algorithms [16]. This checksum is then included in the data that is going to be minted, $DATAMINT$, which includes the URL of the work in an IPFS server and other information. In order to reduce the "gas" fees, The checksum process should result in a much smaller amount of bits than the watermark itself and it aims to provide a validation of the authenticity of the embedded watermark itself. The proposed certification protocol is illustrated in **Figure 4**.

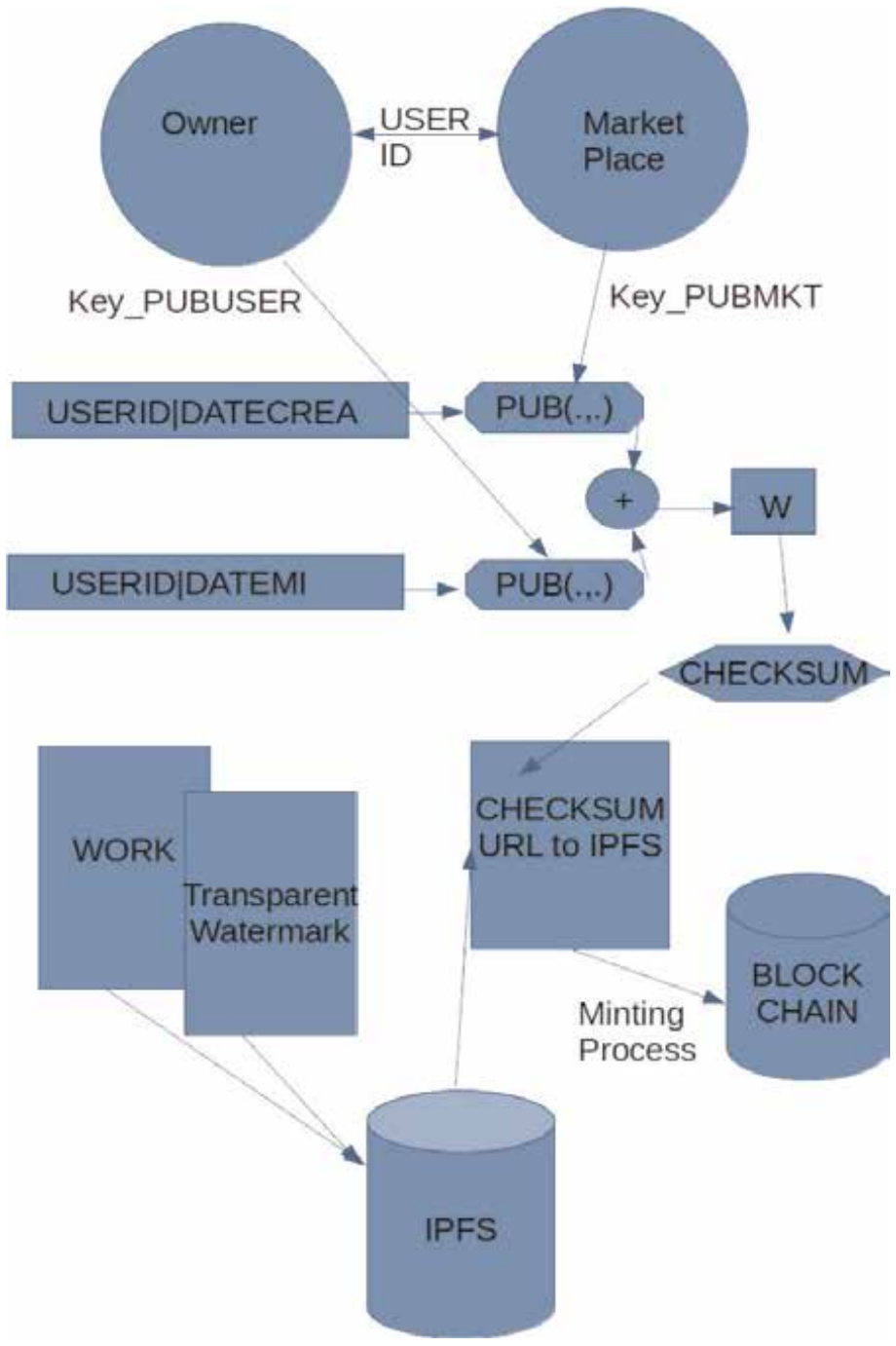

**Figure 4.**
*Proposed Certification Protocol using the owner and a marketplace as entities to validate the ownership.*

### 5.2 Contestant process using the embedded watermark

Consider that the marketplace system, by using forensics tools, finds out that a posted Work is duplicated or very similar. Alternatively, the owner or creator finds out that his work has been stolen and it was also minted into the blockchain after his original work was minted. By extracting the transparent watermarks from the works and using private keys to decrypt the relevant information, one can verify the authorship with the checksum that validates the watermarks into the blockchain, the creation and minting dates along with the users' identification. This information can be used as trusted legal evidence about the contested works. Therefore, the proposed process can be implemented by the marketplace providing a better and more secure service to the artist. Notice that validation depends on the marketplace and the owner's information about the transaction and the work itself. Both entities (owner and marketplace) can verify the corresponding part of the watermark $W = W_1 \oplus W_2$ and validate the ownership. Crossing these two information parts validate the entire process of mitigating a possible fraud from one of these entities. Variations of this protocol can be proposed to increase even more the trust in NFT trade and turning the art market even more valuable. Notice that visible watermarks and multiple transparent techniques can be used with advanced semi-fragile watermarking techniques.

## 6. Conclusions

In this chapter, we discussed how watermarking technology can be employed to increase the security of trading NFTs in this new and multimillionaire market. We propose that transparent embedded watermarks into the original work bring another level of security and do not preclude the use of visible watermarks and the traditional minting process used by current marketplaces. The additional checksum data may increase the costs of minting, however, brings a huge gain in terms of the capacity of securing the authorship of the artistic works in the market. We discuss basic transparent watermarking techniques in order to understand how to generate a watermark to employ with the proposed certification protocol. A certification protocol is discussed in detail and shown to be viable and very interesting to bring more confidence to artistic creators, owners, sellers, and buyers of artistic works.

## Abbreviations

| | |
|---|---|
| NFT | Non Fungile Token |
| URL | Universal Resource Locator |
| SS | Spread Spectrum |
| DCT | Discrete Cosine Transform |
| DWT | Discrete Wavelet Transform |
| LSB | Least Significant Bit |
| XOR | Exclusive OR (binary operation) |
| RSA | Rivest–Shamir–Adleman public-key cryptosystem |
| IPFS | Interplanetary File System |

## Author details

Joceli Mayer
Department of Electrical Engineering, Federal University of Santa Catarina, Florianópolis, Brazil

*Address all correspondence to: mayer@eel.ufsc.br

Chapter 7

# Perspective Chapter: Text Watermark Analysis – Concept, Technique, and Applications

*Preethi Nanjundan and Jossy P. George*

## Abstract

Watermarking is a modern technology in which identifying information is embedded in a data carrier. It is not easy to notice without affecting data usage. A text watermark is an approach to inserting a watermark into text documents. This is an extremely complex undertaking, especially given the scarcity of research in this area. This process has proven to be very complex, especially since there has only been a limited amount of research done in this field. Conducting an in-depth analysis, analysis, and implementation of the evaluation, is essential for its success. The overall aim of this chapter is to develop an understanding of the theory, methods, and applications of text watermarking, with a focus on procedures for defining, embedding, and extracting watermarks, as well as requirements, approaches, and linguistic implications. Detailed examination of the new classification of text watermarks is provided in this chapter as are the integration process and related issues of attacks and language applicability. Research challenges in open and forward-looking research are also explored, with emphasis on information integrity, information accessibility, originality preservation, information security, and sensitive data protection. The topics include sensing, document conversion, cryptographic applications, and language flexibility.

## 1. Introduction

Recent years have seen a dramatic increase in communication via telephone, video, and the Internet. Companies and individuals still exchange paper copies of important documents. The development of a reliable method for authenticating hardcopy documents remains a critical task [1].

In this chapter, we present an innovative method for authenticating documents either electronic or printed documents may be affected combining these methods with traditional ones is recommended In addition to what was mentioned previously. This system is similar to the one proposed in [2], but Noise in the channel is taken into account during the detection process. This is the result of they have a much lower perceptual impact than digital sensors. The binary code of documents is protected by signatures. The system proposed here protects visual content. As a comparison, digital watermarking schemes are capable of transmitting hidden messages [3]. The proposed system classifies documents based on authenticity or nonauthenticity. A major advantage of this system is that it does not require a database. In this case, the information to be compared will be stored. As a result, the proposed method is suitable for this purpose. Self-authentication using text (TSA) is the name given to this technology. In addition, special considerations have been taken. When using a consumer scanner, only a consumer scanner is required. There is a consideration for a printed document. TSA is not relied upon. There are two ways to modify each character: either by modifying the function or by changing the character itself. This can be achieved with very little perceptual impact using text watermarking using techniques [4–6], or visibly, we can increase robustness.

Information on the internet is among the most common information found in today's world. The digital format is having both positive and negative effects on the modern world. The advancement of technologies, medical science, and astronomy are examples. The misuse of these technologies also has a number of negative aspects, such as protecting copyright and manipulating data. As a result of advanced technologies such as the world wide web and high-speed computer networks, unauthorized copying, redistribution, and storing of digital contents has been carried out in many ways. In order to prevent unauthorized copies of digital content, digital content security is crucial [7]. The Internet of Things (IoT) and cloud computing have experienced extensive government and research support at the global level [8]. There are numerous data formats supported by cloud computing, including video, audio, images, and text. However, establishing responsibility and protecting the content are challenging tasks. The data that enables smart cities to function is crucial for sustaining the data infrastructure and enabling the delivery of digital content to citizens. **Figure 1** illustrates this architecture. All data storage, processing, and analysis take place at a central location. Using digital watermarking, you can protect and verify the ownership of digital content. With the right technology, you can embed secret messages in digital content without compromising valuable information. Ownership identification will then be made possible with this information. The different types of digital watermarking are watermarking in text, photos or images, audio, and videos. These three types of watermarking have been the subject of most research. Text watermarks are becoming increasingly popular due to the large number of text documents currently being produced and shared [9].

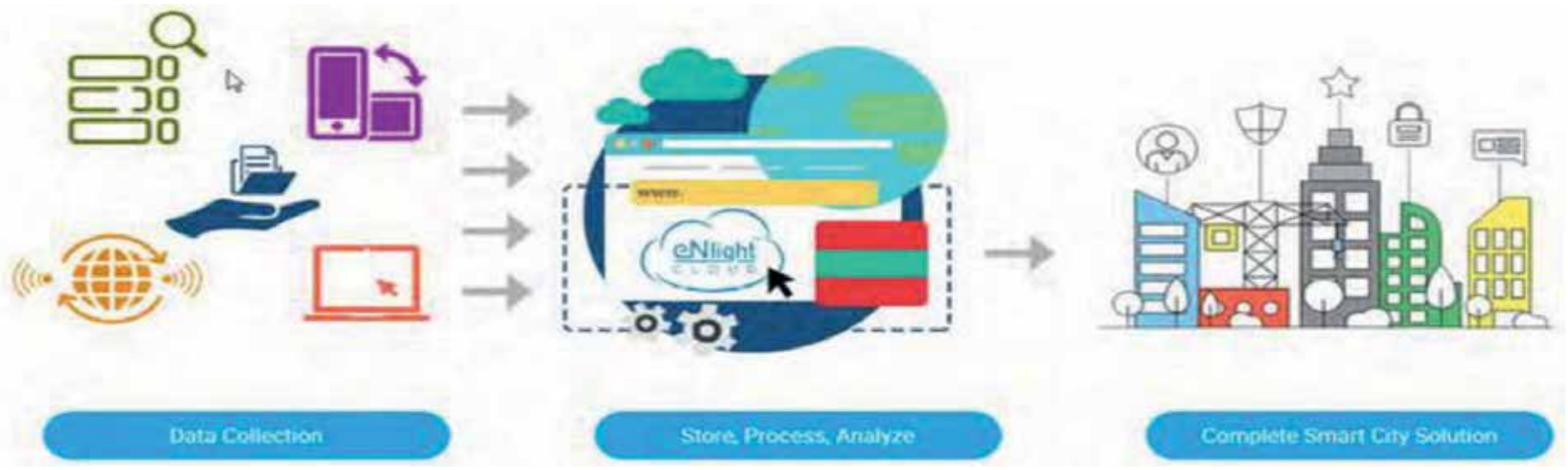

**Figure 1.**
*Smart cities are built according to certain architecture.*

## 2. Technology associated with digital watermarking

The digital watermarking process involves embedding identification information into a data carrier in a way that makes it difficult for third parties to detect, and without it adversely affecting the data. These technologies are often used to protect multimedia data as well as databases and text files. The dynamics and randomness of data make embedded watermarks quite different from those embedded in text files or databases. Data with redundant information and acceptable precision errors are prerequisites to machine learning. Taking into account the range of error tolerance within the database, Boney et al. embedded watermarking in the least important position [10, 11]. Sion et al. proposed a mathematical model based on the statistical property of an array of data (**Figure 2**).

To prevent an attacker from destroying the watermark, attribute data are embedded within it [12, 13]. Furthermore, fingerprints from databases are embedded into watermarking as a means of identifying the information owners and objects distributed [14]. This enables leakers to be identified. Watermarks without secret keys can also be verified using independent component analysis [15]. A number of references are provided for further information [16, 17]. When a fragile watermark is embedded in the tables of databases, data items will be detected in time [18, 19].

Text watermarking uses many generations of methods, which can be categorized into three kinds. There are two types of watermarking: one is based on fine-tuning the document structure, hoping that line spacing will differ from word spacing, and the other is based on subtle differences between line spacing and word spacing. As another type of watermarking, there is text content watermarking, which is based on modifying the content of the text, such as adding white spaces, amending punctuation, etc. Third, watermarking is based on semantic understanding, which can achieve changes by the replacement of synonyms or the transformation of sentences, for instance. Despite the fact that most watermarking studies here focus on static data sets, Big Data's peculiarities, such as high-speed data generation and updating, are not sufficiently addressed (**Figure 3**).

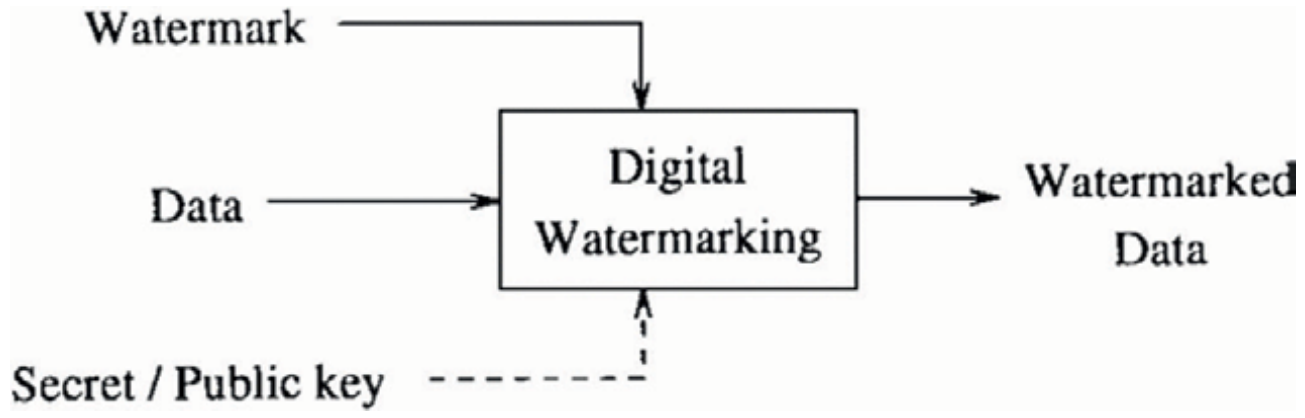

**Figure 2.**
*System for watermarking digital images [12].*

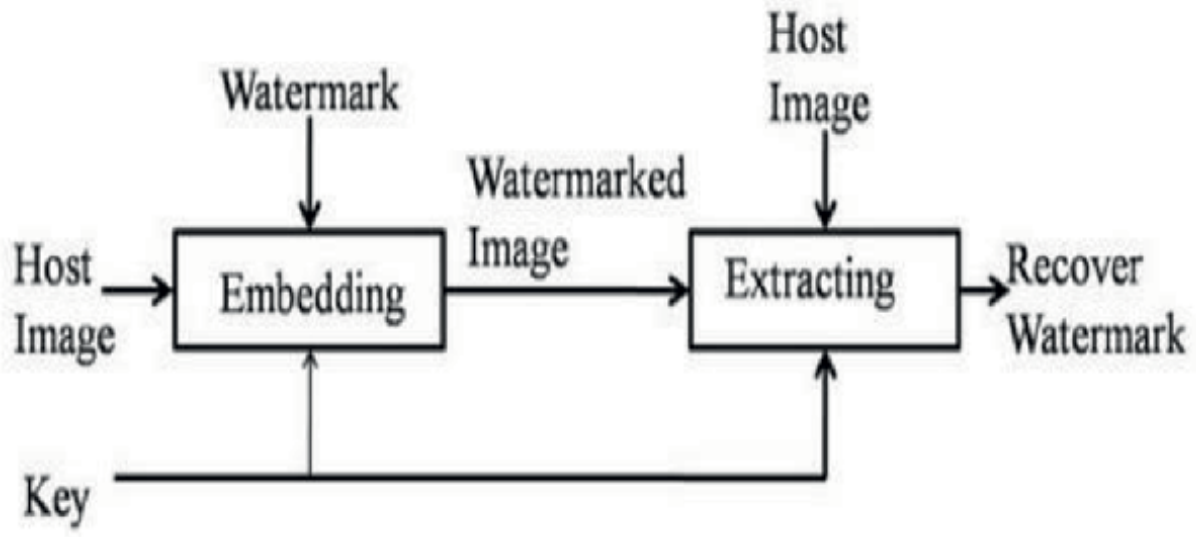

**Figure 3.**
*Watermarking system with a digital signature [20].*

## 3. Analyzing watermarks in text

A digital watermark (digital intellectual property) identifies its owner or originator by providing a unique numerical code. Tracking digital media usage online is done and warnings are sent when unauthorized access or use is possible. A digital watermark is an important part of digital rights management (DRM). A kind of marker is hidden within digital media, such as audio, video, or images, allowing us to determine who owns the copyright to them. By tracking copyright infringement on social media, this technique determines whether a note is authentic in banking. Watermarking is an extremely effective method of securing digital documents in addition to its ability to address distortion, replication, unauthorized access, and security breaches.

There are several ways to classify digital watermarking techniques.

### 3.1 Durability

Digital watermarks that are fragile can no longer be detected if they are altered even slightly. Tamper-proofing is a common method of protecting digital watermarks. Watermarks are commonly used to describe visible changes to a work instead of generalized barcodes.

Digital watermarks, for example, are semi-fragile, which means they resist benign changes but become unrecognizable after malignant changes. The detection of malignant transformations often requires watermarks with semi-fragile properties. The robustness of a watermark depends on how well it resists various types of transformations. Strong watermarks can carry both copying and access control information when used in copy protection applications.

### 3.2 Perception

The term "imperceptible watermarks" refers to those that are virtually indistinguishable from the original signal.

The observable type of watermark is one that can be felt (e.g., network logos, content bugs, code symbols, images that appear opaque). Occasionally, videos and pictures may have transparent/translucent portions for the convenience of the consumer, but these portions degrade the view and degrade the quality of the video.

A perceptual watermarking is not the same as watermarking that uses human perception limitations to appear indistinguishable.

### 3.3 Availability

In general, digital watermarking schemes can be divided into two main categories based on the length of the embedded message:

- A zero-bit message is conceptually sent, and the system is designed to detect whether a watermark is present or absent. Zero-bit or presence watermarking schemes are commonly used for this type of watermark.

- Messages are n-bit-long streams modulated by or and contain watermarks. Watermarking schemes of this type are usually called multiple-bit watermarking or non-zero-bit watermarking.

### 3.4 A method of embedding

The term spread spectrum watermarking refers to digital watermarking methods that are created by modifying signals. Watermarks using spread-spectrum technology may be moderately robust, but they also have poor information capacities as a result of host interference.

Quantization is a type of digital watermarking when the signal is obtained through quantization. The fact that host interference is rejected makes quantization watermarks highly informative despite their low robustness.

A watermark of this type embeds an amplitude modulated signal into an additive modulus, similar to a spread spectrum, but integrated within the spatial domain.

### 3.5 Examining different watermarking algorithms

#### *3.5.1 Requirements*

Watermarking digitally can be applied to a variety of different applications, including:

- The importance of protecting your intellectual property
- Watermarked content can be tracked (different recipients receive different watermarks).
- Monitoring broadcasts (sometimes international agencies watermark their video on television news broadcasts)
- A process for obtaining video authentication
- Screen casting and video editing programs should be purchased in order to get rid of the software.
- Identification cards: their security
- Determining if fraud or tampering has taken place.
- Managing social media content

## 4. Digital watermarking: A life-cycle

A watermarks encrypt information in an audio signal; however, the term is also used in some contexts to distinguish between watermarked and cover signals. The process of embedding the watermark in a host signal is known as encoding the watermark in a host signal. A watermarking process consists of three steps: embedding, attacking, and detecting. Embedding can create a watermarked signal by using an algorithm to take the host and the data.

In general, watermarked digital signals are transmitted or stored to another party. This signal is then modified by another individual. Modifying the digital watermark may be used by a third party to remove the watermark, which may not even be malicious. Copyright protection applications may be affected by this attack. The images or videos can be modified in a variety of ways, such as by lossy compression (which reduces the resolution), cropping, or intentionally adding noise.

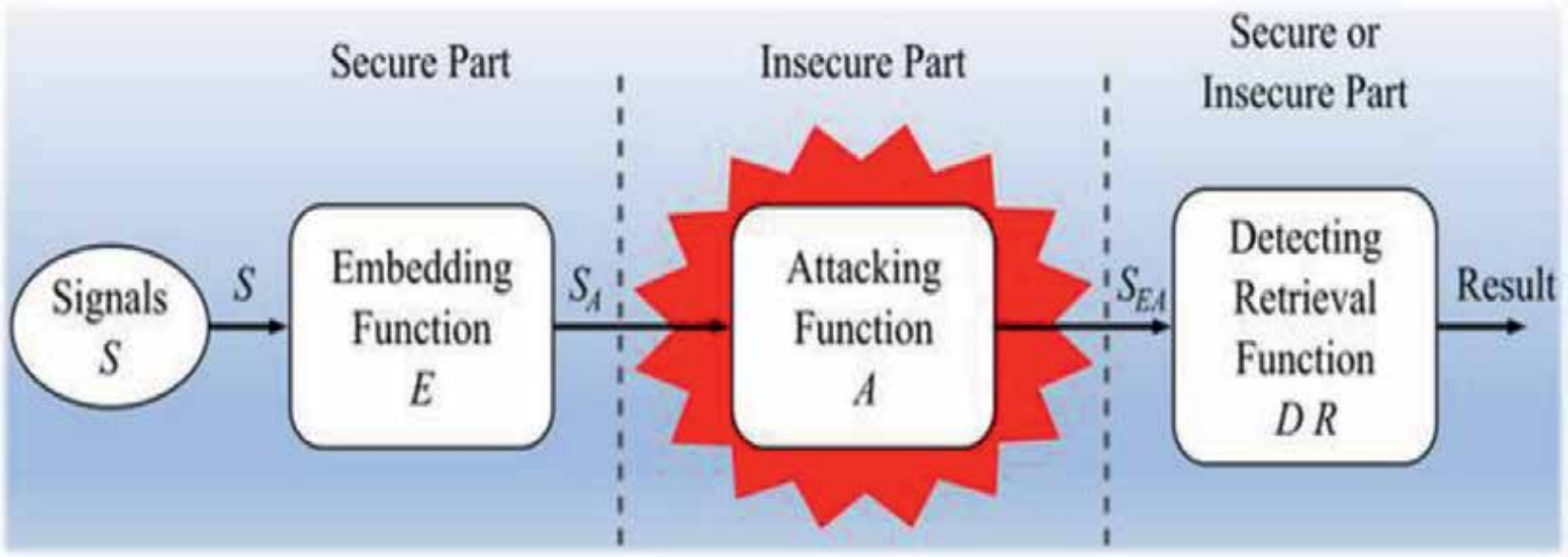

**Figure 4.**
*Life-cycle phases of a digital watermark: embedding, attacking, detecting, and retrieving [21].*

A detection algorithm is used to find a watermark in an attacked signal (also known as extraction). Even a transmission that was unaltered can still contain the watermark. Even robust digital watermarking applications should correctly extract watermarks despite strong changes. The extraction algorithm should fail whenever the signal is modified in fragile digital watermarking (**Figure 4**).

## 5. Watermarking text using digital technology

Watermarking has been a vital research area since 1991 when the concept of text digital watermarking was introduced. As the internet grows and communication spreads globally, a variety of text watermarking techniques have been proposed with the passage of time. Images-based and linguistic-based systems, as well as structural-based and hybrid approaches, are all available [22, 23].

A. Methodologies based on images

The cover text is viewed as an image and embedded with a watermark according to the image-based approach described in [24]. The watermarked logos and images are converted into text strings and the data is generated. Watermarks serve as both a means of verifying ownership and preventing copyright infringement. In spite of the fact that optical character recognition (OCR) is considered safe for formatting attacks, it has limited applicability due to the fact that it ruins hidden information [25]. A technique described by Rizzo et al. [24] encrypts a short piece of text with a hidden watermark while preserving its content strictly. Images cannot be altered in either their content or appearance when converted from the text. Watermarking with blind watermarks helps protect content and ensure visibility. According to the authors of the study, a zero watermarking hybrid approach was used in their study [26, 27]. Watermarks are created by converting images into watermarks and embedding them on book covers. The disadvantage of this technology is that keys generated by certified authorities (CAs) must be stored in a large amount of storage space. In their [28] work, Thongkor and Amornraksa describe the process of watermarking scanned and printed documents with spatial information. For embedding the watermark, the image is composed of white and blue components. To determine whether the proposed technique is efficient, a variety of scanning resolutions, printing materials, and quality levels are analyzed.

B. An interdisciplinary approach to linguistics

The semantic and syntactic approach relies on techniques that emphasize semantics that is used to embed the watermark without altering the meaning of the texts. The idea is to conceal data by using a semantic approach, which replaces words with their synonyms. By using this method, grammatical alternations are used for embedding watermarks without changing the meaning of the original text. The watermarking process involves several language parts, such as verbs, adverbs, nouns, pronouns, adjectives, prepositions, acronyms, and conjunctions. In **Figure 5**, structural and linguistic approaches are compared. Based on integrating Chinese text features, Liu et al. [28] propose a method. Each word has been translated. In addition to measuring entropy, weight is also calculated using entropy. This method does well when considering formatting attacks. An approach to resolving this issue has been proposed by Yingjie et al. [30].
A Watermarking technique based on prose characteristics. Using representative words, one can generate keywords, core verb sets, and proportional features for adjectives. Watermarks are embedded by using verbs, adjectives, nouns, and adverbs. The reposed technique does not embed watermarks well.

C. Approaches based on structural analysis

Those techniques incorporate the essential bits of an In-Text, including its structure and characteristics. Watermarks may be used, for example, as identifiers for documents when they are incorporated into a line. Although this method resolves the problem of document ownership authentication for some types of text documents, it is not applicable to all types of documents. There will be no possibility of hidden information being revealed if there are spaces between words, lines, and paragraphs. The following **Figure 6** illustrates this. The first three lines in **Figure 6** are shifted downward so that the middle line is facing downward. To enhance the textual features preserved by this approach, we also used natural language processing (NLP) techniques and resources. As part of their Arabic watermarking technique, Taha et al. [29] propose the use of Kashida extension characters and extra small whitespaces. This approach does not resist formatting attacks as a result of removing the spaces between words. Ba-Alwi et al. [32] developed a new approach to watermarking zero text based on probabilistic models. A watermark is created by measuring the spacing between characters. In comparison with similar techniques, this method is more robust and performs better in reordering attacks. Zhang et al. [33] proposes a method for encrypting watermarking

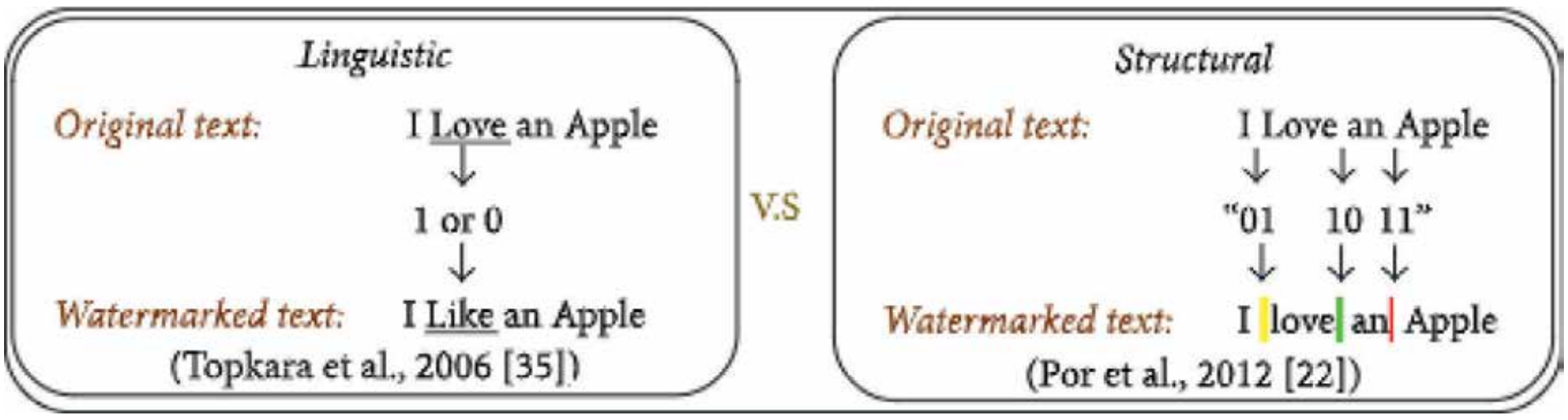

**Figure 5.**
*The comparative analysis of linguistics and structuration [29].*

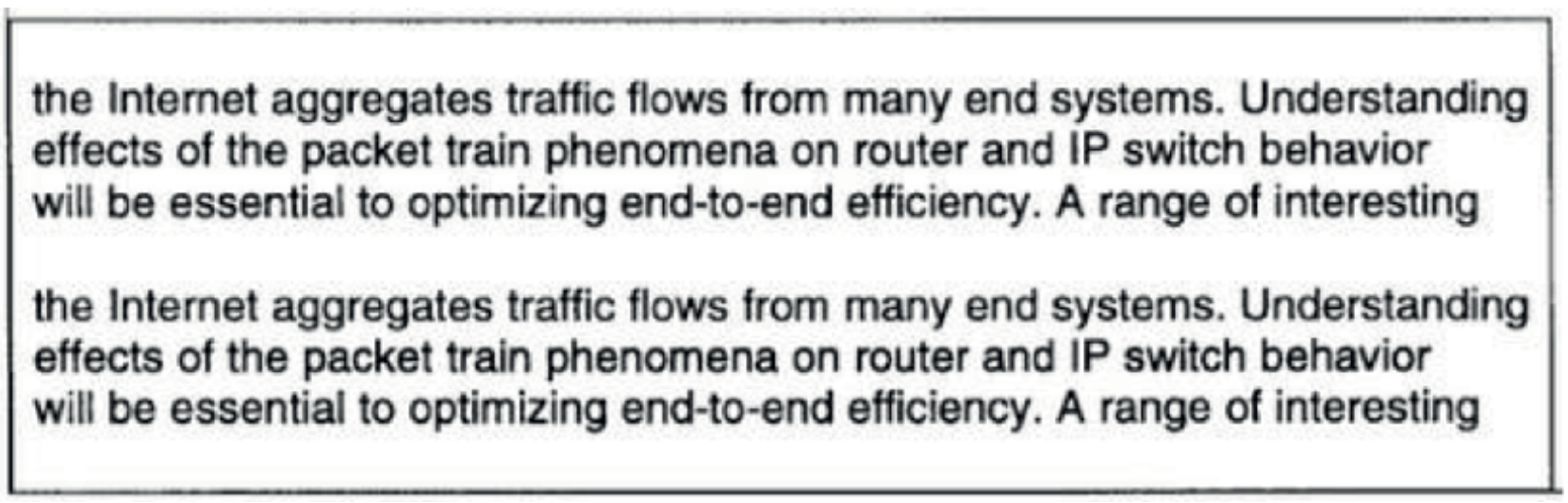

the Internet aggregates traffic flows from many end systems. Understanding
effects of the packet train phenomena on router and IP switch behavior
will be essential to optimizing end-to-end efficiency. A range of interesting

the Internet aggregates traffic flows from many end systems. Understanding
effects of the packet train phenomena on router and IP switch behavior
will be essential to optimizing end-to-end efficiency. A range of interesting information with the Caesar cipher using the user key, then making and packing plain text messages from the encrypted messages. Tests have been performed to prove that the system is undetectable. Whitespace-based approaches are low embedding capable and vulnerable to formatting attacks, according to Liang and Iranmanesh [34]. This method has the disadvantage of requiring a large number of blank spaces to conceal the secret message. By applying the modern technology of information security, Usmonov et al. [35] proposed a technique for protecting data transmitted between logical, physical, and virtual components of IoT systems. A secure online health application is based on the integration of IoT, Big Data, and Cloud convergence, as designed by Suciu et al. [36]. Cloud View Exalead's infrastructure-level information can be used for online and enterprise-based search applications. An approach using Font-Code by Xiao et al. [37, 38] embeds watermarks into font glyphs rather than changing the actual text. This algorithm has the advantage of being robust and imperceptible, but it only works with one font family and is relatively small in capacity. In order to detect the message, a large font size is required, depending on the OCR library.

**Figure 6.**
*An illustration of line-shift coding [31].*

D. Methodologies that combine the best of both worlds

To combine the benefits of different text watermarking techniques, a hybrid approach has been developed. The hybrid approach is considered robust and can be applied to wide-text documents [39]. Elrefaei and Alotaibi [31] proposed a method for handling Arabic text using pseudo-space. This method recovers watermarked letters from strings of connected letters. The proposed method is unnoticeable and robust when used in a formatted document; however, it is not robust when used in a document with retyping. By Hamdan and Hamarsheh [39, 40], Hamdan and Hamarsheh present a new way of hiding text messages in text using Omega network structures. The authors of this study [41, 42] suggested that fragile watermarks be used to safeguard data integrity in the IoT.

## 6. The process of watermarking embedding and extraction

Watermarking is used to discourage illegal copying and prevent digital assets from being distributed [43]. The **Figure 7** below shows an implementation of text digital watermarking. Watermarking is a two-step process that involves embedding and removing the watermark. The document contains a piece of information called a watermark. There are three steps involved in embedding a watermark. Developing a watermark first requires that you include information about its owner, such as the author and publisher. Watermark security is the transformation of a watermark into a binary string or group. The last option is to insert a watermark in a document without having it affect the whole document. **Figure 7** illustrates the process for embedding a watermark. This is accomplished by representing "SM" for the secret message, "T" for the original document, "WD" for a watermarked document and "K" for the key. Watermarked documents can be shared via e-mail, websites, and social media channels. The process of extracting or verifying watermarks reverses watermark embedding. **Figure 6** illustrates how watermarks are extracted by entering the key and the watermarked document and detecting the secret message (**Figures 8** and **9**).

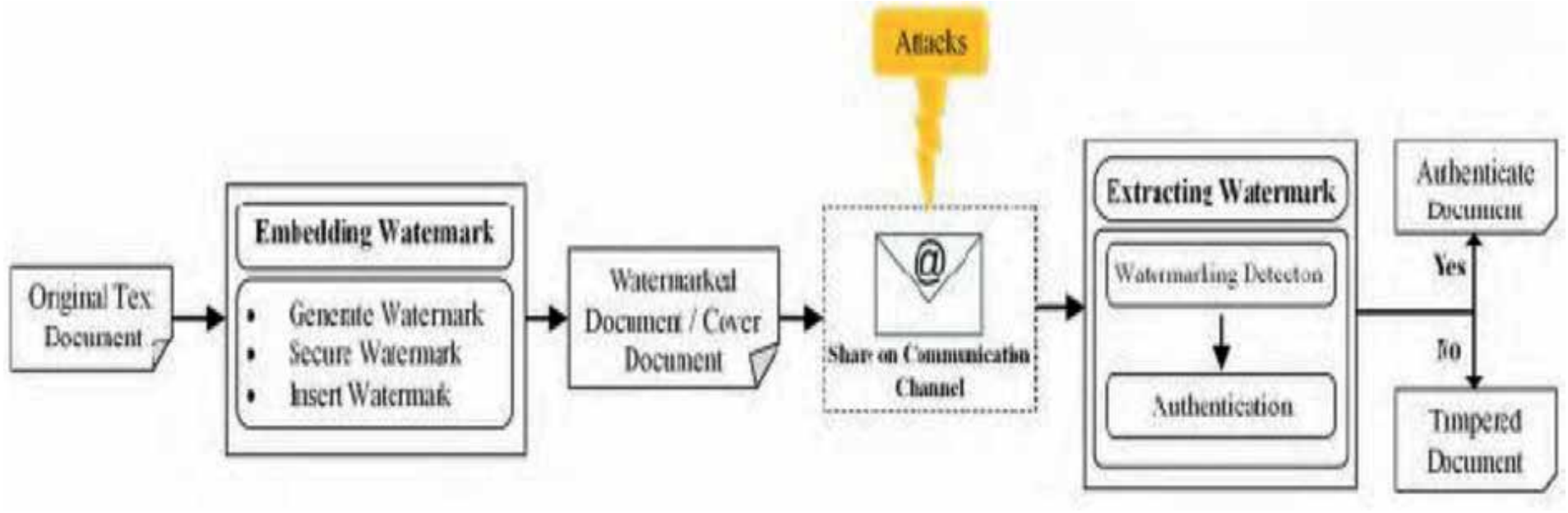

**Figure 7.**
*An overview of digital text watermarking [42].*

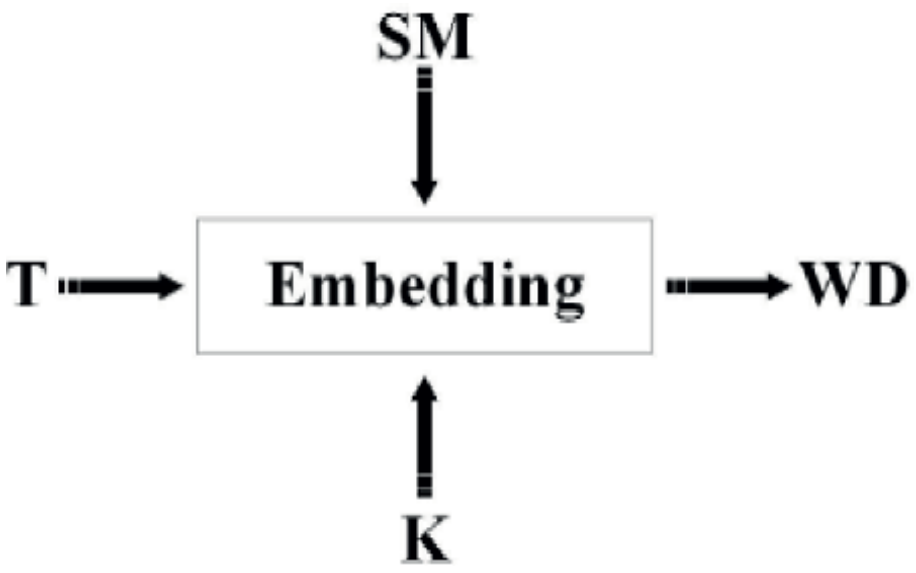

**Figure 8.**
*The process of embedding watermarks [44].*

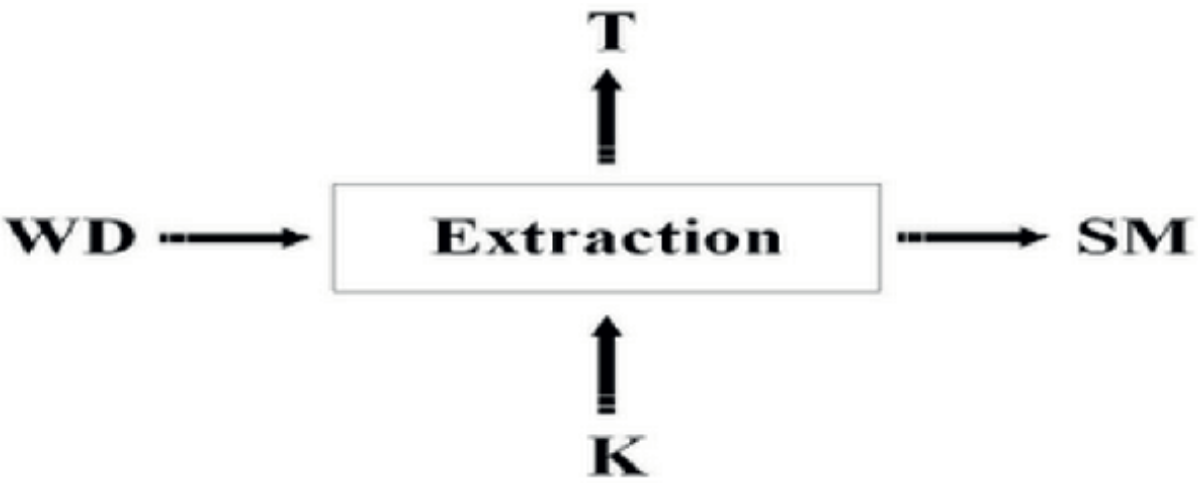

**Figure 9.**
*Process of extracting watermarks [44].*

## 7. Conclusion

Text watermarking provides copyright protection. Generally, this is the most reliable and affordable way, especially for regular users. The algorithm, however, cannot be conquered by a computer hacker. A text document's cost can be kept low to improve watermark protection by preventing them from being attacked. In order to authenticate the digital contents of smart cities, and efficient watermarking algorithm is proposed. A comparison between the proposed technique and previous methods is done in order to evaluate its imperceptibility, security, robustness, and capability. There are many different approaches to this problem, but we also need a method that can be applied to smart cities, IoT devices, and the cloud. Experiments have shown that the proposed algorithm is highly imperceptible and achieves a 95.99 similarity factor. Despite being very robust, this algorithm can detect watermarks with high accuracy despite attacks such as cutting, copying, and pasting, font size, color, and alignment. By comparison, the proposed algorithm is more efficient than previous techniques. This method gives the same results in the cloud computing environment for securing text documents in smart cities. A future extension of the proposed solution could cover the copyright protection of print documents.

## Author details

Preethi Nanjundan* and Jossy P. George
Department of Data Science, Christ University, Pune Lavasa, India

*Address all correspondence to: preethi.n@christuniversity.in

**Chapter 8**

# Application of Computational Intelligence in Visual Quality Optimization Watermarking and Coding Tools to Improve the Medical IoT Platforms Using ECC Cybersecurity Based CoAP Protocol

*Abdelhadi EI Allali, Ilham Morino, Salma AIT Oussous, Siham Beloualid, Ahmed Tamtaoui and Abderrahim Bajit*

## Abstract

To ensure copyright protection and authenticate ownership of media or entities, image watermarking techniques are utilized. This technique entails embedding hidden information about an owner in a specific entity to discover any potential ownership issues. In recent years, several authors have proposed various ways to watermarking. In computational intelligence contexts, however, there are not enough research and comparisons of watermarking approaches. Soft computing techniques are now being applied to help watermarking algorithms perform better. This chapter investigates soft computing-based image watermarking for a medical IoT platform that aims to combat the spread of COVID-19, by allowing a large number of people to simultaneously and securely access their private data, such as photos and QR codes in public places such as stadiums, supermarkets, and events with a large number of participants. Therefore, our platform is composed of QR Code, and RFID identification readers to ensure the validity of a health pass as well as an intelligent facial recognition system to verify the pass's owner. The proposed system uses artificial intelligence, psychovisual coding, CoAP protocol, and security tools such as digital watermarking and ECC encryption to optimize the sending of data captured from citizens wishing to access a given space in terms of execution time, bandwidth, storage space, energy, and memory consumption.

## 1. Introduction

The emergence of the pandemic threatened the existence of humanity, which led scientists to look for solutions to fight against this scourge and reduce its severity. Many solutions have been developed mobile application CovidSafe and CovidScan to check whether the certificate presented by a citizen in the form of a QR code is valid [1], and platforms COVID-19 diagnosis using machine learning from radiography and CT images [2].

Convinced of the harmful influence of this disease on vulnerable people, fighting against the spread of this virus and especially during access to private and public spaces has become a challenge for researchers. Access to these spaces is reserved to people with a valid vaccination pass, people exempt from vaccination, and people with a negative PCR test of fewer than 48 hours.

Any person protected against coronavirus 19 by obtaining the vaccine has a personal code that allows him/her to access a computer service and that shows his/her immunization schedule. This code is still coveted by dishonest people who do not have the complete vaccination scheme to escape controls and access spaces with the help of criminals who offer falsified health passes containing stolen or false QR codes. Even though these acts are criminalized by the states, criminals are constantly innovating in their search for real QR codes.

These irresponsible actions make the task so hard because instead of concentrating on finding a definitive solution to get rid of this dangerous virus, we waste our time looking for solutions to fight against the theft of people's data such as the QR Code. This theft is especially pronounced when a large number of people access public places such as sports stadiums, supermarkets, and events with a large number of participants at the same time. During these times, data management is more difficult in terms of execution time and quality of data sent for processing. IoT platforms have been developed to control access to public places [3].

Upon entering a controlled space an IoT node is required to scan the QR Code, read the tag, capture the citizen's photo and capture the temperature of people to proceed to filter the people who have the right to access from those who do not, and this by controlling the validity of the unique identifier of each one and its belonging to its owner, hence the need for an application that can recognize the face in real-time, processes it, optimize the quality and size of the data. and all this while guaranteeing a fast and secure delivery. Our work aims to use two types of image coding namely visual coding for scanned QR codes, foveal coding for photos captured of people at the entrance of public spaces, Elliptic-curve cryptography (ECC), and watermark to ensure data security.

This work consists of the architecture of our platform based on ECC encryption, CoAP protocol, and technologies used including artificial intelligence, face detection, and recognition in part 2. For part 3, we talk about the psychovisual and foveal coding image using watermarking to evaluate the quality assessment and the execution time of these two coding types. Part 4 presents the comparative results of these coded types of coding/decoding, encryption/decryption, and insertion/extraction of the images and their performances in terms of quality and execution time.

## 2. Medical IoT platform architecture using AI for face detection and recognition

The evolution of IoT platforms is growing, several sectors apply this technique provided that they integrate Artificial Intelligence [4, 5]. The medical sector [6, 7] requires this type of platform to monitor the health status of these patients, so to be able to distinguish different patients we need facial recognition [8, 9].

Our medical IoT platform (**Figure 1**) is to prevent the system from crashing while processing the shipment or citizens from waiting in a long queue or using another person's data to avoid any kind of fraudulent theft. To this end, we used artificial intelligence on the one hand to store the user's personal information, detection of QR codes and images of citizens, face recognition, as well as decision making. On the other hand, we applied visual coding for the QR code and foveal coding for the citizen's faces to study their impact on the image quality and the time to send and process these data.

To access space, a double verification is required; the system verifies the QR code's legitimacy as well as the holder's identification using facial recognition to prevent counterfeiting. There are two methods of personal identification that have been established. The first involves scanning the QR code on the health pass or a PCR test that takes less than 48 hours, and the second is face recognition to verify the individual's identity. After the picture of the person is taken, foveation is used to create a higher-resolution image of the face. The image and QR code are then encoded and decoded; the choice to embed the entire QR code instead of the few bits of the associated information contained in the QR code because it allows us to extract information hidden inside a digital image without distorting the original or losing any data. Authorized recipients can extract not only the embedded message but also the original image, which is an intact and identical bit for bit to the image before the data was inserted. And the most important is that it guarantees and keeps the quality of the image according to the bit rate if the network speed is higher the quality of the image will be with better quality and vice versa. Finally, the images are encrypted with ECC which is a type of encryption that uses an elliptical ECC and a wall cryptosystem that is used in SSL/TLS licenses to encrypt data for devices with limited resources [10]. Moreover, it works with points on an elliptic curve and provides two major benefits;

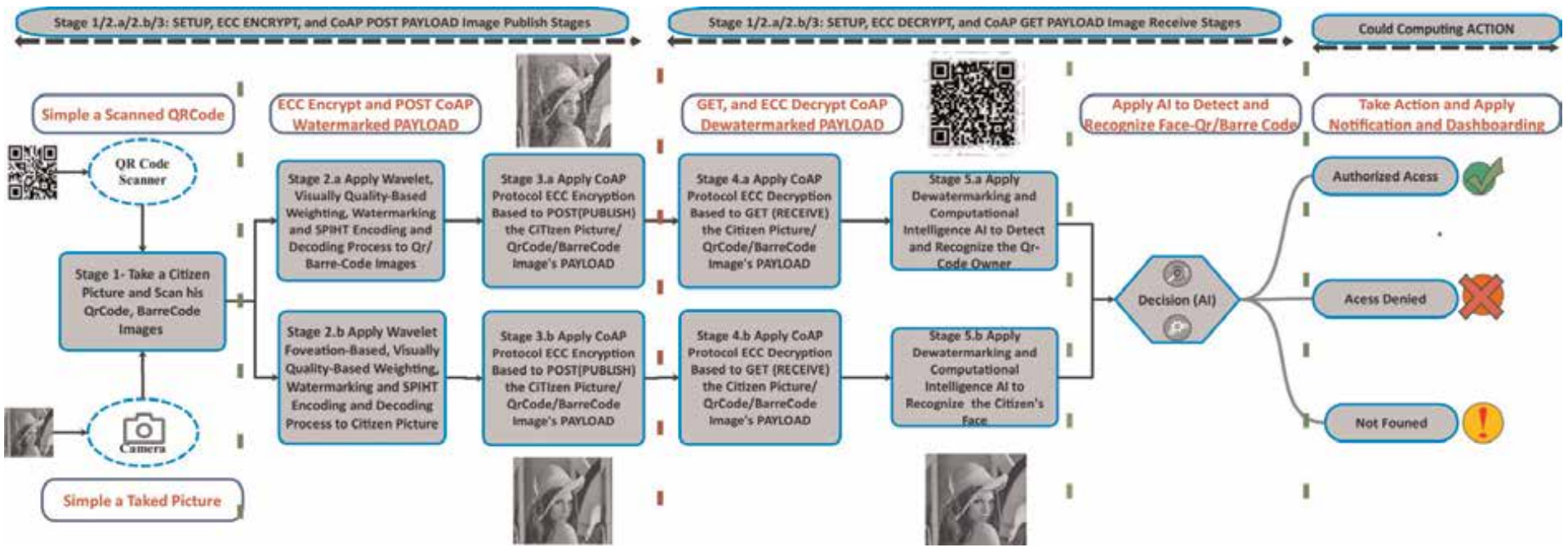

**Figure 1.**
*The medical IoT platform architecture using artificial intelligence.*

security and a short key length. The memory and bandwidth savings, as well as the processing speed and low power usage [11].

Then, the payload is sent via the communication protocol CoAP (Constrained Application Protocol) to the webserver. CoAP is a communication protocol with less memory consumable and time that we used to make nodes and servers communicate (send and receive data exchanged during execution). CoAP is focused on the transmission of tiny messages, which are often sent through UDP (each CoAP message occupies the data section of one UDP datagram). It has a simple binary format. The payload is made up of a 4-byte fixed-size header, a variable-length token, a CoAP option sequence, and a 4-byte fixed-size header. Moreover, CoAP is characterized by its client/server architecture, where the client transmits a method code, such as GET, PUT, POST, or DELETE, to the server [12]. After receiving a request, the server provides a payload and a response code. Finally, a communication layer and a request/response layer are included. The messaging layer is in charge of message redundancy and consistency, whereas the request/response layer is responsible of connectivity and communication. CoAP also provides multicast communication and asynchronous message exchange, as well as confirmable (CON) after it gets the data, unconfirmable (NON) with a unique identification in the event of an unreliable transmission, Acknowledgement (ACK), and resettable messages (RST) [13].

When the data is received by the server, it must be decrypted before it can be read and compared to the database's existing data. The person's entry to the area is granted or denied by a decision system.

Then with the help of the AI, the system begins to process the received information, comparing it to the existing database (Image of the person requesting the access and Image of the real owner of the Qr Code). After comparing the submitted data to the current database, the system decides whether or not to provide the guest access. If the person presenting the QR Code has a valid pass or test, he or she is permitted to access the facility. Access is given if the person is exempt from the health pass and has a fever of no more than 37 degrees. Access is prohibited if someone has a valid QR code but the picture does not match the one previously provided or cannot be identified in the database. Access is disallowed if the person has an expired QR code, a positive PCR test, or is older than 48 hours.

For the AI, the study consists of five stages. The essential phase is the storage of the user's information in the system. The Faster-RCNN architecture is applied for QR code detection. The MTCNN architecture is used for face detection. Finally, the FaceNet architecture is dedicated to face recognition. Deep learning is performed on a database of a few people. For each face image, the MTCNN model produces a fixed-length embedding vector as a unique facial feature (distinguishing characteristics of people). Thus, each person has multiple similarity vectors using Euclidean distance or cosine similarity. The learning process follows the recognition process, namely: image acquisition, detection, and feature extraction. The acquisition phase consists of resizing the input images and normalizing them by removing the average pixel value. Thus, all detected regions of interest (ROI) are scaled to fit the input CNN architecture. In the extraction phase, its role is the feature vectors, to store them in the database. Finally, the detection phase is based on the MTCNN model based on the delimitation boxes of the faces in an image (Landmarks). The MTCNN model includes three processing blocks to perform face detection and tracking. For the first processing block, several candidate windows go through a shallow CNN (P-Net block) and then through a second more complex CNN that consists of refining the windows to reject a large number of windows that do not contain a face (R-Net block). In the third processing block, a robust CNN is used to polish the result and display the landmark positions of the faces.

For FaceNet, it is a facial recognition system, with a unified integration for facial recognition and clustering. It is an architecture that gives an image of a face, extracts high-quality features from the face, and predicts a 128-element vector representation of those features, called face integration. FaceNet directly learns a scene (images or video) from images of faces in a compact Euclidean space where distances correspond directly to a measure of face similarity. Face-Net takes a face embedding as input and predicts the identity of the face that is stored (for recognition). However, a technique that applies a standardized rotation to the face and relies on the facial cues to align the feature vectors. To do the alignment, we first need to find the facial landmarks as fast as possible (speed of execution), so we used two architectures (MTCNN and FaceNet). To extract from the image only the facial features (eyes, face contour, nose, mouth, etc.) as vectors.

Several works [14, 15] use the Faster R-CNN architecture, in our case, it involves extracting the QR code from the image. This architecture is composed of two branches that share convolutional layers. The first branch is a region proposition network that learns a set of window locations, and the second is a classifier that learns to label each window as one of the classes in the training set. We use all the layers in the network that work with object proposals and extract features from the convolutional layers. Build a global image descriptor from the faster activations of the Faster R-CNN layers. The activations of each filter have the same dimension as the number of filters in the convolutional layer. In general, the Faster-RCNN architecture is based on CNN, consisting of a feature extraction network to extract feature maps from the input image. Meanwhile, the convolution layers do not change the size of the image, so usually adopt a basic size, with a set of pad and stride, and each grouping layer reduces the image by half the original size. Faster R-CNN allows us to obtain better feature representations for detection or extraction of Qr-Code from images and improves the performance of spatial analysis and reanalysis.

Our AI platform makes the decision based on the three architectures mentioned earlier: MTCNN, FaceNet, and Faster-RCNN. The extracted features are compared to those stored in the face and QR code database by averaging a similarity metric, in our case we opted for the Euclidean distance. After exploring MTCNN face detectors, including a FaceNet facial recognition package, for verified access.

## 3. Watermarking for foveal and visual image coding to evaluate the quality assessment

Psychovisual coding algorithms consist of optimizing the image quality according to the image complexity. The characteristics of the human visual system (HVS) on the frequency and spatial domain are exploited for best coding results.

Our system is designed according to the scheme shown in **Figures 2** and **3**. The construction steps are the acquisition of the image QR code and face of the citizen, decomposition of the image by applying the wavelet transform (DWT), the psychovisual weighting model, watermark embedding, and the SPIHT scalable coding. The reconstruction steps are the psychovisual inverse weighting model, watermark extraction, and the SPIHT scalable decoding reconstructed image.

It is important to compress the image in order to guarantee a low memory space and a fast image transmission, without degrading the image quality. First, we decompose the original image using discrete wavelet transform. This algorithm consists in cortically splitting the image into a biorthogonal set of wavelets using two filters of type low-pass and high-pass. The use of LPF allows extracting the edges and details of the image, while HPF extracts the most important information seen by the eye.

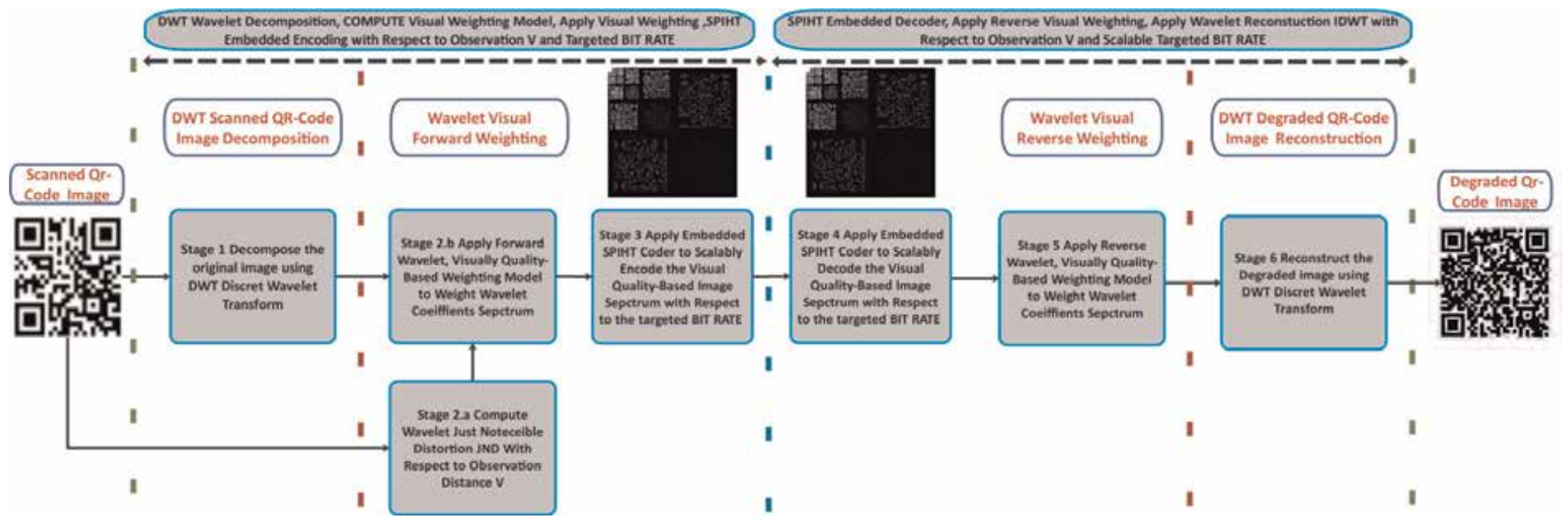

**Figure 2.**
*Visual coding scenario for QR code image.*

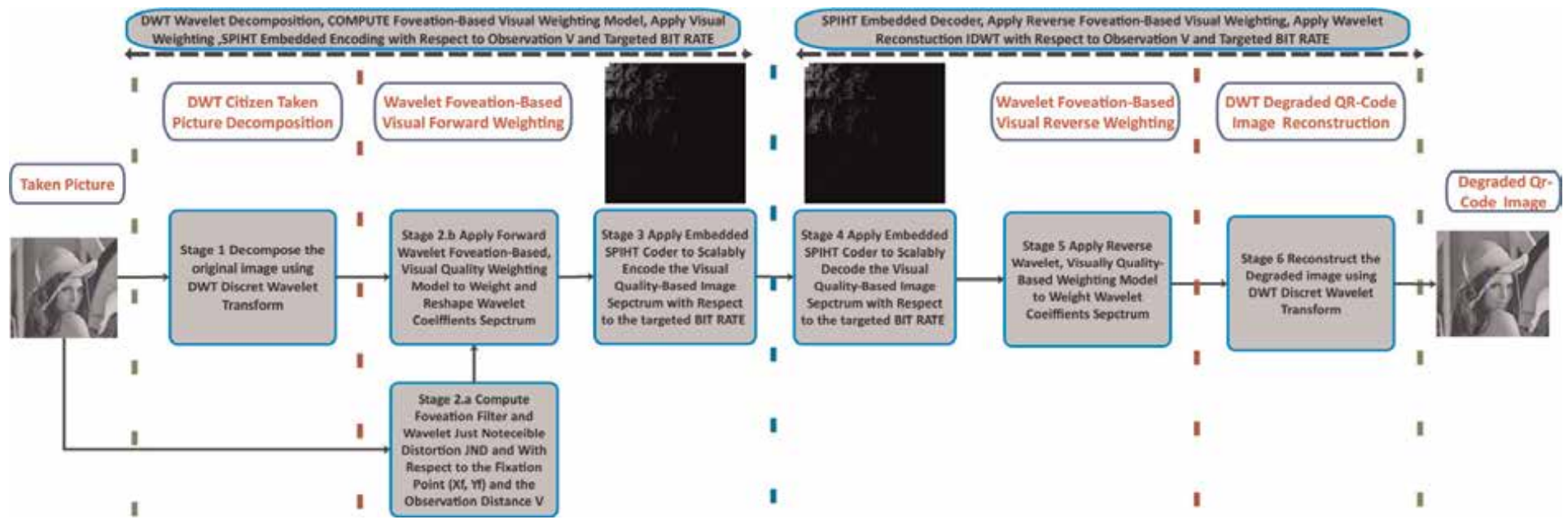

**Figure 3.**
*Foveal coding scenario for Citizen's face image.*

DWT is a transformation used for frequency domain analysis of image [16, 17]. DWT decomposes the image into four non-overlapping multi-resolution sub-bands: LL1 (Approximate or Low-Low sub-band), HL1 (Horizontal or high-Low sub-band), LH1 (Vertical or Low-high sub-band), and HH1 (Diagonal or High-high Sub band). Here, LL1 is a low-frequency component whereas HL1, LH1, and HH1 are high frequency (detail) components.

All wavelets used are based on the Daubechies 9/7 filter for an ideal reconstruction of the image. The benefit of this type of DWT compression is that it ensures fast computation and fewer resources with interesting mathematical properties. Then, a psycho-visual weighting filter is implemented to process the wavelet sub-bands. This model uses the contrast sensitivity filter to discriminate between low frequencies and remove invisible ones. The luminance setting and contrast calibration are adapted according to the perceptual thresholds based on the JND wavelets.

The weighting model combines the following steps; The application of the JND just noticeable distortion model which is established from the adaptation of the luminance and contrast of the image to improve the performance of the perceptual coding. The measurement of the visibility threshold is then based on the JND model. We apply a CSF contrast sensitivity filter that masks invisible frequencies by taking into account the properties of human frequency sensitivity. The luminance mask is then operated on the original wavelet spectrum to adapt the light of the image. The correction factor of the luminance mask is taken into account to varying the luminance of the coded image. The contrast mask is then used according to the perceptual thresholds to calculate the contrast correction, which allows to eliminate the invisible contrast information and to enhance the perceptible information.

The same weighting model is applied to both the visual coding of the QrCode and the foveal coding of the visual image, with an adaptation of the localization of the regions of interest by applying a foveal filter in the case of foveal coding.

Foveation is a lossy filter that reduces the size of the transmitted image by preserving the relevant information and removing all the background from the image that will not be processed by the system. This filter is used to extract the regions of interest and reduces the wavelet coefficients by applying a low-pass filter while focusing on the target region. It is important to determine the parameters of the foveation filter in order to keep the good quality of the image and the needed information. One of the characteristics of the foveation filter is the determination of the frequency spectrum of the area of interest in the function of the distance, when the observation distance increases, the high-frequency areas are higher and higher.

For Watermarking, it is one measure among others to have a good defense against copying [18]. It's a method for embedding data into a multimedia element such as an image, audio, or video file [19]. Current watermarking methods often aim at a certain level of robustness against intrusions that aim at getting rid of the hidden watermark at the cost of destroying the data quality of the media [20]. This chapter presents an image watermarking method based on Discrete Wavelet Transform (DWT). The proposed method is based on a 3-level discrete wavelet transform (DWT). First, the original image of size 512 × 512 is DWT decomposed into the third level using Haar wavelet providing the four sub-bands LL3, LH3, HL3, and HH3 [21]. In the same manner, 3 level DWT is also applied to the watermark image. For this Haar wavelet is used. Digital image Watermarking consists of two processes that are watermark embedding and watermark extracting [22] described below (**Figure 4**).

For watermark embedding, we need a host image and a watermark image, then we begin our process. Firstly, the First level DWT is performed on the host image and the watermark image to decompose it into four sub-bands LL1, HL1, LH1, HH1, and

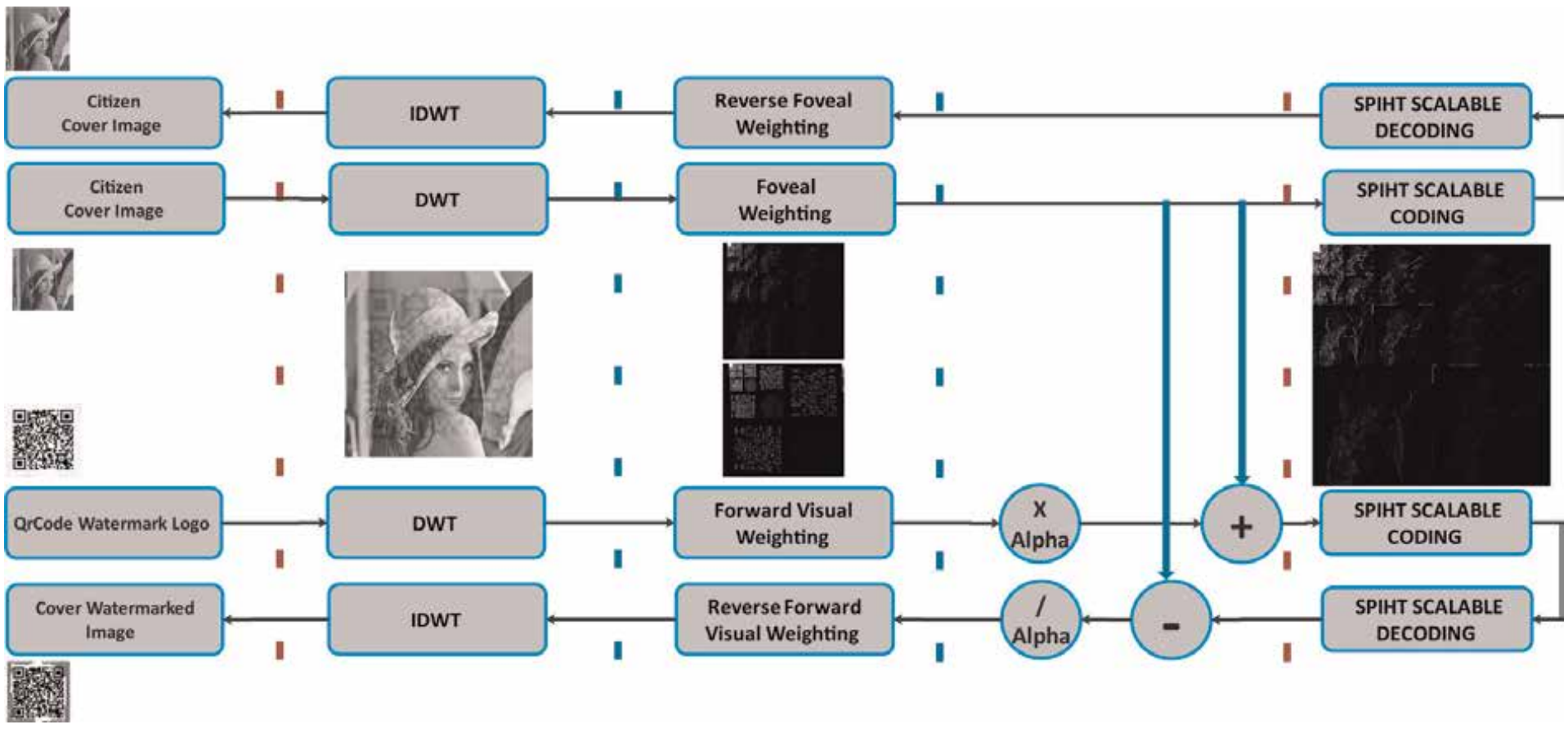

**Figure 4.**
*Watermarked image process.*

wLL1, wHL1, wLH1, wHH1 respectively. Then, the second level DWT is performed on the LL1 and wLL1 sub-band to get four smaller sub-bands LL2, HL2, LH2 et HH2 and wLL2, wHL2, wLH2, wHH2 respectively. While the third level DWT is performed on the LL2 and wLL2 sub-band to get four smaller sub-bands LL3, HL3, LH3, HH3, and wLL3, wHL3, wLH3, wHH3 respectively. Next, A embedding function is used to add the two sub-bands are added with an embedding formula with the value alpha as in is as follows: $newLL3 = LL3 + alpha^* wLL3$. After, the Inverse DWT is performed using the sub-bands newLL3, LH3, HL3, HH3 to get image new LL2. Then, the same Inverse is done using the sub-bands newLL2, LH2, HL2, HH2 to get image new LL1. The last one is performed using the sub-bands newLL2, LH1, HL1, HH1 to get the watermarked image.

For the watermark extracting phase, we start by performing the first level DWT on the host image and the watermarked image to decompose it into four sub-bands LL1, HL1, LH1, HH1, and nLL1, nHL1, nLH1, nHH1 respectively. Then, performing the second level DWT on the LL1 and nLL1 sub-band to get four smaller sub-bands LL2, HL2, LH2, HH2, and nLL2, nHL2, nLH2, and nHH2 respectively. And the third level DWT on the LL2 and nLL2 sub-band to get four smaller sub-bands LL3, HL3, LH3, HH3, and nLL3, nHL3, nLH3, nHH3 respectively. Next, the following extract is performed to get wLL3 with the extraction formulae with the same value of alpha as in embedding $wLL3 = (nLL3 - LL3)/alpha$. After that, we apply inverse DWT on wLL3 with all other subbands (LH, HL, HH) equal to zero to get wLL2. Finally, we repeat the last step 5 two times at each level to get the extracted watermarks (**Figures 5** and **6**).

We developed a visible difference predictor (VDP) metric to evaluate the quality between the reference image and the decoded image. It highlights the set of SVH features by using the wavelet transform to analyze the content of an image. The VDP metric can automatically detect errors in an image that are not visible to the human eye. The principle is to compare the original image and the degraded image by associating each point of the visibility map with a visibility probability. An improved MDVP model is presented in our work [23]. We apply the previously explained weighting model to compare relevant information and neglect unseen information and use a psychometric function to examine the quality factor PS. A mathematical Minkowski summation of all wavelet subbands is then performed to determine the mean opinion score (MOS). The larger the factor determined, the higher the image quality, hence the important role this metric adopts. Similarly for foveal coding, we foveally evaluate the image quality using an FDVP metric improved from the VDP metric by applying a foveal weighting model that uses the foveal filter for the detection of areas of interest [24].

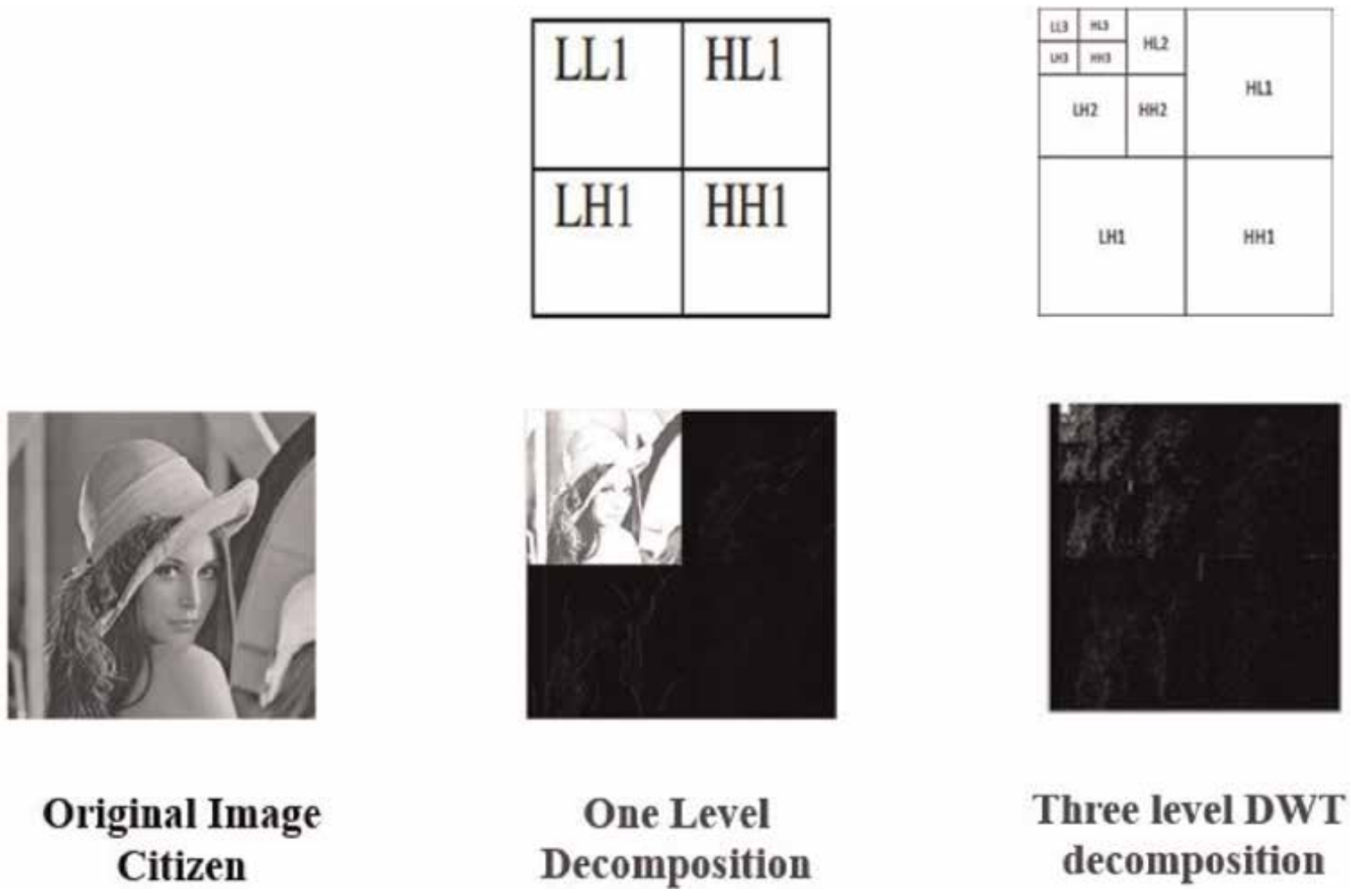

**Figure 5.**
*The DWT compression with face detection and recognition diagram.*

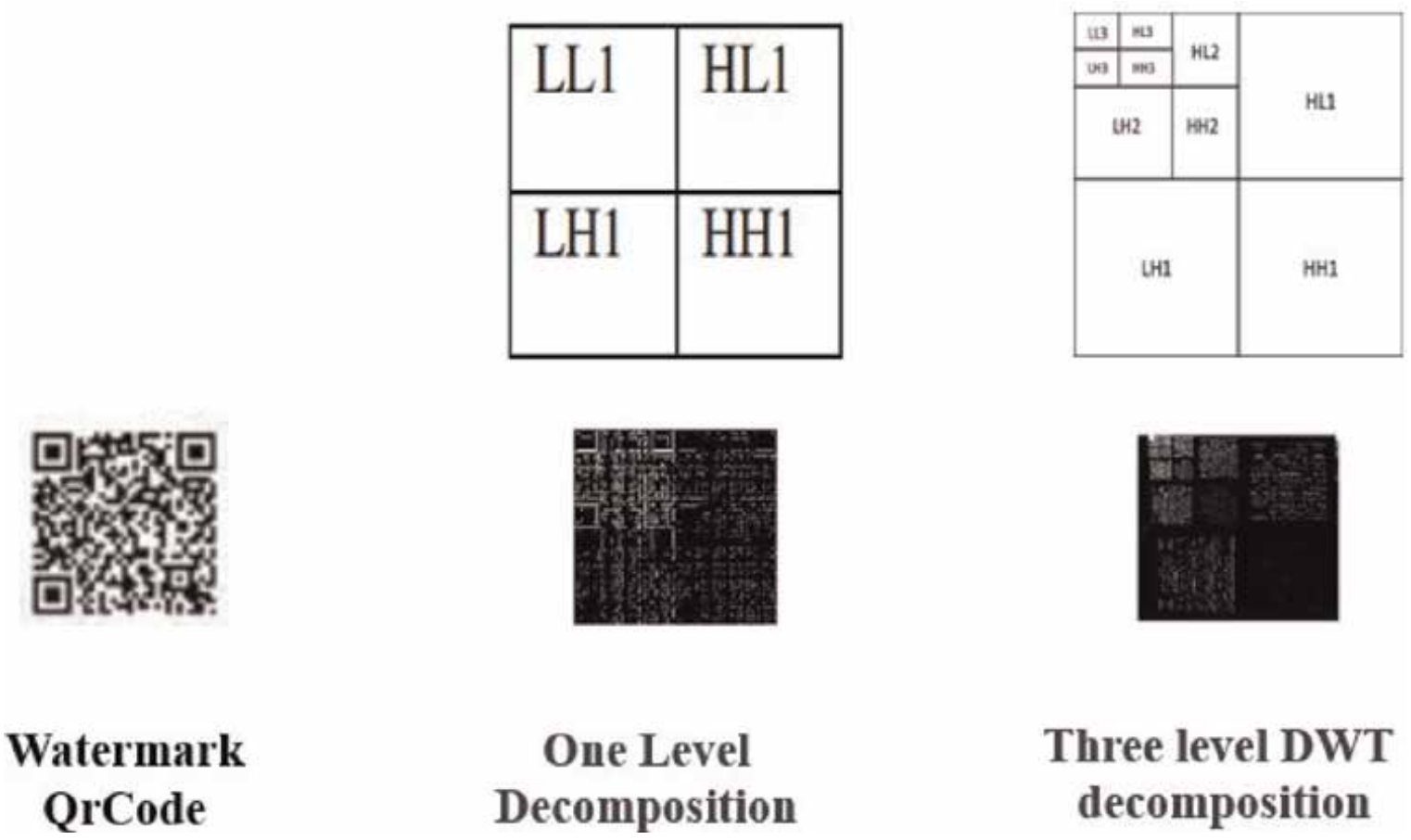

**Figure 6.**
*The DWT compression with QR code embedding diagram.*

The SPIHT progressive encoder is the final phase which has the task of improving the quality of the image progressively and prioritizing the image. The SPIHT encoder is an improved version of the zero tree EZW encoder for lossless compression. The progressive aspect of this encoder is to detect the most relevant information and send it first, transmitting the most significant bits first and then the least significant bits.

## 4. Discussion and results

The transmission of Qr Code and CitizenPicture images classically avoids many problems in terms of size, complexity, loss of time and memory, as well as security, and to ensure their rapid transmission without loss of any of these and specifically without degradation of image quality, We have tried in this work to integrate the classical SPIHT coding, the psychovisual coding, and to highlight the impact of the graphic security based on watermarking and the payload security based on ECC encryption on the execution times.

A broad comparative analysis has been performed which is divided into three evaluations; a qualitative study presented in **Tables 1** and **2**, a subjective study illustrated in **Figures 7** and **8**, and finally the quantitative study displayed in **Tables 3** and **4**.

Starting with **Table 1**, the results obtained concern the different types of coding with different degrees of quality according to four types of binary budgets. Then, the foveal EFIC coding for the CitizenPicture which is in our case Lena test compared it with its reference SPIHT. And the visual coding for QrCode and SPIHT.

From these results, we notice that at the beginning (A. FPS and a. FPS) with a bit rate of 4:1 the images have excellent quality, at the level (B. FPS and b. FPS) with a bit rate of 8:1, the images hold good quality. The images presented in (C. FPS and c. FPS) with bit rate 16:1 have a medium quality, and finally for (D. FPS and d. FPS) with bit rate 32:1 the images retain bad quality. On the other hand, Lena coded with EFIC that focuses on the face and not the whole image provides a better quality comparing it with SPIHT Lena. The same goes for the visual coding of the QrCode EVIC, it is better than SPIHT in terms of perceptual quality.

| | | | |
|---|---|---|---|
| 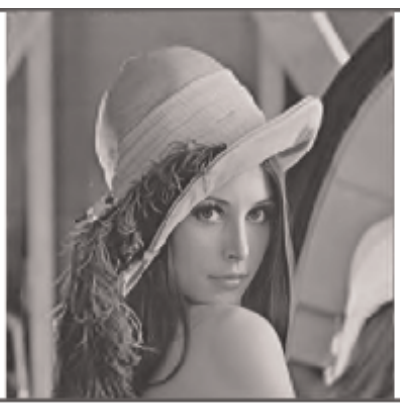 | 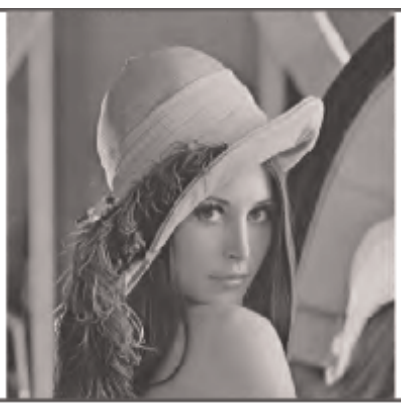 | 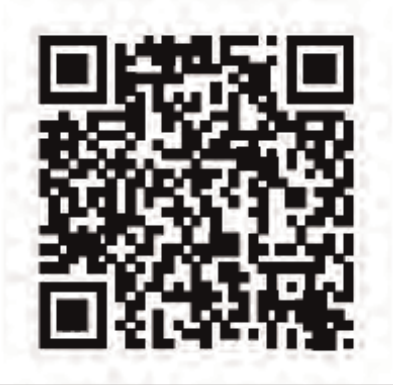 | 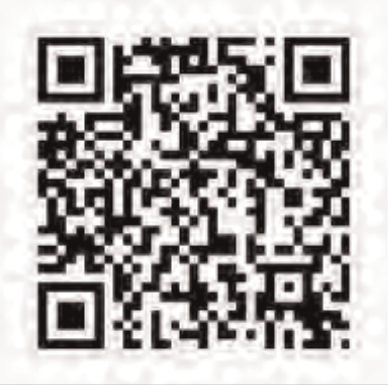 |
| A. $PS_{EFIC} = 0.8392$ | $PS_{SPIHT} = 0.7984$ | a. $PS_{EVIC} = 0.9031$ | $PS_{SPIHT} = 0.7286$ |
| 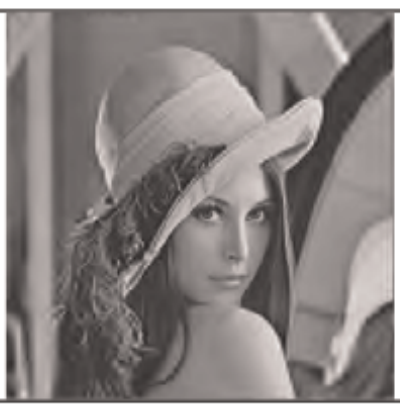 | 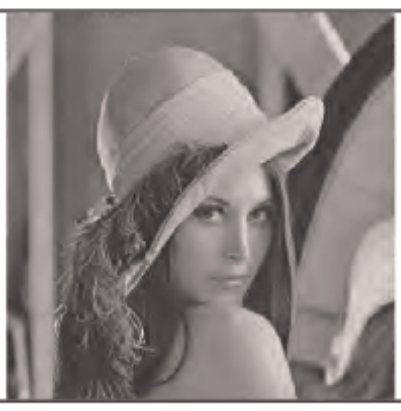 | 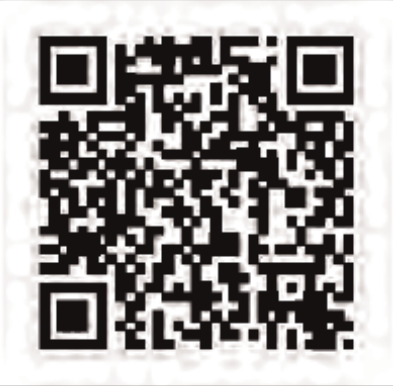 | 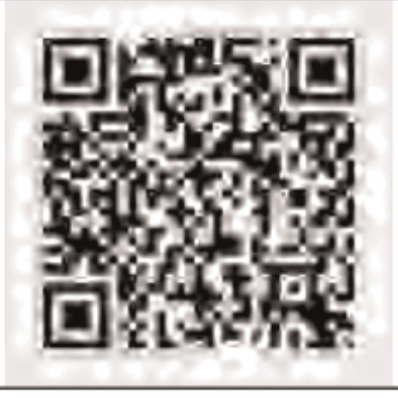 |
| B. $PS_{EFIC} = 0.7299$ | $PS_{SPIHT} = 0.6807$ | b. $PS_{EVIC} = 0.8017$ | $PS_{SPIHT} = 0.5934$ |
| 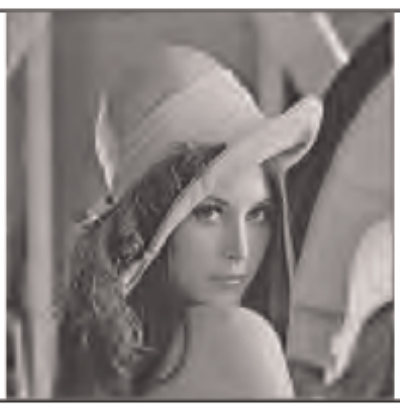 | 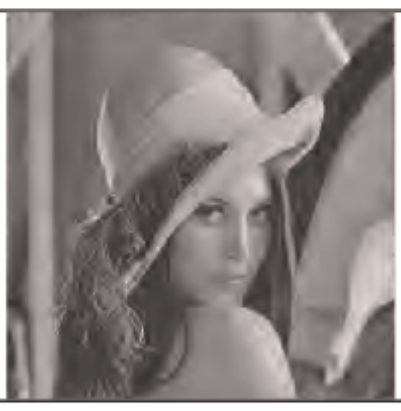 | 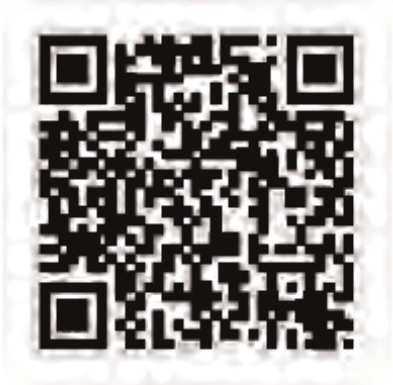 | 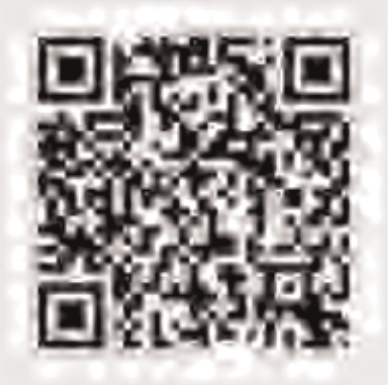 |
| C. $PS_{EFIC} = 0.5785$ | $PS_{SPIHT} = 0.5321$ | c. $PS_{EVIC} = 0.6690$ | $PS_{SPIHT} = 0.4474$ |
| 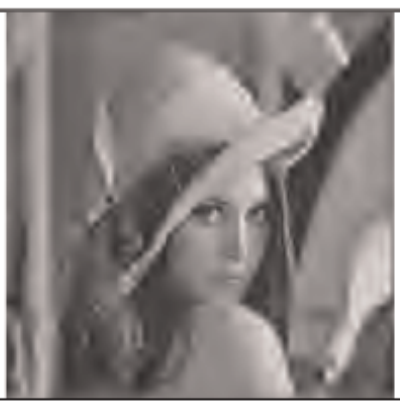 | 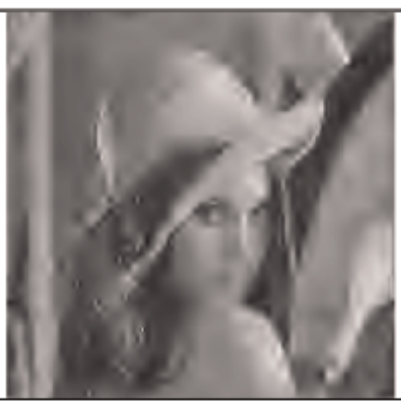 | 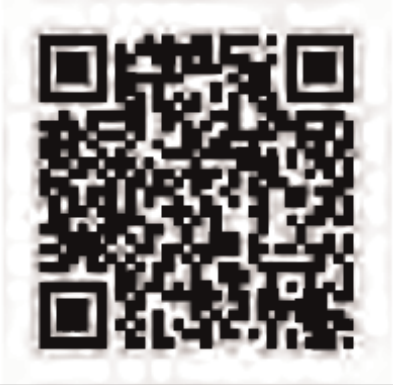 | 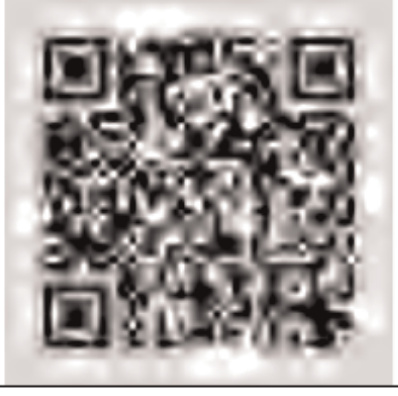 |
| D. $PS_{EFIC} = 0.4130$ | $PS_{SPIHT} = 0.3824$ | d. $PS_{EVIC} = 0.4553$ | $PS_{SPIHT} = 0.2998$ |

**Table 1.**
*Lena EFIC foveal coding images versus the SPIHT coding images and QR code EVIC visual coding images versus its optimized version SPIHT with their quality scores PS using visual WVDP assessor given for varying bit rate and fixed observation distance. The bit rate varies as follow: A. 0.25 bpp, B. 0.15 Bpp, C. 0.0625 bpp, D.0.0313 bpp, a. 0.25 bpp, b. 0.15 bpp, c. 0.0625 bpp, d. 0.0313 bpp.*

Moving on to **Table 2** which illustrates the Watermarking Lena SPIHT (column 1) versus SPIHT deWatermarking QrCode (column 2) and the Watermarking Lena EFIC (column 3) versus EVIC deWatermarking QrCode (column 4) according to different quality and bit rate budgets. From these images, we get four types of bit rate high bite rate which gives excellent quality, medium bite rate which offers good quality, reasonable bite rate which delivers acceptable quality, and inferior bite rate which supplies unsatisfactory quality. On the other hand, we notice that the watermarking Lena EFIC versus the Dewatermarking QrCode EVIC provides an excellent quality when compared with its reference SPIHT.

| | | | |
|---|---|---|---|
| 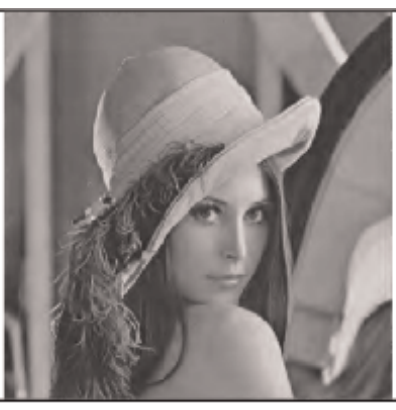 | 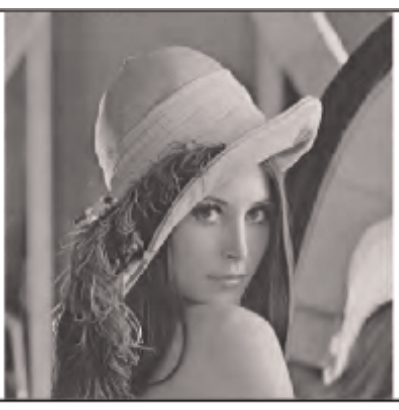 | 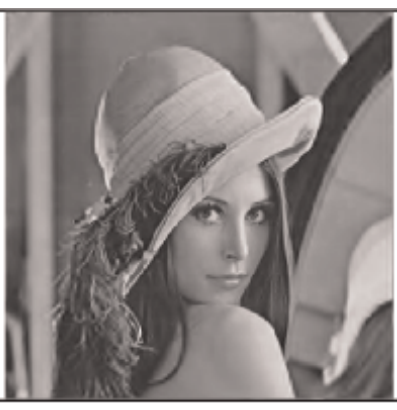 | 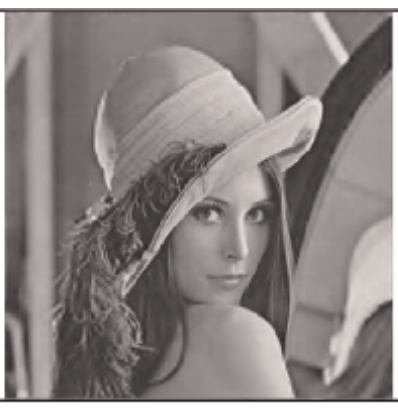 |
| A. REFERENCE CODER SPIHT | REFERENCE CODER SPIHT | A. FOVEATION CODER EFIC | VISUAL CODER EVIC |
| 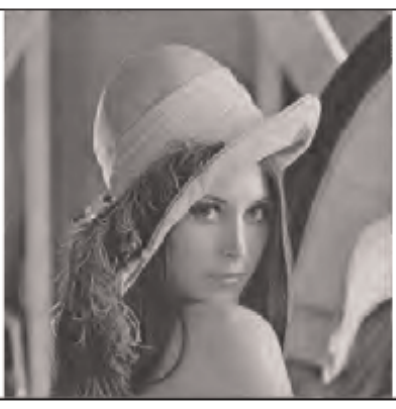 | 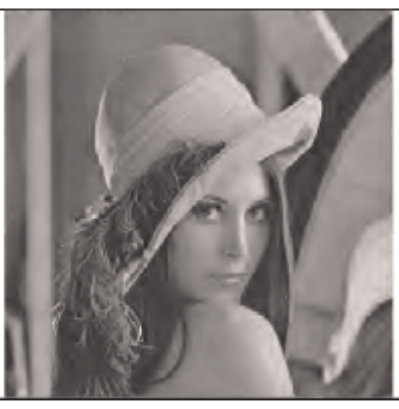 | 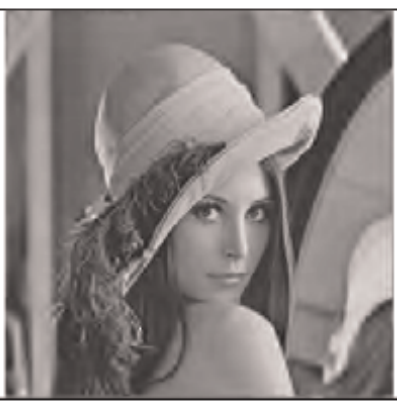 | 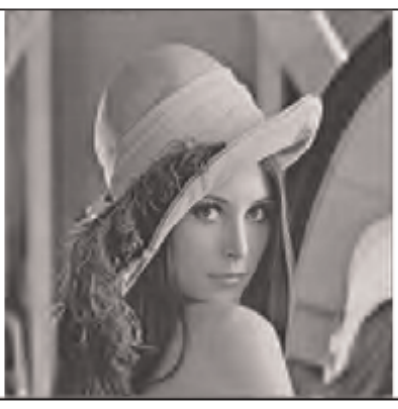 |
| B. REFERENCE CODER SPIHT | REFERENCE CODER SPIHT | B. FOVEATION CODER EFIC | VISUAL CODER EVIC |
| 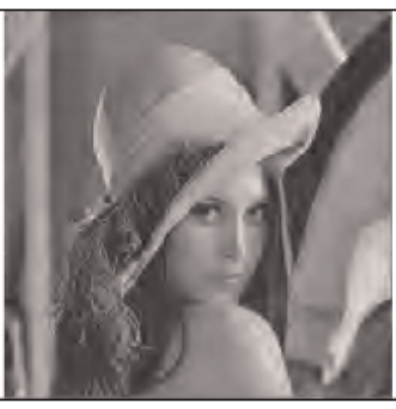 | 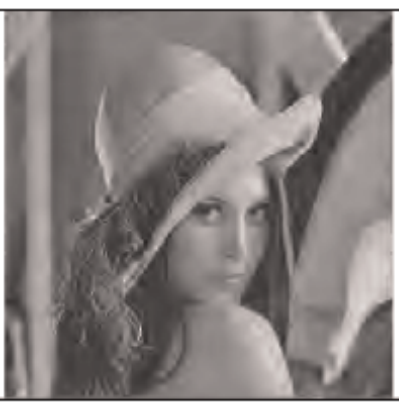 | 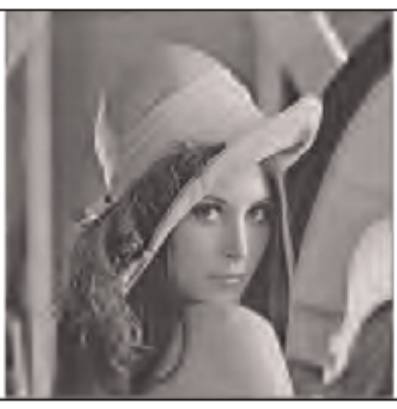 | 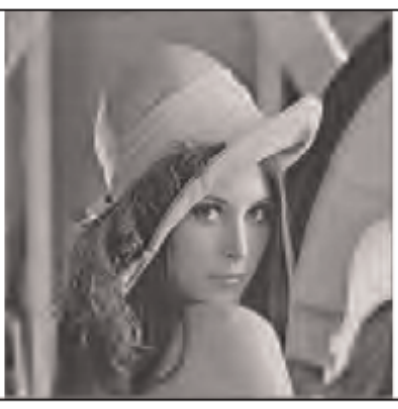 |
| C. REFERENCE CODER SPIHT | REFERENCE CODER SPIHT | C. FOVEATION CODER EFIC | VISUAL CODER EVIC |
| 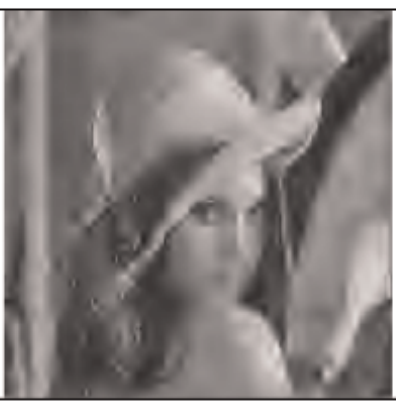 | 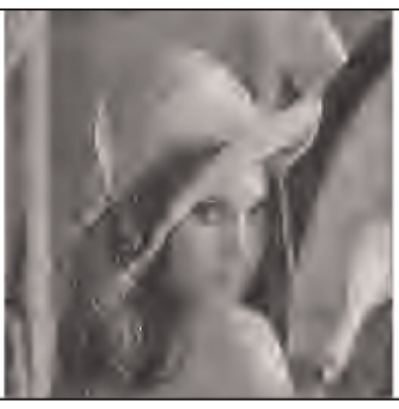 | 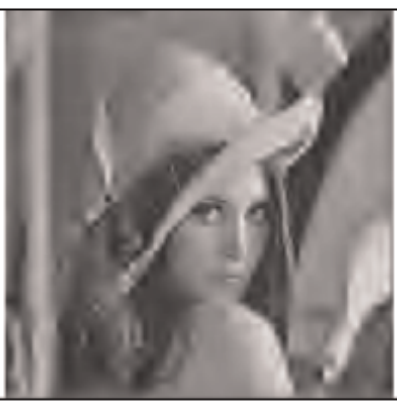 | 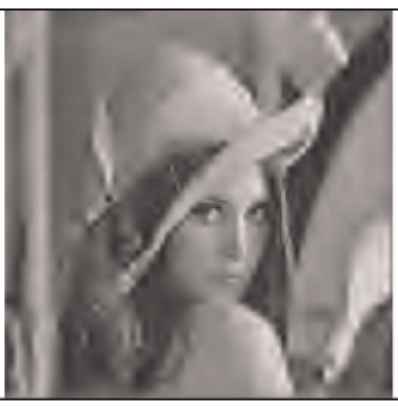 |
| D. REFERENCE CODER SPIHT | REFERENCE CODER SPIHT | D. FOVEATION CODER EFIC | VISUAL CODER EVIC |

**Table 2.**
*The watermarking Lena SPIHT coding images (column 1) versus the SPIHT Dewatermarking QR code images (column 2) and watermarking Lena EFIC foveal coding images (column 3) versus the EVIC version Dewatermarking QR coding (column 4) with their quality scores PS using visual WVDP assessor given for varying bit rate and fixed observation distance. The bit rate varies as follow: A. 0.25 bpp, B. 0.15 Bpp, C. 0.0625 bpp, D.0.0313 bpp.*

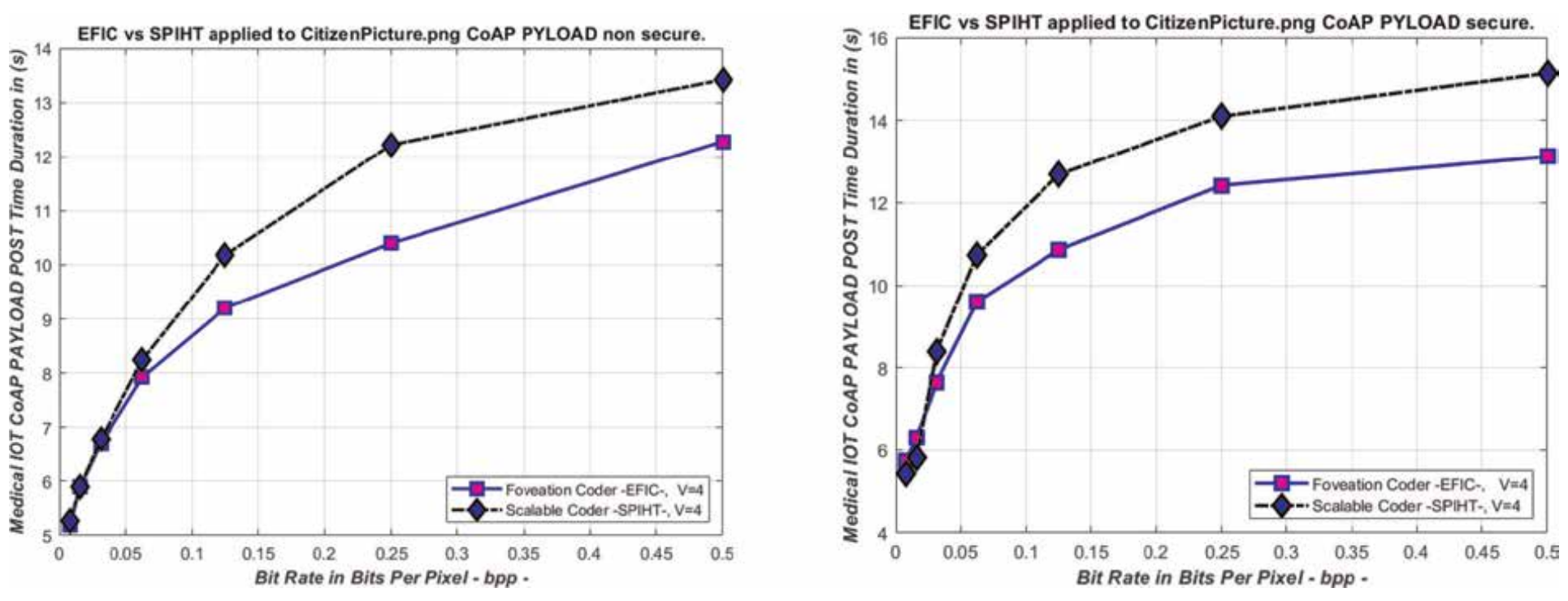

**Figure 7.**
*EFIC vs. SPIHT applied to non-secure CoAP CitzenPicture.Png payload.*

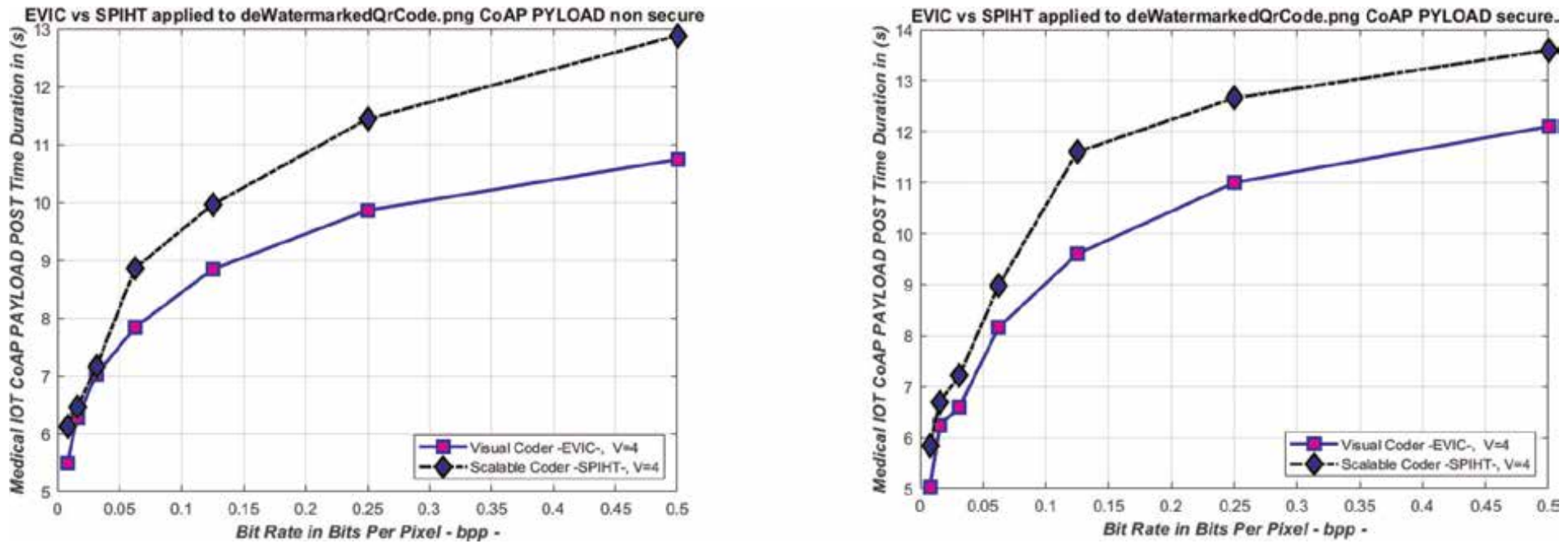

**Figure 8.**
*EVIC vs. SPIHT applied to non-secure CoAP deWatermarkedQrCode.Png payload.*

| Bit Rate (bpp) | 128:1 | 64:1 | 32:1 | 16:1 | 8:1 | 4:1 | 2:1 |
|---|---|---|---|---|---|---|---|
| EFIC vs. SPIHT non secure(%) | 5.51 | 8.24 | 8.83 | 8.75 | 14.35 | 6.33 | 13.34 |
| EFIC vs. SPIHT secure(%) | 1.14 | 0 | 1.17 | 3.88 | 9.62 | 14.89 | 8.49 |

**Table 3.**
*Execution time gain percent EFIC vs. SPIHT non secure for CitzenPicture.Png and secure version with ECC.*

| Bit Rate (bpp) | 128:1 | 64:1 | 32:1 | 16:1 | 8:1 | 4:1 | 2:1 |
|---|---|---|---|---|---|---|---|
| EVIC vs. SPIHT non secure(%) | 14.04 | 6.86 | 8.58 | 9.13 | 23.96 | 22.27 | 17.20 |
| EVIC vs. SPIHT secure(%) | 10.13 | 2.78 | 1.95 | 8.85 | 9.83 | 15.55 | 21.27 |

**Table 4.**
*Execution time gain percent EVIC vs. SPIHT for Dewatermarked QR code non secure and its secure version with ECC.*

So, from **Tables 1** and **2**, we can distinguish that when the binary budget increases we obtain images with good quality whatever the coding. Thus, the psychovisual coding is the best codage compared to SPIHT in terms of quality.

Based on the subjective study illustrated in **Figure 5**, we have curves that present the bite rate execution time consumption for Citizen Picture (Lena) for EFIC versus SPIHT coding in both secured with ECC and unsecured versions. And **Figure 6**, presents the EVIC versus SPIHT coding for the deWatermarked QrCode. From these two figures, we can see that EFIC for the Lena test and EVIC for Dewatermaking consume little time compared to SPIHT.

Let us talk about the security, we can notice that the secured version in both figures is more consuming compared to the non-secured version which is normal when we add the security layer to secure our images but this does not prevent us from saying that ECC does not consume much time if we compare it with non-secured and other security algorithms.

Finally, moving to the quantitative study presented in **Tables 3** and **4**, we have the percentage gain in terms of execution time for CitizenPicture in the non-secure and secure version (**Table 3**), and the gain in execution time EVIC versus SPIHT for Dewatermarked QrCode non-secure and secure (**Table 4**). We notice that the execution is faster when the binary budget is lower than 16:1. So from all these results, we can say that even if the execution time increases when the binary budget is higher than 16:1 we obtain images with good quality but when the budget decreases the quality of the images becomes mediocre and weak.

## 5. Conclusion

In this work, we have processed the QR code images and CitizenPicture captured at the entrance of public or private spaces by decomposing them through the application of the DWT, the psychovisual weighting model, watermark embedding, and the SPIHT embedded coding/ decoding and for the construction of these images we have used the psychovisual weighting model, watermark extraction and the SPIHT embedded coding/decoding.

These visual optimization techniques, based on human visual cortex usage and quality optimization tests, have been used to develop the performance of the Medical IoT platform. Encouraging results were obtained in terms of reduced execution time, storage space, bandwidth and memory load. Adding the watermark to the ECC made it possible to send the citizen's picture containing the QR Code securely with the CoAP protocol.

The use of this Medical IoT platform will help fight the corona virus pandemic by allowing a large number of people to simultaneously access a space with full security of their private data such as their CitizenPicture and QR Code.

## Author details

Abdelhadi EI Allali[1*†], Ilham Morino[1†], Salma AIT Oussous[1†], Siham Beloualid[1†], Ahmed Tamtaoui[2†] and Abderrahim Bajit[1†]

1 Ibn Tofail University, Laboratory of Advanced Systems Engineering (ISA), National School of Applied Sciences, Kenitra, Morocco

2 SC Department, Mohammed V University, Laboratory of Advanced Systems National Institute of Posts and Telecommunications, Rabat, Morocco

*Address all correspondence to: aelallali@gmail.com

† These authors contributed equally.

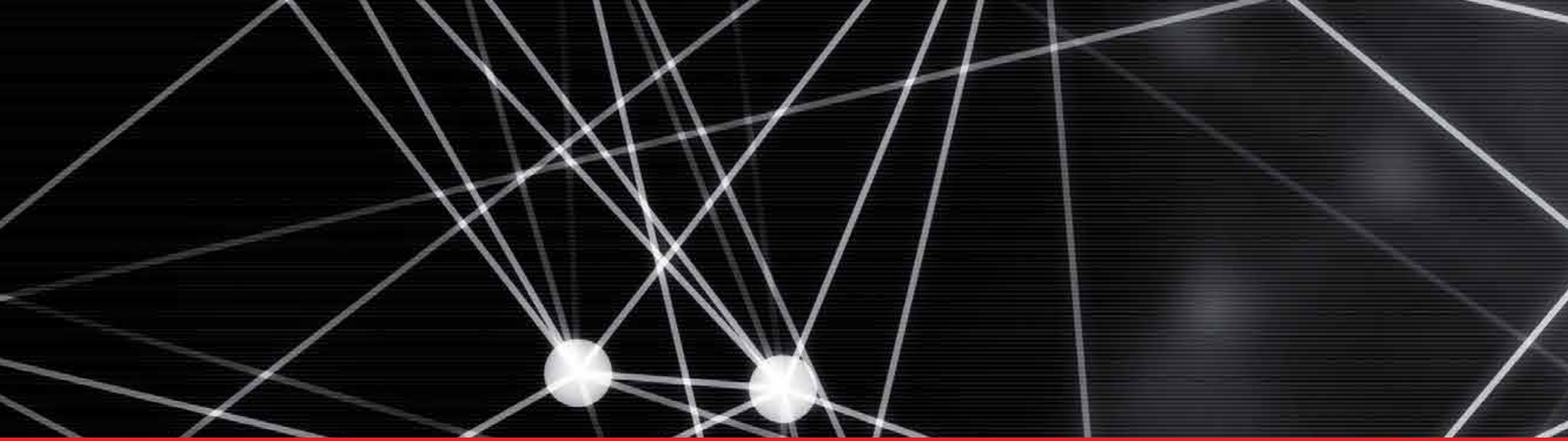

*Edited by Jaydip Sen and Joceli Mayer*

In the era of generative artificial intelligence and the Internet of Things (IoT) and while there is explosive growth in the volume of data and the associated need for processing, analysis, and storage, several new challenges have arisen in identifying spurious and fake information and protecting the privacy of sensitive data. This has led to an increasing demand for more robust and resilient schemes for authentication, integrity protection, encryption, non-repudiation, and privacy preservation of data. This book presents some of the state-of-the-art research in the field of cryptography and security in computing and communications. It is a useful resource for researchers, engineers, practitioners, and graduate and doctoral students in the field of cryptography, network security, data privacy issues, and machine learning applications in the security and privacy in the context of the IoT.

IntechOpen

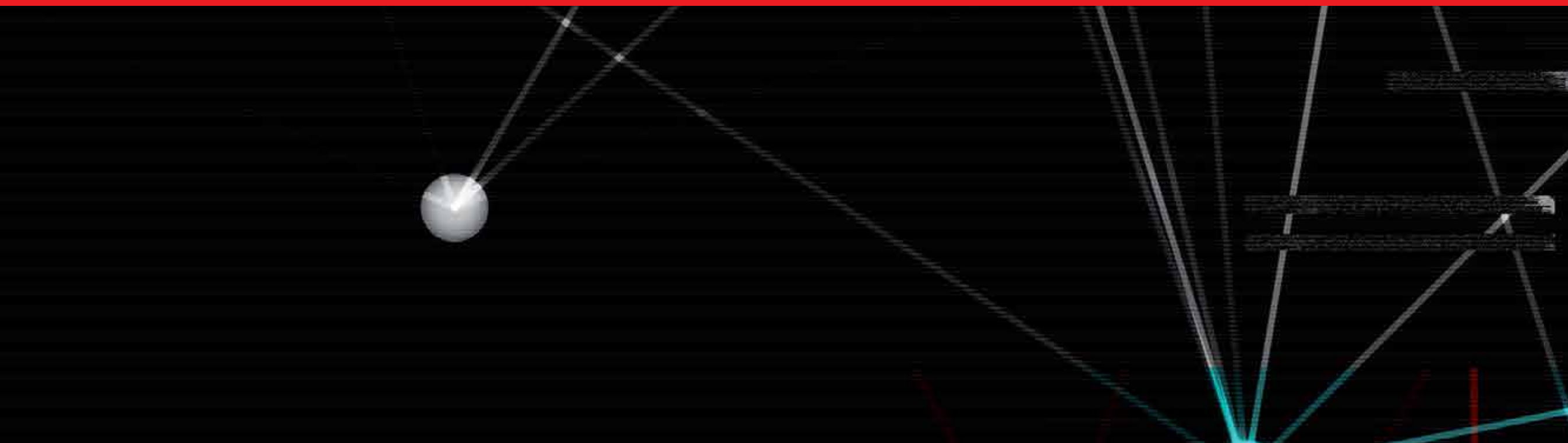